\documentclass[11pt,a4paper]{article}
\pdfoutput=1
\usepackage{jheppub}
\DeclareMathOperator{\tr}{tr} 
\DeclareMathOperator{\Spin}{Spin} 
\DeclareMathOperator{\SL}{SL} 
\DeclareMathOperator{\Li}{Li} 

\title{Q-balls of Quasi-particles in a $(2,0)$-theory model of the Fractional Quantum Hall Effect}

\author[a]{Ori~J.~Ganor,}
\author[b]{Yoon~Pyo~Hong,}
\author[a]{Nathan~Moore,}
\author[a]{Hao-Yu~Sun,}
\author[c]{Hai~Siong~Tan,}
\author[a]{and Nesty~R.~Torres-Chicon}

\affiliation[a]{
Department of Physics,
  University of California,\\
Berkeley, CA 94720, U.S.A.}
\affiliation[b]{School of Physics, 
Korea Institute for Advanced Study,\\
Seoul 130-722, Korea}
\affiliation[c]{
Institute for Theoretical Physics, University of Amsterdam,\\
Science Park 904, Postbus 94485, 1090 GL, Amsterdam, The Netherlands
}
\emailAdd{origa@socrates.berkeley.edu}
\emailAdd{yph@kias.re.kr}
\emailAdd{nmoore@socrates.berkeley.edu}
\emailAdd{hkdavidsun@berkeley.edu}
\emailAdd{H.S.Tan@uva.nl} 
\emailAdd{ntorres@berkeley.edu}

\abstract{
A toy model of the fractional quantum Hall effect appears as part of the low-energy description of the Coulomb branch of the $A_1$ $(2,0)$-theory formulated on $(S^1\times\mathbb{R}^2)/\mathbb{Z}_k$, where the generator of $\mathbb{Z}_k$ acts as a combination of translation on $S^1$ and rotation by $2\pi/k$ on $\mathbb{R}^2$. At low energy the configuration is described in terms of a 4+1D Super-Yang-Mills theory on a cone ($\mathbb{R}^2/\mathbb{Z}_k$) with additional 2+1D degrees of freedom at the tip of the cone that include fractionally charged particles. These fractionally charged ``quasi-particles'' are BPS strings of the $(2,0)$-theory wrapped on short cycles. We analyze the large $k$ limit, where a smooth cigar-geometry provides an alternative description. In this framework a W-boson can be modeled as a bound state of $k$ quasi-particles. The W-boson becomes a Q-ball, and it can be described as a soliton solution of Bogomolnyi monopole equations on a certain auxiliary curved space. We show that axisymmetric solutions of these equations correspond to singular maps from $AdS_3$ to $AdS_2$, and we present some numerical results and an asymptotic expansion.
}
\subheader{
Preprint:
KIAS-P14055, UCB-PTH-14/34
}

\begin{document}
\maketitle
\flushbottom

\newcommand{\secref}[1]{\S\ref{#1}}
\newcommand{\figref}[1]{Figure~\ref{#1}}
\newcommand{\appref}[1]{Appendix~\ref{#1}}
\newcommand{\apprefrange}[2]{Appendices~\ref{#1}-\ref{#2}}
\newcommand{\tabref}[1]{Table~\ref{#1}}

\newcommand\rep[1]{{\bf {#1}}} 

\newcommand\SUSY[1]{{${\mathcal{N}}={#1}$}}  
\newcommand\px[1]{{\partial_{#1}}}

\def\be{\begin{equation}}
\def\ee{\end{equation}}
\def\bear{\begin{eqnarray}}
\def\eear{\end{eqnarray}}
\def\nn{\nonumber}

\newcommand\bra[1]{{\left\langle{#1}\right\rvert}} 
\newcommand\ket[1]{{\left\lvert{#1}\right\rangle}} 

\newcommand{\C}{\mathbb{C}}
\newcommand{\R}{\mathbb{R}}
\newcommand{\Z}{\mathbb{Z}}
\newcommand{\CP}{\mathbb{CP}}

\def\SO{{SO}}
\def\SU{{SU}}

\def\Mst{M_{\text{st}}} 
\def\gst{g_{\text{st}}} 
\def\gIIB{g_{\text{IIB}}} 
\def\lst{\ell_{\text{st}}} 
\def\lP{\ell_{\text{P}}} 

\def\xa{{\mathbf{a}}} 
\def\xb{{\mathbf{b}}} 
\def\xc{{\mathbf{c}}} 
\def\xd{{\mathbf{d}}} 

\def\Mf{{\mathbf{M}}} 
\def\lvk{{\mathbf{k}}} 
\def\xR{{R}}
\def\zC{{\mathbf{z}}} 
\def\bzC{{\overline{\zC}}} 
\def\miniC{{S^1_m}} 
\def\gtw{{\gamma}} 
\def\Zf{{Z}} 

\def\gYM{g_{\text{ym}}}

\def\SOzC{{\SO(2)_\zC}} 
\def\Ztwop{{\Z_2'}} 
\def\Ztwoc{{\Z_2''}} 
\def\SORg{{\SO(2)_\gtw}} 
\def\SORp{{\SO(2)_r}} 
\def\SOthreeR{{\SU(2)_R}} 

\def\qzC{{q_\zC}} 
\def\qRg{{q_\gtw}} 
\def\qRp{{q_r}} 
\def\qs{{s}} 
\def\qJ{{q_J}}

\def\Mqp{{M_{\text{\tiny qp}}}} 

\def\Vph{{v}} 
\def\StringVph{{\tilde{V}}} 


\def\zz{{\mathbf{z}}}
\def\zone{{\mathfrak{1}}}
\def\ztwo{{\mathfrak{2}}}
\def\zthree{{\mathfrak{3}}}
\def\bzz{{\overline{\zz}}}
\def\bzone{{\overline{\zone}}}
\def\bztwo{{\overline{\ztwo}}}
\def\bzthree{{\overline{\zthree}}}

\def\sa{{\mathbf{a}}} 
\def\sb{{\mathbf{b}}} 
\def\sc{{\mathbf{c}}} 
\def\sd{{\mathbf{d}}} 

\def\sA{{\mathbf{A}}} 
\def\sB{{\mathbf{B}}} 
\def\sC{{\mathbf{C}}} 
\def\sD{{\mathbf{D}}} 

\def\ten{{\natural}} 

\def\aB{{\mathfrak{a}}} 

\def\tzC{{\tilde{\zC}}} 
\def\cfR{{\widetilde{\alpha}}} 

\def\Cig{{\Upsilon}} 
\def\Mxxr{{\mathcal{W}}} 

\def\zE{{\zeta}} 
\def\bzE{{\overline{\zeta}}} 
\def\rE{{\rho}} 

\def\fK{{K}} 
\def\fL{{L}}
\def\fY{{Y}}
\def\fM{{M}}
\def\fN{{N}}

\def\vx{{\mathbf{p}}} 
\def\vy{{\mathbf{q}}} 

\def\fKL{{A}} 
\def\fKYL{{B}} 

\def\ErnstE{{\mathcal E}} 
\def\bErnstE{{\overline{\ErnstE}}} 
\def\reEE{\Ff}
\def\imEE{\Fpsi}
\def\gE{{\mathcal G}} 
\def\tgE{{\widetilde{\mathcal G}}} 

\def\Ff{{f}} 
\def\Fpsi{{\chi}} 

\def\HBz{{\xi}}
\def\bHBz{{\overline{\HBz}}}
\def\HBM{{\mathbf{M}}}
\def\bHBM{{\overline{\HBM}}}
\def\HBN{{\mathbf{N}}}
\def\bHBN{{\overline{\HBN}}}
\def\HBq{{\mathbf{W}}}
\def\bHBq{{\overline{\HBq}}}

\def\HBg{{\gamma}}

\def\xsF{{F}} 
\def\xsK{{K}} 

\def\bxi{{\overline{\xi}}}

\def\uA{{\mathcal A}} 
\def\uF{{\mathcal F}} 
\def\uD{{\mathcal D}} 

\def\ttR{{\tilde{\mathfrak R}}} 
\def\ttth{{\tilde{\theta}}} 

\def\Eikf{{\tilde{\psi}}} 

\def\tVph{{\tilde{\Vph}}} 
\def\dtorz{{\mathbf{w}}} 
\def\dtorzpr{{\mathbf{u}}} 

\def\impr{{\mathbf{\tilde{r}}}} 
\def\impz{{\mathbf{\tilde{z}}}} 

\def\LFXi{{\Xi}} 

\def\nR{{\mathbf{R}}}
\def\nZ{{\mathbf{Z}}}

\def\TPhi{{\Phi}}
\def\SPsi{{\Psi}}
\def\Ca{a}

\def\su{{\mathbf{u}}}
\def\sv{{\mathbf{v}}}

\def\uv{{\alpha}}
\def\upv{{\beta}}

\def\rAdS{{\mathfrak{r}}}
\def\xAdS{{\mathfrak{u}}}

\def\PSk{{\mathbf{k}}}
\def\PSh{{\mathbf{h}}}
\def\tPSk{{\tilde{\PSk}}}
\def\tPSh{{\tilde{\PSh}}}
\def\lkxk{{\xi}}

\def\Fp{{\mathbf{P}}}
\def\Fq{{\mathbf{Q}}}
\def\Fs{{\mathbf{S}}}
\def\Ft{{\mathbf{T}}}

\def\nXi{{\Xi}}

\def\sqau{{\alpha}}
\def\sqav{{\beta}}
\def\FXuu{{\mathbf{A}}}
\def\FXuv{{\mathbf{B}}}
\def\FXvv{{\mathbf{C}}}

\def\Moduli{{\mathfrak{m}}}

\def\uvpm{{\xi}}
\def\FXp{{\mathbf{G}}}
\def\FXm{{\mathbf{H}}}
\def\wpm{{\mathbf{w}}}

\def\Fsg{{\varphi}}

\def\psG{{\mathfrak{G}}} 

\def\wnu{{\mathfrak{x}}}
\def\wnv{{\mathfrak{y}}}
\def\wn{{\mathfrak{w}}}

\def\rq{{\mathfrak{p}}}
\def\rEq{{\mathfrak{q}}}

\def\uup{{\tilde{x}}}
\def\uum{{\tilde{y}}}
\def\uuz{{\tilde{z}}}

\def\gP{{P}}
\def\gH{{H}}

\def\sx{{\mathbf{x}}}
\def\sy{{\mathbf{y}}}

\def\PolEq{{\Xi}}
\def\Ap{{\mathcal{A}}}
\def\Bd{{\mathcal{B}}}

\def\slp{{\xi}}
\def\slm{{\eta}}

\def\sxy{{\mathbf{p}}}
\def\sxby{{\mathbf{d}}}

\def\wx{{\xi}}
\def\wy{{\eta}}

\def\swp{{\sigma}}
\def\swm{{\tau}}

\def\fT{{\varphi}}
\def\wt{{\mathbf{t}}}

\def\ks{{\sigma}}
\def\kt{{\tau}}
\def\rs{{\zeta}}

\def\kq{{\varsigma}}
\def\invCa{{b}}
\def\vb{{\ell}}

\def\fMZ{{\mathcal{P}}}
\def\gMZ{{\mathcal{S}}}
\def\hMZ{{\mathcal{T}}}
\def\pMZ{{\mathcal{M}}}
\def\qMZ{{\mathcal{K}}}
\def\vMZ{{\mathcal{W}}}

\def\xU{{\mathbf U}}
\def\xV{{\mathbf V}}

\def\EqX{{\mathcal{X}}}

\def\nU{{\mathfrak U}}
\def\nV{{\mathfrak V}}
\def\Id{{\mathbf{I}}}

\def\intA{{{\mathbf A}_{\text{int}}}} 
\def\extA{{\mathbf A}} 

\def\Manif{{\mathcal M}}

\def\ScPhi{{\Phi}}
\def\Cigth{{\overline{\theta}}} 
\def\AdSchi{{\mu}}
\def\AdSth{{\alpha}}

\def\tfMZ{{\tilde{\fMZ}}}
\def\tgMZ{{\tilde{\gMZ}}}
\def\thMZ{{\tilde{\hMZ}}}
\def\tqMZ{{\tilde{\qMZ}}}
\def\tvMZ{{\tilde{\vMZ}}}

\def\Eex{{\mathcal E}} 
\def\Estatic{{\mathcal E}_{\text{stat}}} 
\def\Energy{{\mathcal E}} 

\def\ResPhi{{\widetilde{\Phi}}} 
\def\Ebps{{{\mathcal E}_{\text{BPS}}}} 
\def\lambdaBPS{{\lambda}}
\def\SmoothedR{{{\mathcal R}}}
\def\SmoothedRCb{{\mathcal R}_\Cb}
\def\Cb{{b}}
\def\MaxUVn{{N}}
\def\Uex{{\mathcal U}}
\def\Ubps{{{\mathcal U}_\text{BPS}}}

\def\Fa{{\mathbf{f}}}
\def\Ga{{\mathbf{g}}}

\def\hPS{{H}}
\def\kPS{{K}}
\def\psR{{u}}

\def\tFa{{\widetilde{\Fa}}}

\def\APhiSpace{{\mathcal N}} 
\def\ModSpace{{\mathcal M}} 

\def\tauIsig{{\xi}} 
\def\btauIsig{{\overline{\tauIsig}}} 

\def\prx{x'} 


\section{Introduction}
\label{sec:Intro}

The fractional quantum Hall effect (FQHE) with filling-factor $1/\lvk$ ($\lvk\in\Z$) appears in 2+1D condensed matter systems whose low-energy effective degrees of freedom can be described by the Chern-Simons action
\be\label{eqn:CS}
I = \frac{\lvk}{4\pi}\int \intA\wedge d\intA + \frac{1}{2\pi}\int\extA\wedge d\intA
\,.
\ee
Here, $\extA$ is the electromagnetic gauge field, and $\intA$ is a 2+1D $U(1)$ gauge field that describes the low-energy internal degrees of freedom of the system. It is related to the electromagnetic current by $j = {}^*d\intA.$ 
Excited states of the system may include quasi-particle excitations that are charged under the gauge symmetry associated with $\intA$.  Such quasi-particles with one unit of $\intA$-charge will have $1/\lvk$ electromagnetic charge.

The goal of this paper is to construct an integrally charged particle as a bound state of quasi-particles using a particularly intuitive string-theoretic toy model of the FQHE.
Over the past two decades several realizations of the integer and fractional quantum Hall effects in string theory have been constructed \cite{Brodie:2000yz}-\cite{Lippert:2014XX}. Generally speaking, these constructions engineer the Chern-Simons action \eqref{eqn:CS} as a low-energy effective description of a $(d+2)$-dimensional brane compactified on a $d$-dimensional space, possibly in the presence of suitable fluxes, to yield the requisite $2+1$D effective description.
In the present paper, we will begin by constructing an FQHE model by compactifying the 5+1D $(2,0)$-theory.
Our system is a special case of a general class of $2+1$D theories obtained from the $(2,0)$-theory by taking three of the dimensions to be a nontrivial manifold. (We note that a beautiful framework for understanding such compactifications has been developed in \cite{Dimofte:2011ju}-\cite{Gadde:2014wma}.)
We will focus on a particular aspect of the system which is the dynamics of the quasi-particles that in the condensed-matter system can arise from impurities. As we will see, the quasi-particles and their relationship to the integrally charged particles have a simple geometrical interpretation in terms of the $(2,0)$ theory, as follows. In our construction, the geometry of the extra dimensions will have long $1$-cycles and short $1$-cycles, the short ones being $1/\lvk$ the size of the long ones. The quasi-particles will be realized as BPS strings of the $(2,0)$ theory wound around short $1$-cycles, while the integrally charged particles will be realized as strings wound around long $1$-cycles.

We are especially interested in the limit $\lvk\gg 1$, where the filling fraction becomes extremely small.
This is the strong-coupling limit of the condensed-matter system, and as we will see, our model has a dual description where quasi-particles are elementary and the integrally charged particles can be described as classical solitons, or rather Q-balls, in terms of the fundamental quasi-particle fields. We will show that solutions to the equations of motion describing these solitons correspond to certain singular harmonic maps from $AdS_3$ to $AdS_2$.

The paper is organized as follows.
In \secref{sec:twozcomp} we describe the $(2,0)$ theory setting for our model.
In \secref{sec:quasip} we study the quasi-particles, which are BPS strings, and we calculate their quantum numbers.
In \secref{sec:largek} we study the large $\lvk$ limit and write down the semiclassical action of the system.
In \secref{sec:boundstates} we develop the differential equations that describe the integrally charged particles as solitons of the fundamental quasi-particle fields in the large $\lvk$ limit. We show that they can be mapped to the equations describing a magnetic monopole on a 3D space with metric $ds^2 = x_3^2 (dx_1^2+dx_2^2+dx_3^2).$
In \secref{sec:analysisBPS} we analyze the soliton equations in more detail and show the connection to harmonic maps from $AdS_3$ to $AdS_2$.
The equations are not integrable in the standard sense, and we were unable to solve them in closed form, but we were able to make several additional observations: (i) we present an expansion up to second order in the inverse of the distance from the ``center'' of the solution to the origin; (ii) using a rather complicated transformation we can recast the equations in terms of a single ``potential'' function;  and (iii) we plot an example of a numerical solution. Points (ii)-(iii) are explored in \apprefrange{app:singlePhi}{app:Numerical}.

\section{The $(2,0)$ theory on $(\R^2\times S^1)/\Z_\lvk$}
\label{sec:twozcomp}
Our starting point is the 5+1D $A_1$ $(2,0)$-theory on $\R^{2,1}\times\Mf_3$, where $\R^{2,1}$ is 2+1D Minkowski space and $\Mf_3\simeq (\R^2\times S^1)/\Z_\lvk$ is the flat, noncompact, smooth three-dimensional manifold defined as the quotient of $\R^2\times S^1$ by the isometry that acts as a simultaneous rotation of $\R^2$ by an angle $2\pi/\lvk$, and a translation of $S^1$ by $1/\lvk$ of its circumference. The $A_1$ $(2,0)$-theory is the low-energy limit of either type-IIB on $\R^4/\Z_2$ \cite{Witten:1995zh} or of $2$ M$5$-branes \cite{Strominger:1995ac} (after decoupling of the center of mass variables).
We are interested in the low-energy description of the Coulomb branch of the theory, and in particular in the low-energy degrees of freedom that are localized near the origin of $\R^2$. The fractional quantum Hall effect, as we shall see, naturally appears in this context.  We will now expand on the details.
(See \cite{Witten:1997kz} for a related study of M-theory and type-II string theory in this geometry  and \cite{Bergman:2001rw}-\cite{Dhokarh:2008ki} for the study of effects on other kinds of branes in a similar geometry.)

\subsection{The geometry}
\label{subsec:geom}

The space $\Mf_3$ can be constructed as a quotient of $\R^3$ as follows.
We parameterize $\R^3$ by $x_3, x_4, x_5$ and set $\zC\equiv x_4 + i x_5$.
Then, $\Mf_3$ is defined by the equivalence relation
\be\label{eqn:Mfsim}
(x_3,\,\zC)\sim (x_3 + 2\pi\xR,\, \zC e^{-2\pi i/\lvk})\,,
\qquad\qquad
\text{[defining relation of $\Mf_3$]}
\ee
where $\xR$ is a constant parameter that sets the scale, and $\lvk>1$ is an integer.
The Euclidean metric on $\Mf_3$ is given by
$$
ds^2 =dx_3^2 + dx_4^2 + dx_5^2 = dx_3^2 + |d\zC|^2\,.
$$
For future reference we define the $(2\lvk)^{th}$ root of unity:
\be\label{eqn:omegadef}
\omega \equiv e^{\pi i/\lvk}\,.
\ee
We also  set
$$
\zC = r e^{i\theta}\,,
$$
so that \eqref{eqn:Mfsim} can be written as
\be\label{eqn:Mfsim2}
(x_3,r, \theta)\sim (x_3 + 2\pi\xR, r, \theta-\tfrac{2\pi}{\lvk})\,.
\ee
The $\zC=0$ locus [i.e., the set of points $(x_3,0)$ with arbitrary $x_3$] forms an $S^1$ of radius $\xR$ that we will call {\it the minicircle} and denote by $\miniC.$
The space $\Mf_3\setminus\miniC$ (which is $\Mf_3$ with the minicircle excluded) is a circle-bundle over a cone (with the origin $\{0\}$ excluded):
\be\label{eqn:fib}
\begin{array}{ccc}
S^1 & \longrightarrow & \Mf_3\setminus\miniC \\
       &                  & \downarrow \\
      &                     & \C/\Z_\lvk\setminus\{0\} \\
\end{array}
\ee
The cone $\C/\Z_\lvk$ is parameterized by $\zC$, subject to the equivalence relation $\zC\sim\omega^2\zC$. In polar coordinates the cone is parameterized by $(r,\theta)$ with $0<r<\infty$ and $0\le\theta<2\pi/\lvk.$ ($\theta$ is understood to have period $2\pi/\lvk$ when describing the cone.) The projection $\Mf_3\rightarrow\C/\Z_\lvk$ is given by $(x_3,\zC)\mapsto\zC$. For a given $\zC\neq 0$, the fiber $S^1$ of the fibration \eqref{eqn:fib} over $\zC\simeq\omega^2\zC$ is given by all points $(x_3,\zC)$ with $0\le x_3<2\pi\lvk\xR$. The equivalence \eqref{eqn:Mfsim} then implies $(x_3+2\pi\lvk\xR,\zC)\sim (x_3,\zC)$, and so this $S^1$ has radius $\lvk\xR$.

\vskip 12pt
\begin{figure}[t]
\begin{picture}(400,200)
\put(25,10){\begin{picture}(190,180)
\thicklines
\put(50,180){(a)}
\put(0,0){\vector(0,1){180}}\put(5,170){$x_3$}

\put(0,0){\vector(1,0){190}}\put(170,4){$x_4$}

\thinlines
\multiput(5,30)(15,0){12}{\line(1,0){10}}
\put(180,30){\vector(1,0){10}}

\multiput(0,0)(14,7){11}{\line(2,1){10}}
\put(154,77){\vector(2,1){10}}

\thicklines
\put(0,30){\vector(2,1){164}}

\put (-25,27){$2\pi\xR$}
\put(100,30){\circle*{8}}
\put(89,45){\circle*{8}}

\put(-1,0){\line(0,1){30}}
\put(0,0){\line(0,1){30}}
\put(1,0){\line(0,1){30}}
\put(0,0){\circle*{4}}
\put(0,30){\circle*{4}}

\put(98,-10){$r$}

\thinlines
\put(42,95){\vector(-1,-2){40}}
\put(40,108){\it the}
\put(40,100){\it minicircle}

\qbezier(150,30)(150,65)(130,95)
\put(150,55){$2\pi/\lvk$}
\end{picture}}

\put(228,0){\line(0,1){200}}

\put(235,10){\begin{picture}(190,180)
\thicklines
\put(50,180){(b)}

\put(0,0){\vector(2,1){164}}
\put(0,0){\vector(1,0){190}}
\put(100,0){\circle*{8}}
\thinlines
\qbezier(150,0)(150,35)(130,65)
\put(160,35){$\C/\Z_\lvk$}

\put(100,0){\line(0,1){180}}
\put(100,180){\circle*{8}}
\put(105,100){{\it generic}}
\put(105,90){{\it fiber}}

\end{picture}}

\end{picture}
\caption{
(a) The geometry of $\Mf_3\simeq (\C\times S^1)/\Z_\lvk$:
in the coordinate system $(x_4+i x_5, x_3)$, the point $(r,0)$ is identified with $(r e^{-2\pi i/\lvk},2\pi\xR)$ and $(r, 2\pi\lvk\xR)$; The large dots indicate equivalent points;
(b) The fibration $\Mf_3\rightarrow\C/\Z_\lvk$ with the generic fiber that is of radius $\lvk\xR$.
}
\label{fig:Mf3}
\end{figure}
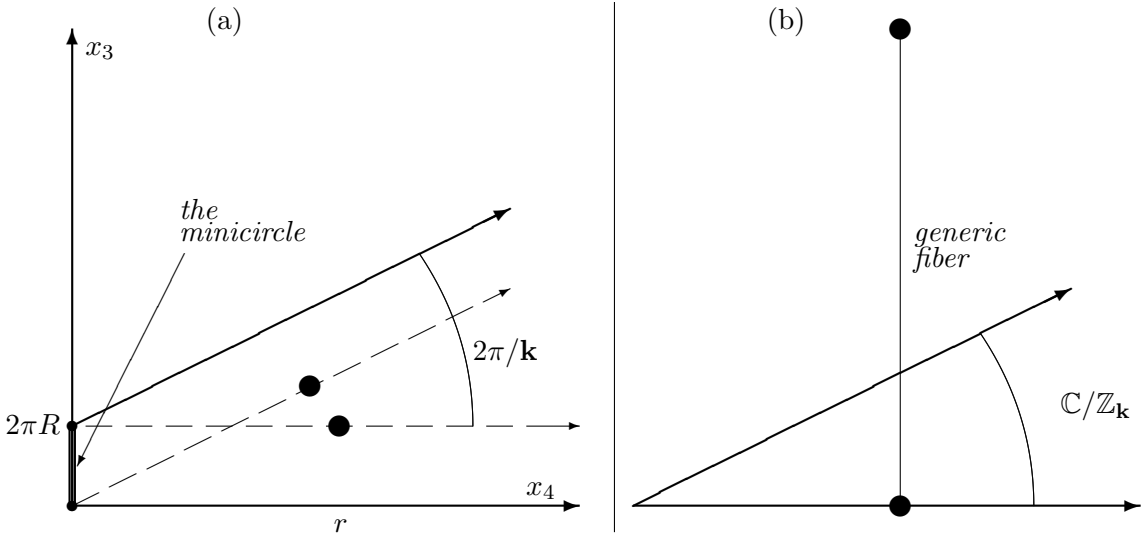

In order to preserve half of the $16$ supersymmetries we augment \eqref{eqn:Mfsim} by an appropriate R-symmetry twist as follows. Let $\Spin(5)\simeq Sp(2)$ be the R-symmetry of the $(2,0)$-theory. In the M$5$-brane realization of the $(2,0)$-theory \cite{Strominger:1995ac}, $\Spin(5)$ is the group of rotations (acting on spinors) in the five directions transverse to the M$5$-branes, which we take to be $6,\dots,10$. We now split them into the subsets $6,7$ and $8,9,10$. This corresponds to the rotation subgroup $[\Spin(3)\times\Spin(2)]/\Z_2\subset\Spin(5)$. Let $\gtw\in\Spin(5)$ correspond to a $2\pi/\lvk$ rotation in the $6,7$ plane. We then augment the RHS of the geometrical identification \eqref{eqn:Mfsim} by an R-symmetry transformation $\gtw$. The setting now preserves $8$ supersymmetries.


We now go to the Coulomb branch of the $(2,0)$-theory by separating the two M$5$-branes of \secref{subsec:geom} in the M-theory direction $x_{10}$. This breaks $\Spin(3)$ to an $\SO(2)$ subgroup (corresponding to rotations in directions $8,9$) which we denote by $\SORp$. 
On the Coulomb branch of the $(2,0)$-theory there is a BPS string whose tension we denote by $\StringVph$. 


At energies $E\ll 1/\lvk\xR$, sufficiently far away from $\miniC$, the dynamics of the $(2,0)$-theory on $\R^{2,1}\times\Mf_3$ reduces to $\SU(2)$ 4+1D Super-Yang-Mills theory on $\R^{2,1}\times(\C/\Z_k)$. The coupling constant is given by
\be\label{eqn:gYM}
\frac{4\pi^2}{\gYM^2} = \frac{1}{\lvk\xR}\,.
\ee
All fields are functions of the coordinates $(x_0, x_1, x_2, r,\theta)$, but the periodicity $\theta\sim\theta+2\pi/\lvk$ is modified in two ways: 
\begin{itemize}
\item
The shift by $2\pi\xR$ in $x_3$, expressed in \eqref{eqn:Mfsim2}, implies that as we cross the $\theta=2\pi/\lvk$ ray a translation by $2\pi\xR$ in $x_3$ is needed in order to patch smoothly with the $\theta=0$ ray. Since $x_3$-momentum corresponds to conserved instanton charge in the low-energy SYM, we find that we have to add to the standard SYM action an additional term
\be\label{eqn:FFonray}
\frac{1}{16\lvk\pi}\int\limits_{\theta=0}\tr(F\wedge F)\,,
\ee
where the integral is performed on the ray at $\theta=0$.
\item
the R-symmetry twist $\gtw$ introduces phases in the relation between values of fields at $\theta=0$ and at $\theta=2\pi/\lvk$.
Of the five (gauge group adjoint-valued) scalar fields $\Phi^6,\dots,\Phi^{10}$ (corresponding to M$5$-brane fluctuations in directions $6,\dots,10$) the last three $\Phi^8,\Phi^9,\Phi^{10}$ are neutral under $\gamma$ and hence periodic in $\theta$, while the combination $\Zf\equiv\Phi^6+i\Phi^7$ satisfies
\be\label{eqn:Zftwbc}
\Zf(x_0,x_1,x_2,r,\theta+\tfrac{2\pi}{\lvk})=\omega^2\Omega^{-1}\Zf(x_0,x_1,x_2,r,\theta)\Omega\,.
\ee
where we have included an arbitrary gauge transformation $\Omega(x_0,x_1,x_2,r)\in\SU(2)$.
The gluinos have similar boundary conditions with appropriate $\exp(\pm\pi/\lvk)$ phases.
\end{itemize}
At the origin, $\zC=0$, which is the tip of the cone $\C/\Z_k$, boundary conditions need to be specified and additional 2+1D degrees of freedom need to be added.
These degrees of freedom and their interactions with the bulk SYM fields are the main focus of this paper and will be discussed in \secref{subsec:FQHE}.
But at this point we can make a quick observation.
When a BPS string of the $(2,0)$-theory wraps the $S^1$ of \eqref{eqn:fib} we get the $W$-boson of the effective 4+1D SYM.
The circle has radius $\lvk\xR$ and so the mass of the $W$-boson is $2\pi\lvk\xR\StringVph$.
On the other hand, the BPS string can also wrap the minicircle $\miniC$ whose radius is only $\xR$. (A similar effect has been pointed out in \cite{Witten:1997kz} in the context of type-IIA string theory on this same geometry.)
The resulting particle in 2+1D has mass $2\pi\xR\StringVph$ which is $1/\lvk$ of the mass of the $W$-boson.
Its charge is also $1/\lvk$ of the charge of the W-boson.
This is our first hint that we are dealing with a system that exhibits a fractional quantum Hall effect (FQHE).
We will soon see that indeed a BPS string that wraps $\miniC$ can be identified with a {\it quasiparticle} of FQHE.

\subsection{Symmetries}
\label{subsec:symm}

Now, let us discuss the symmetries of the theory at a generic point on the Coulomb branch.
The continuous isometries of $\Mf_3$ are generated by translations of $x_3$ and rotations of the $\zC$-plane.
We denote the latter by $\SOzC$ and normalize the respective charge so that the differential $d\zC$ has charge $+1$.
The isometry group of $\Mf_3$ also contains a discrete $\Z_2$ factor generated by the orientation-preserving isometry
$$
(x_3,\zC)\mapsto(-x_3,\bzC).
$$
This by itself does not preserve our setting because it converts the R-symmetry twist $\gtw$ to $\gtw^{-1}$.
To cure this problem, we introduce an extra reflection $x_7\rightarrow -x_7$ in the plane on which $\gtw$ acts, and finally, in order to preserve parity we also introduce one more reflection in a transverse direction, say, $x_{10}\rightarrow -x_{10}$. Altogether, we define the discrete symmetry $\Ztwop$ to be generated by
\be\label{eqn:Ztwop}
(x_0, x_1, x_2, x_3, \zC, x_6, x_7, x_8, x_9, x_{10})\mapsto
(x_0, x_1, x_2, -x_3, \bzC, x_6, -x_7, x_8, x_9, -x_{10})\,.
\qquad\qquad
\text{[$\Ztwop$]}
\ee
The $\SO(2)$ subgroup of the R-symmetry that corresponds to rotations in the $6-7$ plane will be referred to as $\SORg$ and normalized so that $\Phi^6+i\Phi^7$ has charge $+1$. The $\SU(2)=\Spin(3)$ subgroup of the R-symmetry that corresponds to rotations in the $8,9,10$ directions will be referred to as $\SOthreeR$. For future reference we also denote the $\SO(2)$ subgroup of rotations in the $8,9$ plane by $\SORp$.

The parity symmetry of M-theory \cite{Horava:1995qa}, which acts as reflection on an odd number of dimensions combined with a reversal of the $3$-form gauge field ($C_3\rightarrow -C_3$) can also be used to construct a symmetry of our background. We define $\Ztwoc$ as the discrete symmetry generated by the reflection that acts as
$$
x_{10}\rightarrow -x_{10}, \quad
C_3\rightarrow -C_3\,.
\qquad\qquad
\text{[$\Ztwoc$]}
$$
This symmetry preserves the M$5$-brane configuration  and the twist.
We summarize the symmetries in the following table:
\vskip 12pt
\begin{tabular}{ll}
\hline
$\SOzC$ & rotations of the $\zC$ ($x_4-x_5$) plane; \\
$\SORg$ & rotations of the $x_6-x_7$ plane; \\
$\SOthreeR$ & rotations of the $x_8, x_9, x_{10}$ plane; \\
$\SORp$ & rotations of the $x_8, x_9$ plane; \\
$\Ztwop$ & reflection in directions $x_3, x_5, x_7, x_{10}$; \\
$\Ztwoc$ & reflection in direction $x_{10}$ (and $C_3\rightarrow -C_3$); \\
\hline
\end{tabular}
\vskip 12pt
We denote the conserved charges associated with $\SOzC$, $\SORg$, and $\SORp$ by $\qzC$, $\qRg$, and $\qRp$, respectively. These are the spins in the $4-5$, $6-7$, and $8-9$ planes. 
The supersymmetry generators are also charged under these groups, and the background preserves those supercharges for which $\qzC+\qRg=0$. 
These observations will become useful in \secref{sec:quasip}, where we will study the quantum numbers of the quasi-particles.

\subsection{Relation to D$3$-$(p,q)5$-brane systems}
\label{subsec:D3-5}

As we have seen in \secref{subsec:geom},
following dimensional reduction on the $S^1$ fiber of \eqref{eqn:fib}, we get a low-energy description in terms of 4+1D SYM on the cone $\C/\Z_\lvk$, interacting with additional (as yet unknown, but to be described below) degrees of freedom at the tip of the cone (at $x_4=x_5=0$). These additional degrees of freedom are three-dimensional and can be expressed in terms of $SU(2)$ Chern-Simons theory coupled to the IR limit of a $U(1)$ gauge theory with two charged hypermultiplets (with \SUSY{4} supersymmetry in $2+1$D). The latter is the self-mirror theory introduced in \cite{Intriligator:1996ex}, and 
named $T(SU(2))$ by Gaiotto and Witten \cite{Gaiotto:2008ak}. The arguments leading to the identification of the degrees of freedom at the tip of the cone were presented, in a somewhat different but related context, in \cite{Ganor:2010md}.
The idea is to relate the local degrees of freedom of M-theory on the geometry of \secref{subsec:geom} to those of a $(p,q)$ $5$-brane of type-IIB, as originally done in \cite{Witten:1997kz}, and then map our two M$5$-branes to two D$3$-branes, to obtain the problem of two D$3$-branes ending on a $(p,q)$ $5$-brane. This problem was solved in \cite{Gaiotto:2008ak} in terms of $T(SU(2))$ (and see also \cite{Kitao:1998mf} for previous work on this subject, and \cite{Hashimoto:2014vpa} for generalizations with less supersymmetry). The Gaiotto-Witten solution thus also furnishes the solution to our problem. On the Coulomb branch, the gauge part of the system reduces to $U(1)$ Chern-Simons theory interacting with $T(U(1))$, which reproduces \eqref{eqn:CS}.
Although the details of the argument will not be needed for the rest of this paper, we will review them below for completeness.
More details can be found in \cite{Ganor:2010md}.

Our geometry in directions $3,\ldots, 7$ is of the form $(S^1\times\C^2)/\Z_\lvk$, and leads to a $(1,\lvk)$ $5$-brane according to \cite{Witten:1997kz}.  This was demonstrated in \cite{Witten:1997kz} by replacing $\C^2$ with a Taub-NUT space, whose metric can be written as
\be\label{eqn:dsTN}
ds^2 =
\tilde{R}^2\left(1+\frac{\tilde{R}}{2\tilde{r}}\right)^{-1} (dy +\sin^2(\frac{\tilde{\theta}}{2})\,d\tilde{\phi})^2
+\left(1+\frac{\tilde{R}}{2\tilde{r}}\right)
\lbrack d\tilde{r}^2 + \tilde{r}^2 (d\tilde{\theta}^2+\sin^2\tilde{\theta}\,d\tilde{\phi}^2)\rbrack
\,,
\ee
where $y$ is a periodic coordinate with range $0\le y<2\pi.$
We then introduce the $S^1$, parameterized by $x_3$ as in \eqref{eqn:Mfsim}.
The plane $\C$ that appears in \eqref{eqn:Mfsim} is now embedded in the $\C^2$ tangent space of the Taub-NUT space at the origin $\tilde{r}=0$, and is recovered in the limit $\tilde{R}\rightarrow\infty$. In that limit, and with a change of variables $\tilde{r}=r^2/\tilde{R}$, we can identify the $\C$ plane of \eqref{eqn:Mfsim} as a plane at constant $(\tilde{\theta},\tilde{\phi})$ (say $\tilde{\theta}=\pi/2$ and $\tilde{\phi}=0$), and the $\zC\equiv x_4 + i x_5$ coordinate of \eqref{eqn:Mfsim} is identified with 
$$
\zC = r e^{i y} = \sqrt{\tilde{R}\tilde{r}}\, e^{i y}.
$$
In this limit ($\tilde{R}\rightarrow\infty$), the $x_6, x_7$ plane is identified with a plane transverse to the $\zC$-plane, which we can take to be given by $\tilde{\theta}=\pi/2$ and $\tilde{\phi}=\pi$.
We now return to the finite $\tilde{R}$ geometry, and impose the $\Z_\lvk$ equivalence of \eqref{eqn:Mfsim} by setting
$$
(x_3, y, \tilde{r}, \tilde{\theta}, \tilde{\phi}) \sim
(x_3+2\pi\xR, y-\tfrac{2\pi}{\lvk}, \tilde{r}, \tilde{\theta}, \tilde{\phi})\,.
$$
We then wrap two M$5$-branes on the $(\tilde{\theta}=\tfrac{\pi}{2}, \tilde{\phi}=0)$ subspace of this $5$-dimensional geometry. In the limit $\tilde{R}\rightarrow\infty$ this reproduces the setting of \secref{subsec:geom}.

The technique that Witten employed in \cite{Witten:1997kz} is to convert the Taub-NUT geometry to a D$6$-brane by reduction on the $y$-circle from M-theory to type-IIA, and then apply T-duality on the $x_3$-circle to get type-IIB with a complex string coupling constant of the form
$$
\tau_{\text{IIB}} = \frac{2\pi i}{\gIIB} -\frac{1}{\lvk}\,.
$$
This turns out to be strongly coupled ($\gIIB\rightarrow\infty$) in the limit $\tilde{R}\rightarrow\infty$, but it can, in turn, be converted to weak coupling with an $\SL(2,\Z)$ transformation
$$
\tau_{\text{IIB}}\rightarrow\tau_{\text{IIB}}'=\frac{\tau_{\text{IIB}}}{\lvk\tau_{\text{IIB}}+1}
=\tfrac{1}{\lvk}+\frac{i\gIIB}{2\pi\lvk^2}\rightarrow \tfrac{1}{\lvk}+i\infty\,.
$$
As explained in \cite{Witten:1997kz}, the combined transformations convert the Taub-NUT geometry to a $5$-brane of $(p,q)$-type $(1,\lvk)$ (where $\lvk$ is the NS$5$-charge and $1$ is the D$5$-charge). It also converts the M$5$-branes to D$3$-branes.
The boundary degrees of freedom where the two D$3$-branes end on the  $(1,\lvk)$ $5$-brane were found in \cite{Gaiotto:2008ak} as follows. Let $A$ denote the boundary $2+1$D value of the $SU(2)$ gauge field  of the D$3$-branes (with the superpartners left implicit).
Using the identity
$$
\begin{pmatrix} 1 \\ \lvk \\ \end{pmatrix}
=
\begin{pmatrix} 0 & 1 \\ -1 & 0 \\ \end{pmatrix}
\begin{pmatrix} 1 & (-\lvk) \\ 0 & 1 \\ \end{pmatrix}
\begin{pmatrix} 0 \\ 1 \\ \end{pmatrix}\,,
$$
we see that we can obtain a $(1,\lvk)$ $5$-brane from an NS$5$-brane by applying an $\SL(2,\Z)$ transformation that acts as $\tau\rightarrow\tau-\lvk$, followed by another transformation that acts as $\tau\rightarrow -1/\tau$.
Each transformation can be implemented on the boundary conditions. The $\tau\rightarrow\tau-\lvk$ transformation introduces a level-$\lvk$ Chern-Simons theory expressed in terms of an ancillary $SU(2)$ gauge field that we denote by $A'$, and the $\tau\rightarrow -1/\tau$ (S-duality) transformation introduces $2+1$D degrees of freedom, named $T(SU(2))$ by Gaoitto and Witten, that couple to both the $A$ and $A'$ gauge fields. 
$T(SU(2))$ was identified with the Intriligator-Seiberg theory \cite{Intriligator:1996ex} that is defined as the low-energy limit of \SUSY{4} $U(1)$ gauge theory coupled to two hypermultiplets. The theory has a classical $SU(2)$ flavor symmetry (which will ultimately couple to, say, the gauge field $A$), and it also has a $U(1)$ global symmetry under which only magnetic operators are charged, and this symmetry is enhanced to $SU(2)$ in the (strongly coupled) low-energy limit.
This $SU(2)$ is then coupled to $A'$.
It is also not hard to check that $A$ is the $r\rightarrow 0$ limit of the $4+1$D gauge field on the cone.
To see this, consider the $T^2$ formed by varying $(x_3,y)$ for fixed $\tilde{r},\tilde{\theta},$ and $\tilde{\phi}$. The $\SL(2,\Z)$ transformation $\begin{pmatrix} 1 & 0 \\ \lvk & 1 \\ \end{pmatrix}$ converts $1$-cycle from $(0,0)$ to $(2\pi\xR,-2\pi/\lvk)$ into the $1$-cycle from $(0,0)$ to $(2\pi\lvk\xR,0)$, and this is precisely the $1$-cycle used in the reduction from the $(2,0)$-theory to 4+1D SYM.

\subsection{Appearance of the fractional quantum Hall effect}
\label{subsec:FQHE}

On the Coulomb branch the $SU(2)$ gauge group of 4+1D SYM is broken to $U(1)$.
At energies below the breaking scale, the $SU(2)$ gauge fields $A$ and $A'$ reduce to $U(1)$ gauge fields which we denote by $\extA$ and $\intA$. The theory $T(SU(2))$ reduces to $T(U(1))$ which is described by the action \cite{Gaiotto:2008ak} $(1/2\pi)\int\extA\wedge d\intA$. The total gauge part of the action at the tip of the cone is therefore given by \eqref{eqn:CS}.
As we have already seen, the BPS strings that wrap the minicircle $\miniC$ have fractional charge $1/\lvk$ under the bulk $\extA$, which we have now identified as the unbroken $U(1)$ gauge field of the bulk 4+1D SYM.
If we slowly move such a string away from the tip, we get a string that, in the $(x_3,y)$ coordinates of \secref{subsec:D3-5}, wraps the $1$-cycle from $(0,0)$ to $(2\pi\xR,-2\pi/\lvk)$. This implies that it has one unit of charge under $\intA$, which lends credence to the proposal of identifying such a string with a quasi-particle of FQHE.
The quasi-particle is confined to $\R^{2,1}$, because everywhere else a wound string is longer than the BPS bound $2\pi\xR$.

Following the breaking of $SU(2)$ to $U(1)$, the bulk 4+1D $W$-boson gets a mass. The $W$-boson corresponds to a $(2,0)$-string wound around the $S^1$ fiber of \eqref{eqn:fib}, and the homotopy class of the bulk $S^1$ fiber is $\lvk$ times the homotopy class of the minicircle $\miniC$. It is therefore clear that, in principle, we should be able to design a process in which a bulk $W$-boson reaches the tip of the cone and breaks-up into $\lvk$ strings that wrap the minicircle:
\be\label{eqn:Wtokqp}
W\longrightarrow \text{$\lvk$ quasi-particles.}
\ee
Alternatively, it should be possible to describe the $W$-boson as a bound state of $\lvk$ quasi-particles.
In \secref{sec:largek}-\secref{sec:analysisBPS}, we will show how this works in the limit of large $\lvk$.
Before we proceed to the analysis, which is the main focus of our paper, let us compute the spin quantum numbers of the quasi-particles.

\section{Quasi-particles}
\label{sec:quasip}

The quasi-particle is obtained by wrapping the $(2,0)$ BPS string on the minicircle $\miniC$.
Its quantum numbers can be deduced by quantizing the zero-modes of the low-energy fermions that live on the BPS string.
Let us begin by reviewing the low-energy fermionic degrees of freedom on a BPS string.
We assume that the M$5$-branes are in directions $0,\dots,5$, separated in direction $10$, and the BPS string is in direction $x_3$. We first ignore the equivalence \eqref{eqn:Mfsim} and the R-symmetry twist. For simplicity we will now refer to rotation groups as $\SO(m)$ instead of $\Spin(m)$. Thus, the VEV breaks the R-symmetry to $\SO(4)_R\subset\SO(5)_R$, and the presence of the string breaks the Lorentz group down to $\SO(1,1)\times\SO(4)$. We will denote the last factor by $SO(4)_T$, and we will describe representations of  $\SO(1,1)\times\SO(4)_T\times\SO(4)_R$ as
$(\rep{r_1},\rep{r_2},\rep{r_3},\rep{r_4})_{\rep{s}}$,
where $(\rep{r_1},\rep{r_2})$ is a representation of $\SO(4)_T\sim\SU(2)\times\SU(2)$, $(\rep{r_3},\rep{r_4})$ is a representation of $\SO(4)_R\sim\SU(2)\times\SU(2)$, and $\rep{s}$ is an $\SO(1,1)$ charge (spin).
The representation of the unbroken supersymmetry charges is the same as the supersymmetry that is preserved by an M$2$-brane ending on an M$5$-brane. If the M$2$-brane is in directions $0,3,10$ and the M$5$-brane is in directions $0,1,2,3,4,5$ then a preserved SUSY parameter $\epsilon$ satisfies
\be\label{eqn:SUSYM5M2}
\epsilon 
= \Gamma^{03\ten}\epsilon 
= \Gamma^{012345}\epsilon,
\ee
where we denote $\ten\equiv 10$, to avoid ambiguity.
The SUSY parameter therefore transforms as
$$
(\rep{2},\rep{1},\rep{2},\rep{1})_{\rep{+\frac{1}{2}}}
\oplus
(\rep{1},\rep{2},\rep{1},\rep{2})_{\rep{-\frac{1}{2}}}
\,.
$$
On the worldsheet of the BPS string there are $4$ scalars $X^A$ ($A=1,2,4,5$) that correspond to translations of the string in transverse directions. These are in the representation
$(\rep{2},\rep{2},\rep{1},\rep{1})_{\rep{0}}.$
In addition, there are fermions in
\be\label{eqn:bpsstf}
(\rep{2},\rep{1},\rep{1},\rep{2})_{\rep{+\frac{1}{2}}}
\oplus
(\rep{1},\rep{2},\rep{2},\rep{1})_{\rep{-\frac{1}{2}}}
\,.
\ee
Now, consider this theory on $\R^{2,1}\times\Mf_3$ and let the BPS string be at rest at $x_1=x_2=0$. It thus breaks the Lorentz group $\SO(2,1)$ to the rotation group $\SO(2)$ in the $x_1-x_2$ plane, which we denote by $SO(2)_J$. The representations appearing in the brackets of \eqref{eqn:bpsstf} refer to $\SO(4)_T\times\SO(4)_R,$ but in our setting, according to the discussion above, we have to reduce $\SO(4)_T\rightarrow\SO(2)_J\times\SOzC$ and $\SO(4)_R\rightarrow\SORg\times\SORp.$
Thus, denoting representations as
\be\label{eqn:qlegend}
(\qJ,\qzC,\qRg,\qRp)_\qs\,,
\ee
we decompose the left-moving spinors of \eqref{eqn:bpsstf} as
\be\label{eqn:bpsJTRR-L}
(+\tfrac{1}{2},+\tfrac{1}{2},-\tfrac{1}{2},+\tfrac{1}{2}
)_{\rep{+\tfrac{1}{2}}}
\oplus
(+\tfrac{1}{2},+\tfrac{1}{2},+\tfrac{1}{2},-\tfrac{1}{2}
)_{\rep{+\tfrac{1}{2}}}
\oplus
(-\tfrac{1}{2},-\tfrac{1}{2},-\tfrac{1}{2},+\tfrac{1}{2}
)_{\rep{+\tfrac{1}{2}}}
\oplus
(-\tfrac{1}{2},-\tfrac{1}{2},+\tfrac{1}{2},-\tfrac{1}{2}
)_{\rep{+\tfrac{1}{2}}}
\ee
and the right-movers as
\be\label{eqn:bpsJTRR-R}
(+\tfrac{1}{2},-\tfrac{1}{2},+\tfrac{1}{2},+\tfrac{1}{2}
)_{\rep{-\tfrac{1}{2}}}
\oplus
(+\tfrac{1}{2},-\tfrac{1}{2},-\tfrac{1}{2},-\tfrac{1}{2}
)_{\rep{-\tfrac{1}{2}}}
\oplus
(-\tfrac{1}{2},+\tfrac{1}{2},+\tfrac{1}{2},+\tfrac{1}{2}
)_{\rep{-\tfrac{1}{2}}}
\oplus
(-\tfrac{1}{2},+\tfrac{1}{2},-\tfrac{1}{2},-\tfrac{1}{2}
)_{\rep{-\tfrac{1}{2}}}
\ee
These modes can be described by fermionic fields on the string worldsheet, which are functions of $(x_0,x_3).$
To get the quantum numbers of the lowest-energy multiplet we need to find the zero-modes of these fermionic fields.
For that, we need to know the boundary conditions of these fields in the $x_3$ direction.
Due to the rotation by $2\pi/\lvk$ in the $x_4-x_5$ and $x_6-x_7$ planes that were introduced in \secref{subsec:geom}, there are nontrivial phases in the boundary conditions of some of the fields that appear in \eqref{eqn:bpsJTRR-L}-\eqref{eqn:bpsJTRR-R}. The boundary conditions on a field $\psi(x_0, x_3)$ with charges $\qzC$ and $\qRg$ are
\be\label{eqn:psizCRg}
\psi(x_0,x_3+2\pi\xR) = 
\omega^{2(\qzC+\qRg)}
\psi(x_0,x_3).
\ee
The only zero modes are therefore of those modes with $\qzC+\qRg=0.$
These have quantum numbers
\be\label{eqn:zmqn}
(+\tfrac{1}{2},+\tfrac{1}{2},-\tfrac{1}{2},+\tfrac{1}{2}
)_{\rep{+\tfrac{1}{2}}}
\oplus
(-\tfrac{1}{2},-\tfrac{1}{2},+\tfrac{1}{2},-\tfrac{1}{2}
)_{\rep{+\tfrac{1}{2}}}
\oplus
(+\tfrac{1}{2},-\tfrac{1}{2},+\tfrac{1}{2},+\tfrac{1}{2}
)_{\rep{-\tfrac{1}{2}}}
\oplus
(-\tfrac{1}{2},+\tfrac{1}{2},-\tfrac{1}{2},-\tfrac{1}{2}
)_{\rep{-\tfrac{1}{2}}}
\ee
Quantizing these modes gives a multiplet with quantum numbers
\be\label{eqn:zmqnx}
\begin{array}{cllllcllll}
&(\qJ^{(0)}-\tfrac{1}{2},&\qzC^{(0)},&\qRg^{(0)},&\qRp^{(0)}-\tfrac{1}{2}),
 & \qquad
&(\qJ^{(0)},&\qzC^{(0)}+\tfrac{1}{2},&\qRg^{(0)}-\tfrac{1}{2},&\qRp^{(0)}),
\\
&(\qJ^{(0)},&\qzC^{(0)}-\tfrac{1}{2},&\qRg^{(0)}+\tfrac{1}{2},&\qRp^{(0)}) ,
&\qquad
&(\qJ^{(0)}+\tfrac{1}{2},&\qzC^{(0)},&\qRg^{(0)},&\qRp^{(0)}+\tfrac{1}{2}),
\\
\end{array}
\ee
where the charges $\qJ^{(0)}$, $\qzC^{(0)}$, $\qRg^{(0)}$, $\qRp^{(0)}$ still need to be determined.
To determine them, consider the discrete symmetry $\Ztwop$, defined in \secref{subsec:symm}. It preserves the setting and the BPS particle but does not commute with all the charges  $\qJ$, $\qzC$, $\qRg$, $\qRp$ . It acts on the charges as follows:
$$
\qJ\rightarrow \qJ\,,\quad
\qzC\rightarrow -\qzC\,,\quad
\qRg\rightarrow -\qRg\,,\quad
\qRp\rightarrow \qRp\,.\qquad
\qquad\text{[generator of $\Ztwop$]}
$$
 The constants  $\qJ^{(0)}$, $\qzC^{(0)}$, $\qRg^{(0)}$, $\qRp^{(0)}$ must therefore be chosen so that the charges \eqref{eqn:zmqnx} will be invariant, as a set, under $\Ztwop$.  In other words, $\Ztwop$ is allowed to permute the states in  \eqref{eqn:zmqnx}, but must convert an allowed state to an allowed state.
This is only possible if both $\qzC^{(0)}$ and $\qRg^{(0)}$ vanish.
The BPS states are therefore in a multiplet with quantum numbers given by:
$$
(\qJ^{(0)}-\tfrac{1}{2},0,0,\qRp^{(0)}-\tfrac{1}{2})\oplus
(\qJ^{(0)},+\tfrac{1}{2},-\tfrac{1}{2},\qRp^{(0)})\oplus
(\qJ^{(0)},-\tfrac{1}{2},+\tfrac{1}{2},\qRp^{(0)})\oplus
(\qJ^{(0)}+\tfrac{1}{2},0,0,\qRp^{(0)}+\tfrac{1}{2})\,.
$$
Note that the setting of \eqref{eqn:Mfsim2} can be defined for any value of $\lvk$, not necessarily an integer (as suggested in \cite{Witten:1997kz}). 
We can then easily determine $\qJ^{(0)}$ and $\qRp^{(0)}$ in the limit $\lvk\rightarrow\infty$ at which the multiplet must become part of the multiplet of the wrapped string of the $(2,0)$-theory. This determines the charges up to an overall sign (which can be determined arbitrarily and flipped with a parity transformation).
So we pick $\qJ^{(0)}=-\qRp^{(0)}=\tfrac{1}{2}$ and find the following multiplet structure:
\be\label{eqn:zmqn}
(0,0,0,-1)\oplus
(+\tfrac{1}{2},+\tfrac{1}{2},-\tfrac{1}{2},-\tfrac{1}{2})\oplus
(+\tfrac{1}{2},-\tfrac{1}{2},+\tfrac{1}{2},-\tfrac{1}{2})\oplus
(1,0,0,0)\,,
\qquad(\lvk\rightarrow\infty)
\ee
This is as far as we can go with an analysis of the quantum mechanics of the zero modes.
We can do better by considering the full 1+1D low-energy effective action on a string wrapped on the minicircle whose worldsheet is in directions $(x_0, x_3)$. This is a 1+1D CFT of $4$ free bosons together with $4$ free left-moving and $4$ free right-moving fermions in the representations given by \eqref{eqn:bpsJTRR-L}-\eqref{eqn:bpsJTRR-R}. Half of the fermionic fields have twisted boundary conditions with nontrivial phases, according to \eqref{eqn:psizCRg}, and the other half have periodic boundary conditions, whose zero modes we quantized above. The CFT of the $4$ fermionic fields ($2$ left-moving and $2$ right-moving) whose boundary conditions include nontrivial phases has a unique ground state, but quantum corrections lead to corrections to the $\qJ$ and $\qRp$ quantum numbers of this ground state. That, in turn, leads to $\frac{1}{\lvk}$ corrections to the $\qJ$ and $\qRp$ charges, as we will now explain.\footnote{
The $1/\lvk$ correction to the spin discussed below was missed in an earlier version of this paper. We corrected this part of \secref{sec:quasip} following a related observation in \cite{Dedushenko:2014nya}.}
 We recall from basic 1+1D conformal field theory that a free complex left-moving fermion satisfying the boundary condition $\psi(x_0+x_3+2\pi\xR)=e^{2\pi i\nu}\psi(x_0+x_3)$ with $0<\nu<1$, and charged under a global $U(1)$ symmetry such that $\psi$ has charge $q$ and $\overline{\psi}$ has charge $-q$, has a unique ground state with charge $(\tfrac{1}{2}-\nu)q$. For a right-moving fermion with boundary condition $\psi(x_0-x_3-2\pi\xR)=e^{2\pi i\nu}\psi(x_0-x_3)$ the ground state charge is $(\nu-\tfrac{1}{2})q$. For $\nu=0$ (periodic Ramond-Ramond boundary conditions) there are two ground states with charge $\pm\tfrac{1}{2}q$. The charge assignments of the fermions were calculated in \eqref{eqn:bpsJTRR-L}-\eqref{eqn:bpsJTRR-R}. We set $q=\qRp$ or $q=\qJ$ and according to \eqref{eqn:psizCRg}, we need to set $\nu=\tfrac{1}{\lvk}(\qzC+\qRg)$. The bosonic fields with twisted boundary conditions have neither $\qRp$ nor $\qJ$ charge, and so do not contribute to the ground state charge. Combining the modes in \eqref{eqn:bpsJTRR-L}-\eqref{eqn:bpsJTRR-R}, we find that the left-moving sector of the CFT has ground states of $\qJ$ charge $\pm\tfrac{1}{4}+\tfrac{1}{2}(\tfrac{1}{2}-\tfrac{1}{\lvk})$ and the right-moving sector has ground states of $\qJ$ charge $\pm\tfrac{1}{4}-\tfrac{1}{2}(\tfrac{1}{\lvk}-\tfrac{1}{2})$.
For $\qRp$ we find that the left-moving sector of the CFT has ground states of charge $\pm\tfrac{1}{4}-\tfrac{1}{2}(\tfrac{1}{2}-\tfrac{1}{\lvk})$ and the right-moving sector has ground states of charge $\pm\tfrac{1}{4}+\tfrac{1}{2}(\tfrac{1}{\lvk}-\tfrac{1}{2})$. Altogether, we find the quantum-corrected quasi-particle quantum numbers:
\be\label{eqn:zmqn}
(-\tfrac{1}{\lvk},0,0,-1+\tfrac{1}{\lvk})\oplus
(+\tfrac{1}{2}-\tfrac{1}{\lvk},+\tfrac{1}{2},-\tfrac{1}{2},-\tfrac{1}{2}+\tfrac{1}{\lvk})\oplus
(+\tfrac{1}{2}-\tfrac{1}{\lvk},-\tfrac{1}{2},+\tfrac{1}{2},-\tfrac{1}{2}+\tfrac{1}{\lvk})\oplus
(1-\tfrac{1}{\lvk},0,0,\tfrac{1}{\lvk})\,,
\ee


As a corollary, we can immediately restrict the types of processes described in \eqref{eqn:Wtokqp}.
Let us write down the  $\qJ$, $\qzC$, $\qRg$, and $\qRp$ quantum numbers of the $W$-boson supermultiplet.
The bosons (vectors and scalars) are in
\be\label{eqn:WmultB}
(\pm 1, 0, 0, 0)\oplus
(0,\pm 1, 0, 0)\oplus
(0,0,\pm 1, 0)\oplus
(0,0,0,\pm 1)\,.
\ee
and the gluinos are in
\be\label{eqn:WmultF}
(\pm\tfrac{1}{2},\pm\tfrac{1}{2},\pm\tfrac{1}{2},\pm\tfrac{1}{2})\,.\qquad
\text{[even number of $(-\tfrac{1}{2})$'s]}
\ee
Starting with, say, a $W$-boson with charges $(-1,0,0,0)$, consider a process such as
\be\label{eqn:W2kqp}
\text{$W$-boson}\longrightarrow\text{$\lvk$ quasi-particles.}
\ee
By examining $\qRp$ charge conservation, we see that
out of the $\lvk$ quasi-particles either (i) $(\lvk-1)$ quasi-particles are of charge $(1-\tfrac{1}{\lvk},0,0,\tfrac{1}{\lvk})$, and one is of charge $(-\tfrac{1}{\lvk},0,0,-1+\tfrac{1}{\lvk})$, or (ii) $(\lvk-2)$ are of charge $(1-\tfrac{1}{\lvk},0,0,\tfrac{1}{\lvk})$, one is of charge $(+\tfrac{1}{2}-\tfrac{1}{\lvk},+\tfrac{1}{2},-\tfrac{1}{2},-\tfrac{1}{2}+\tfrac{1}{\lvk})$, and one is of charge $(+\tfrac{1}{2}-\tfrac{1}{\lvk},-\tfrac{1}{2},+\tfrac{1}{2},-\tfrac{1}{2}+\tfrac{1}{\lvk})$. Therefore, examining the $\qJ$ charge, we see that $(\lvk-1)$ units of orbital angular momentum need to convert into spin. We therefore expect that if the typical product quasi-particle's velocity $u$ in the $x_1-x_2$ plane is small, the amplitude will be suppressed by a factor of $u^{\lvk-1}$.

The process \eqref{eqn:W2kqp} also suggests that the $W$ boson can be viewed as a bound state of $\lvk$ quasi-particles.
This is similar to the well-known result in FQHE theory that in some contexts the electron can be regarded as a bound state of $\lvk$ fractionally charged edge-states. The edge-states are the low-energy excitations of the Chern-Simons theory that reside on the boundary, or on impurities in the bulk.
In this analogy, our quasi-particles correspond to external impurities that couple to the Chern-Simons theory gauge field. The fractional corrections of $\frac{1}{\lvk}$ that we found for the spin of the quasi-particles are consistent with the well-known anyonic properties of quasi-particles of the FQHE.

Our goal is to develop a concrete description of the $W$-boson as a composite of $\lvk$ quasi-particles.
For this purpose we will first need to switch to a dual formulation of the low-energy theory whereby the quasi-particles are fundamental.

\section{The large $\lvk$ limit}
\label{sec:largek}

A weakly-coupled dual formulation of our system can be constructed in the limit $\lvk\rightarrow\infty$.
In FQHE terminology, this is the {\it small filling fraction} regime which in ordinary systems corresponds to very strong interactions.
More insight can be gained in this limit by choosing a different fibration structure for $\Mf_3$ than the one represented in \eqref{eqn:fib}. While \eqref{eqn:fib} is convenient to work with, because the fibers are of constant size and are geodesics, the fibration is singular at the origin $\zC=0$ --- indeed the tip of the cone is singular, and the fiber over $\zC=0$ is smaller by a factor of $\lvk$ from the generic one.

Instead, in this section we will represent $\Mf_3$ as a smooth fibration in another way.
The base is the well-known cigar geometry and the fiber corresponds to a loop at constant $|\zC|$.
(See also \cite{Nekrasov:2010ka,Hellerman:2012rd} for other uses of this technique.)
We will then reduce the $(2,0)$-theory to $4+1$D SYM along this fiber.
The fiber's size varies and the base's geometry is curved, but nevertheless this representation is very useful, as we shall see momentarily.
(See for example \cite{Kim:2011mv,Linander:2011jy} for recent discussions of dimensional reductions of this type.)

\subsection{Cigar geometry}
\label{subsec:cigar}

To arrive at the alternative fibration we change variables on $\Mf_3$ from $(x_3,\zC)$  to $x_3$ and
\be\label{eqn:tzC}
\tzC\equiv \exp\left(\frac{ix_3}{\lvk\xR}\right)\zC\equiv r e^{i\Cigth}\,.
\ee

We then write the metric as
\be\label{eqn:CircFib}
ds^2 = dx_3^2 + |d\zC|^2=
\cfR(dx_3-\frac{r^2}{\lvk\xR\cfR}d\Cigth)^2
+dr^2+\cfR^{-1} r^2 d\Cigth^2
\,,\qquad
(\cfR\equiv 1+\frac{r^2}{\lvk^2\xR^2})
\ee
This metric describes a circle fibration over a cigar-like base with metric
\be\label{eqn:dsB}
ds_B^2= dr^2 + \cfR^{-1} r^2 d\Cigth^2 = dr^2 + (\frac{\lvk^2\xR^2 r^2}{\lvk^2\xR^2+r^2})d\Cigth^2
\,.
\ee
We denote the cigar space by $\Cig$.
Note that the cigar-metric is smooth everywhere and for $r\gg\lvk\xR$ it behaves like a cylinder $\R_{+}\times S^1$, where $S^1$ has radius $\lvk\xR$.
The ``global angular form'' of the circle fibration is
\be\label{eqn:gaf}
\chi\equiv dx_3-\frac{r^2}{\lvk\xR\cfR}d\Cigth \equiv dx_3 -\xR\aB\,,
\ee
where we have defined the $1$-form
\be\label{eqn:aBdef}
\aB\equiv \frac{r^2}{\lvk\xR^2\cfR}d\Cigth = (\frac{\lvk r^2}{\lvk^2\xR^2 + r^2})d\Cigth\,.
\ee
In this context, $\aB$ is a $U(1)$ gauge field on the cigar with associated field-strength
$$
d\aB= -\frac{1}{\xR}d\chi=\frac{2\lvk^3\xR^2 r}{(\lvk^2\xR^2 + r^2)^2}dr\wedge d\Cigth\,.
$$
The total magnetic flux of the gauge field $\aB$ is $\int_B d\aB = 2\pi\lvk$.

\begin{figure}[t]
\begin{picture}(400,210)
\put(10,90){\begin{picture}(390,200)
\thicklines
\qbezier(0,0)(0,60)(200,60)
\qbezier(0,0)(0,-60)(200,-60)
\put(200,-60){\line(1,0){150}}
\put(200,60){\line(1,0){150}}

\put(260,0){\vector(1,0){40}}
\put(305,-2){$r$}

\qbezier(260,0)(260,20)(270,40)
\put(270,40){\vector(1,2){5}}
\put(278,47){$\Cigth$}

\thinlines
\qbezier(320,0)(320,20)(330,40)
\qbezier(330,40)(340,60)(350,60)

\qbezier(320,0)(320,-20)(330,-40)
\qbezier(330,-40)(340,-60)(350,-60)

\qbezier(380,0)(380,20)(370,40)
\qbezier(370,40)(360,60)(350,60)

\qbezier(380,0)(380,-20)(370,-40)
\qbezier(370,-40)(360,-60)(350,-60)

\thinlines
\multiput(60,-90)(0,40){5}{\qbezier(0,0)(5,5)(5,10)\qbezier(5,10)(5,15)(0,20)}
\multiput(60,-70)(0,40){4}{\qbezier(0,0)(-5,5)(-5,10)\qbezier(-5,10)(-5,15)(0,20)}
\put(53,95){$r\sim\lvk\xR$}
\put(40,110){curvature scale}

\thinlines
\multiput(220,-90)(0,40){5}{\qbezier(0,0)(5,5)(5,10)\qbezier(5,10)(5,15)(0,20)}
\multiput(220,-70)(0,40){4}{\qbezier(0,0)(-5,5)(-5,10)\qbezier(-5,10)(-5,15)(0,20)}
\put(213,95){$r\sim\lvk^2\xR$}
\put(200,110){$\gYM$ becomes large}

\put(10,60){\vector(1,0){10}}
\put(20,60){\vector(-1,0){10}}
\put(10,58){\line(0,1){4}}
\put(20,58){\line(0,1){4}}
\put( 1,68){$\gYM^2\sim\xR$}

\end{picture}}
\end{picture}
\caption{
The cigar geometry with the typical scales indicated.
The curvature of the cigar sets the length scale $\lvk\xR$, and the $4+1$D SYM coupling constant sets the length scale $\gYM^2$.
}
\label{fig:cigar}
\end{figure}
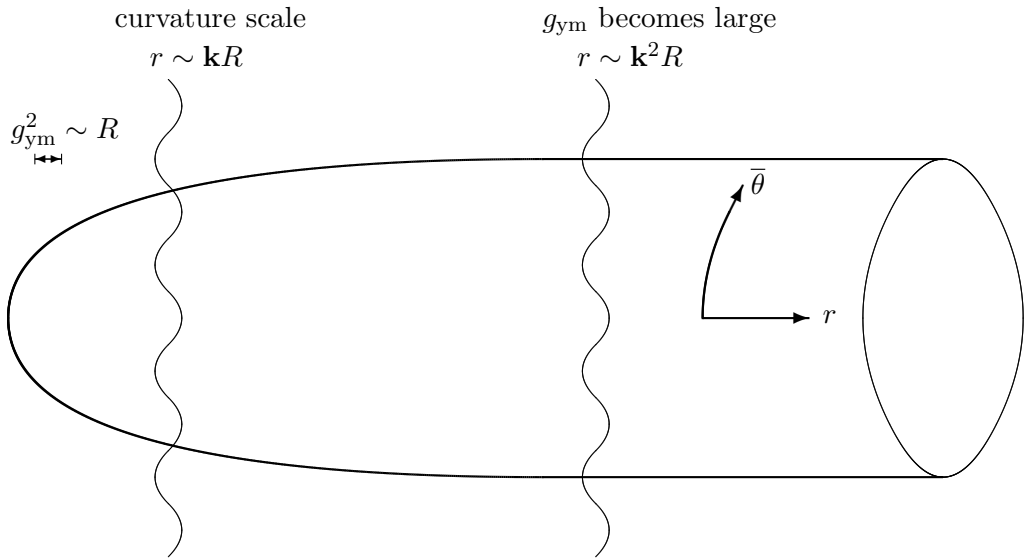

An anti-self-dual field $H=-{}^*H$ on $\Mf_3\times\R^{2,1}$ can be reduced along the fibers of the circle fibration \eqref{eqn:CircFib} to obtain a $4+1$D gauge field strength $f$  on $\Cig\times\R^{2,1}$ as follows:
\be\label{eqn:Hf}
H = \bigl(dx_3-\frac{r^2}{\lvk\xR\cfR}d\Cigth\bigr)\wedge f 
-\cfR^{-\frac{1}{2}}({}^*f).
\ee
Here ${}^*f$ is the 4+1D Hodge dual of the $2$-form $f$ on $\Cig\times\R^{2,1}.$
The coupling constant of the effective $4+1$D super Yang-Mills theory for $f$ is 
\be\label{eqn:gYMcfR}
\gYM^2 = (2\pi)^2\cfR^{1/2}\xR=(2\pi)^2\left(1+\frac{r^2}{\lvk^2\xR^2}\right)^{\frac{1}{2}}\xR\,.
\ee
The coupling constant $\gYM^2$ has dimensions of length and can be compared to the length scale set by the order of magnitude of the curvature of the cigar metric at the origin -- this length-scale is $\lvk\xR$. For $r\sim\lvk\xR$ we find $\gYM^2\ll \lvk\xR$ (in the large $\lvk$ limit), and so the Yang-Mills theory is weakly coupled on length scales of the order of the curvature. The Yang-Mills theory becomes strongly coupled only when the two scales become comparable, which happens for $r\sim\lvk^2\xR$, and therefore for large $\lvk$ our low-energy semi-classical $4+1$D SYM approximation is valid, because the strongly coupled region $r\gg\lvk^2\xR$ is pushed to $r\rightarrow\infty$. The various length scales are depicted in \figref{fig:cigar}.

\subsection{Equations of motion}
\label{subsec:eom}
The bosonic fields of our maximally supersymmetric 4+1D SYM are the $SU(2)$ gauge field and $5$ adjoint-valued scalars.
The scalars correspond to the relative motion in directions $x_6,\dots,x_{10}$ of the M$5$-branes (which become D$4$-branes after dimensional reduction on direction $x_3$).
We will be interested in supersymmetric solutions where only the scalar corresponding to direction $x_{10}$ can be nonzero.
We will therefore ignore the remaining $4$ scalars, as well as the fermions, and we will denote the scalar associated with direction $x_{10}$ by $\ScPhi$.
The boundary conditions at infinity are
$$
\ScPhi\rightarrow\begin{pmatrix} \tfrac{1}{2}\Vph & 0 \\ 0 & -\tfrac{1}{2}\Vph \\ \end{pmatrix}\qquad
\text{(up to a gauge transformation),}
$$
where $\Vph\equiv 2\pi\xR\StringVph$, and $\StringVph$ is the tension of the BPS string defined in \secref{subsec:geom}.

We convert to polar coordinates in the $x_1-x_2$ plane by
\be\label{eqn:rEdef}
\rE\equiv\sqrt{x_1^2+x_2^2}\,,\qquad
x_1 + i x_2 = \rE e^{i\fT}\,.
\ee
The 4+1D SYM theory is therefore formulated on a space with 4+1D metric 
$$
ds^2 = -dt^2 + dr^2 + \cfR^{-1} r^2 d\Cigth^2 + d\rE^2 + \rE^2 d\fT^2\,.
$$
The action contains three terms,
\be\label{eqn:Itot}
I_{\text{bosonic}}=I_\ScPhi + I_{\text{YM}} + I_{\theta}\,,
\ee
where $I_\ScPhi$ is the action of the scalar field, $I_{\text{YM}}$ is the standard Yang-Mills action with variable coupling constant, and $I_{\theta}$ is the 4+1D $\theta$-term that arises due to the nonzero connection $\aB$ [see \eqref{eqn:aBdef}].
We will only consider $\Cigth$-independent field configurations. For such configurations the explicit expressions for the terms in the action are
\bear
I_\ScPhi &=& \tfrac{1}{8\pi^2\xR}\tr\int\bigl\lbrack
(D_0\ScPhi)^2
-(D_\rE\ScPhi)^2
-\tfrac{1}{\rE^2}(D_\fT\ScPhi)^2
-(D_r\ScPhi)^2
\bigr\rbrack
 r \rE dr d\rE d\fT dt
\,,\label{eqn:Iphi}\\
I_{\text{YM}} &=&
\tfrac{1}{8\pi^2\xR}\tr\int
\tfrac{1}{\cfR}
\bigl(
F_{0r}^2
+F_{0\rE}^2
+\tfrac{1}{\rE^2}F_{0\fT}^2
-F_{r\rE}^2
-\tfrac{1}{\rE^2}F_{r\fT}^2
-\tfrac{1}{\rE^2}F_{\rE\fT}^2
\bigr)
 r \rE dr d\rE d\fT dt
\,,\label{eqn:IYM}\\
I_{\theta} &=&
\tfrac{1}{4\pi^2\xR}
\tr\int\tfrac{r^2}{\lvk\xR\cfR}\bigl(
F_{0r}F_{\rE\fT}
-F_{0\rE}F_{r\fT}
+F_{0\fT}F_{r\rE}
\bigr)
dr d\rE d\fT dt\,,
\label{eqn:Itheta}
\eear
where $D_\mu\ScPhi = \partial_\mu\ScPhi + i[A_\mu,\ScPhi]$ is the covariant derivative of an adjoint-valued field.
The equations of motion are
\bear
0 &=&
D^\beta F_{\alpha\beta} + D_r F_{\alpha r} -\tfrac{1}{r} F_{\alpha r} 
-i\cfR\lbrack D_\alpha\ScPhi,\ScPhi\rbrack -(\tfrac{\cfR'}{\cfR})F_{\alpha r}
-\tfrac{\cfR}{2r}(\tfrac{r^2}{\lvk\xR\cfR})'\epsilon_{\alpha\beta\gamma}F^{\beta\gamma}
\,,\label{eqn:EOMf}\\
0 &=& D^\beta F_{r\beta}-i\cfR\lbrack D_r\ScPhi,\ScPhi\rbrack
\,,\label{eqn:EOMfr}\\
0 &=&D^\alpha D_\alpha\ScPhi +D_r D_r\ScPhi +\tfrac{1}{r}D_r\ScPhi
\,,\label{eqn:EOMphi}
\eear
where $\alpha,\beta=0,1,2$ are lowered and raised with the Minkowski metric $ds^2 = -dt^2 + dx_1^2 + dx_2^2 = -dt^2 +d\rE^2 +\rE^2d\fT^2$, the notation $(\cdots)'$ denotes a derivative with respect to $r$, and $\epsilon_{\alpha\beta\gamma}$ is the Levi-Civita tensor.

We note that the term $-i\cfR\lbrack D_\alpha\ScPhi,\ScPhi\rbrack$ in \eqref{eqn:EOMf} leads to a quadratic potential in the $r$ direction for $A_\alpha$, when $\ScPhi$ gets a nonzero VEV. The ground states of this ``harmonic-oscillator'' are the $(\pm 1,0,0,0)$-charged states in \eqref{eqn:WmultB}, which have spin $\pm 1$ in the $x_1 - x_2$ plane. The next term in \eqref{eqn:WmultB}, with charges $(0,\pm 1,0,0)$, describes states with $x_4-x_5$ spin and corresponds to the ground states of the $(A_r, A_\Cigth)$ field components. Note that $A_r$ gets an $r$-dependent potential by a similar mechanism through \eqref{eqn:EOMfr}.
The $A_\Cigth$ component was set to zero in our analysis, so its equation of motion does not appear in \eqref{eqn:EOMf}-\eqref{eqn:EOMphi}.
The remaining terms in \eqref{eqn:WmultB} correspond to excitations of scalar field components that we also set to zero.

\section{Integrally charged particles as bound states of quasi-particles}
\label{sec:boundstates}

We now have two alternative descriptions of the low-energy limit in terms of 4+1D SYM.
In the first description, studied in \secref{sec:twozcomp}, the 4+1D SYM theory is formulated on a cone, with extra degrees of freedom at the tip.
In the second description, studied in \secref{sec:largek}, the 4+1D SYM theory is formulated on a cigar geometry.
The latter description is most suitable in the large $\lvk$ limit, as we have seen at the end of \secref{subsec:cigar}.
The quasi-particles that we studied in \secref{sec:quasip} are the fundamental fields of 4+1D SYM in the cigar-setting.
We have seen that $\lvk$ quasi-particles can form a bound state that is free to move into the bulk of the cone.
Let us now identify this state in the cigar-setting.

\begin{figure}[t]
\begin{picture}(400,210)
\put(10,90){\begin{picture}(390,200)
\thicklines
\qbezier(0,0)(0,60)(200,60)
\qbezier(0,0)(0,-60)(200,-60)
\put(200,-60){\line(1,0){150}}
\put(200,60){\line(1,0){150}}

\put(260,0){\vector(1,0){40}}
\put(305,-2){$r$}

\qbezier(260,0)(260,20)(270,40)
\put(270,40){\vector(1,2){5}}
\put(278,47){$\Cigth$}

\thicklines
\qbezier(220,0)(220,20)(230,40)
\qbezier(230,40)(240,60)(250,60)
\qbezier(220,0)(220,-20)(230,-40)
\qbezier(230,-40)(240,-60)(250,-60)
\thinlines
\put(215,0){\line(1,0){10}}
\put(217,2){\line(1,0){6}}
\put(215,5){\line(1,0){10}}
\put(217,7){\line(1,0){6}}
\put(216,10){\line(1,0){10}}
\put(218,12){\line(1,0){6}}
\put(217,15){\line(1,0){10}}
\put(219,17){\line(1,0){6}}
\put(218,20){\line(1,0){10}}
\put(220,22){\line(1,0){6}}
\put(219,25){\line(1,0){10}}
\put(222,27){\line(1,0){6}}
\put(221,30){\line(1,0){10}}
\put(224,32){\line(1,0){6}}
\put(223,35){\line(1,0){10}}
\put(226,37){\line(1,0){6}}
\put(225,40){\line(1,0){10}}
\put(228,42){\line(1,0){6}}
\put(228,45){\line(1,0){10}}
\put(231,47){\line(1,0){6}}
\put(231,50){\line(1,0){10}}
\put(235,52){\line(1,0){6}}
\put(235,55){\line(1,0){10}}
\put(239,57){\line(1,0){6}}
\put(241,58){\line(1,0){6}}

\put(215,0){\line(1,0){10}}
\put(217,-2){\line(1,0){6}}
\put(215,-5){\line(1,0){10}}
\put(217,-7){\line(1,0){6}}
\put(216,-10){\line(1,0){10}}
\put(218,-12){\line(1,0){6}}
\put(217,-15){\line(1,0){10}}
\put(219,-17){\line(1,0){6}}
\put(218,-20){\line(1,0){10}}
\put(220,-22){\line(1,0){6}}
\put(219,-25){\line(1,0){10}}
\put(222,-27){\line(1,0){6}}
\put(221,-30){\line(1,0){10}}
\put(224,-32){\line(1,0){6}}
\put(223,-35){\line(1,0){10}}
\put(226,-37){\line(1,0){6}}
\put(225,-40){\line(1,0){10}}
\put(228,-42){\line(1,0){6}}
\put(228,-45){\line(1,0){10}}
\put(231,-47){\line(1,0){6}}
\put(231,-50){\line(1,0){10}}
\put(235,-52){\line(1,0){6}}
\put(235,-55){\line(1,0){10}}
\put(239,-57){\line(1,0){6}}
\put(241,-58){\line(1,0){6}}

\thinlines
\qbezier(320,0)(320,20)(330,40)
\qbezier(330,40)(340,60)(350,60)

\qbezier(320,0)(320,-20)(330,-40)
\qbezier(330,-40)(340,-60)(350,-60)

\qbezier(380,0)(380,20)(370,40)
\qbezier(370,40)(360,60)(350,60)

\qbezier(380,0)(380,-20)(370,-40)
\qbezier(370,-40)(360,-60)(350,-60)

\put(185,30){\vector(2,-1){30}}
\put(117,32){thin monopole}
\put(248,-70){$\Ca$}

\end{picture}}
\end{picture}
\caption{
In the limit $\Vph \Ca^2\gg 1$ the soliton is approximately described by the Prasad-Sommerfield solution (of width $1/\Vph \Ca$) near $r=\Ca$ and $x_1=x_2=0$. Note that directions $x_1, x_2$ are not drawn since they are perpendicular to the $r,\theta$ directions.
}
\label{fig:thinring}
\end{figure}
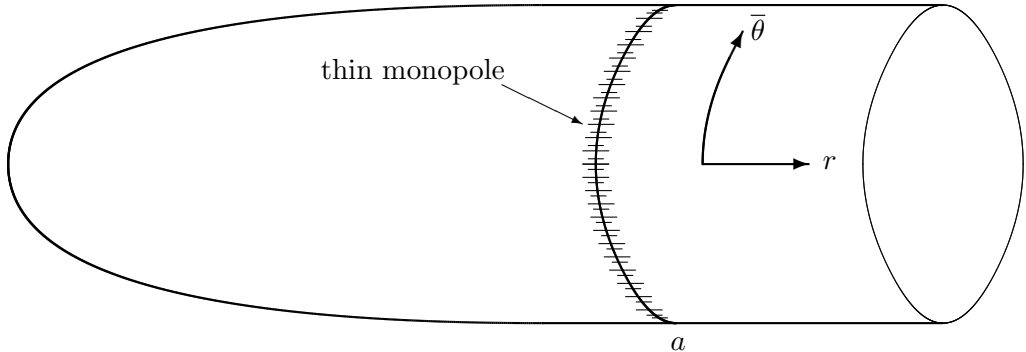

From the perspective of the $(2,0)$-theory, the bound state is a string wrapped on the fiber of \eqref{eqn:fib}. Let us consider such a wrapped string at the cone base point given by coordinates $r=\Ca$ and $\theta=x_1=x_2=0$, with variable $x_3$.
In the cigar variables, this reduces to a string at fixed $r=\Ca$ and $x_1=x_2=0$ but variable $\Cigth$.
Recall that on the Coulomb branch of $SU(2)$ 4+1D SYM, the monopole is a 1+1D object -- a monopole-string. The bound state of $\lvk$ quasi-particles is therefore associated with a monopole-string wrapped around the $\Cigth$-circle of the cigar at $r=\Ca$, as depicted in \figref{fig:thinring}.
Thanks to the $\theta$-term \eqref{eqn:Itheta}, the monopole-string gains $\lvk$ units of charge, as required.

In flat space, a monopole-string is described by the Prasad-Sommerfield solution \cite{Prasad:1975kr}. In our case, the Prasad-Sommerfield solution is a good approximation if the thickness of the monopole is small compared to the typical scale $\lvk\xR$ over which the coupling constant varies, and also small compared to $\Ca$.
In this case, setting
$\dtorz\equiv\sqrt{(r-\Ca)^2+\rE^2}$,
we find the gauge invariant magnitude of the scalar field near the core $r=\Ca$ to be given by \cite{Prasad:1975kr}:
\be\label{eqn:PSsolpr}
|\ScPhi|\equiv\sqrt{2\tr(\ScPhi^2)}  = \tVph\coth(\tVph\dtorz)-\frac{1}{\dtorz}\,,
\ee
where 
$\tVph\equiv(1+\tfrac{\Ca^2}{\lvk^2\xR^2})^{1/2}\Vph$
is the effective VEV of the normalized scalar field $\cfR^{1/2}\ScPhi$ at the core ($r=\Ca$) of the monopole.
The ``thickness'' of the Prasad-Sommerfield solution is of the order of $1/\tVph$, and the condition that the monopole should be ``thin'' becomes $\Ca\gg 1/\Vph$.
If this condition is not met, the Prasad-Sommerfield solution does not provide a good approximation for the particle that corresponds to a $(2,0)$-string wrapped on the generic fiber (of size $\lvk\xR$) of \eqref{eqn:fib}.
Nevertheless, this is a BPS state with charge $\lvk$, which can be described in the large $\lvk$ limit by a soliton solution to the equations of motion \eqref{eqn:EOMf}-\eqref{eqn:EOMphi}. The solution describes a Q-ball, and we expect the position $\Ca$ to be a free parameter. In the next subsection we present the BPS equations that this soliton satisfies.

\subsection{BPS equations}
\label{subsec:BPS}
As we will derive in \secref{subsec:DeriveBPS},
the BPS equations that describe stationary solutions that preserve the same amount of supersymmetry as a $(2,0)$-string wrapped on a fiber of \eqref{eqn:fib} are
\be\label{eqn:phifBPSnon}
D_r\ScPhi =  \tfrac{\lvk\xR}{r}F_{12} = F_{0r}\,,\qquad
D_1\ScPhi =  \tfrac{\lvk\xR}{r}F_{2r} = F_{01}\,,\qquad
D_2\ScPhi =  -\tfrac{\lvk\xR}{r}F_{1r} = F_{02}\,.
\ee
Assuming that $A_r, A_1, A_2$ are time independent, we find
$D_\mu\ScPhi = F_{0\mu}= -D_\mu A_0$ (for $\mu=1,2,r$), which is solved by $\ScPhi=-A_0$.
So the equations are reduced to
\be\label{eqn:Drphi}
D_r\ScPhi =  \tfrac{\lvk\xR}{r}F_{12}\,,\qquad
D_1\ScPhi =  \tfrac{\lvk\xR}{r}F_{2r}\,,\qquad
D_2\ScPhi =  -\tfrac{\lvk\xR}{r}F_{1r}\,,\qquad
\ScPhi=-A_0\,.
\ee
These equations imply the equations of motion \eqref{eqn:EOMf}-\eqref{eqn:EOMphi}.
In fact, for a stationary configuration (all fields are $t$-independent), using the Bianchi identity for the gauge field, we can rewrite the action \eqref{eqn:Itot} as:
\bear
\lefteqn{
I_\ScPhi + I_{\text{YM}} + I_{\theta} =
}\nn\\ &&
\tfrac{1}{8\pi^2\xR}\tr\int\Bigl\{ [A_0,\ScPhi]^2
-\bigl\lbrack
\tfrac{1}{\rE^2} (\tfrac{\lvk\xR\rE}{r}F_{r\rE}-D_\fT\ScPhi)^2
+(\tfrac{\lvk\xR}{r\rE}F_{r\fT}+D_\rE\ScPhi)^2
+(\tfrac{\lvk\xR}{r\rE}F_{\rE\fT}-D_r\ScPhi)^2\bigr\rbrack
\nn\\ &+&
\tfrac{1}{\cfR}\bigl\lbrack
\tfrac{1}{\rE^2}(\tfrac{\lvk\xR\rE}{r}F_{r\rE}-F_{0\fT})^2
+(\tfrac{\lvk\xR}{r\rE}F_{r\fT}+F_{0\rE})^2
+(\tfrac{\lvk\xR}{r\rE}F_{\rE\fT}-F_{0r})^2
\bigr\rbrack
\Bigr\}r\rE dr d\rE d\fT
\nn\\
&+&
\tfrac{\lvk}{4\pi^2}\int\bigl\{
\partial_\rE\tr\bigl\lbrack F_{r\fT}(\ScPhi+A_0)\bigr\rbrack
+\partial_\fT\tr\bigl\lbrack F_{\rE r}(\ScPhi+A_0)\bigr\rbrack
+\partial_r\tr\bigl\lbrack F_{\fT\rE}(\ScPhi+A_0)\bigr\rbrack
\bigr\}
dr d\rE d\fT\,.
\nn\\ &&
\label{eqn:IsumSquares}
\eear
The expressions of the form $(\cdots)^2$ on the $2^{nd}$ and $3^{rd}$ lines of \eqref{eqn:IsumSquares} are squares of combinations that vanish if \eqref{eqn:Drphi} holds, while the $4^{th}$ line is a total derivative, so a configuration that satisfies \eqref{eqn:Drphi} is therefore a saddle point of the action.

The nonzero $A_0$ in the solution \eqref{eqn:Drphi} is consistent with the configuration being a Q-ball \cite{Coleman:1985ki}. $A_0$ can be gauged away at the expense of creating time-varying phases for the other fields, but we will not do so.

We can rewrite the first three equations of \eqref{eqn:Drphi} as the Prasad-Sommerfield \cite{Prasad:1975kr} equations
\be\label{eqn:DiphiF}
D_i\ResPhi = B_i\,,
\ee
where
\be\label{eqn:defResPhiB}
\ResPhi\equiv\tfrac{1}{\lvk\xR}\ScPhi\,,\qquad
B_i\equiv \tfrac{1}{2\sqrt{g}}g_{ij}\epsilon^{jkl}F_{kl}\,,
\ee
are defined on a $3$D auxiliary space $\Mxxr$ parameterized by $x_1,x_2,r$, with metric $g_{ij}$ given by
\be\label{eqn:effds}
ds^2 =g_{ij} dx^i dx^j
=r^2(dr^2 + dx_1^2 + dx_2^2)
=r^2(dr^2 + d\rE^2 + \rE^2 d\fT^2).
\ee
In \secref{subsec:hmapsAdS} we will show that the problem of finding an axisymmetric ($\fT$-independent) BPS soliton can be converted to the problem of finding a harmonic map from the $AdS_3$ space with metric
$$
ds^2 = \frac{1}{r^2}(dr^2 + d\rE^2 + \rE^2d\fT^2)
$$
to $AdS_2$, with a certain singular behavior along a Dirac-like string at $\rE=0$ and $0<r<\Ca$.

\subsection{Energy}
\label{subsec:Energy}
The energy of a general solution of the equations of motion [not necessarily stationary and not necessarily obeying \eqref{eqn:Drphi}] is given by
\bear
\Energy &=&
\tfrac{1}{8\pi^2\xR}\tr\int\bigl\lbrack (D_0\ScPhi)^2 + (D_r\ScPhi)^2 + (D_\rE\ScPhi)^2 +\tfrac{1}{\rE^2}(D_\fT\ScPhi)^2
\bigr\rbrack r\rE dr d\rE d\fT
\nn\\
&+&
\tfrac{1}{8\pi^2\xR}\tr\int\cfR^{-1}\bigl\lbrack
F_{0r}^2 + F_{0\rE}^2 +\tfrac{1}{\rE^2}F_{0\fT}^2 + F_{r\rE}^2 +\tfrac{1}{\rE^2}F_{r\fT}^2 + \tfrac{1}{\rE^2}F_{\rE\fT}^2
\bigr\rbrack r\rE dr d\rE d\fT\,.
\eear
Using the equations of motion \eqref{eqn:EOMf}-\eqref{eqn:EOMphi}, it is not hard to check that if $\Pi_\Phi$, $\Pi_{A_r}$, $\Pi_{A_\rE}$, and $\Pi_{A_\fT}$ are the canonical momenta dual to the fields $\Phi$, $A_r$, $A_\rE$, $A_\fT$, then the Hamiltonian is related to $\Energy$ by a total derivative:
\bear
\lefteqn{
H \equiv \tr\int\left(\Pi_\Phi\partial_0\Phi + \Pi_{A_r}\partial_0 A_r + \Pi_{A_\rE}\partial_0 A_\rE + \Pi_{A_\fT}\partial_0 A_\fT\right) dr d\rE d\fT 
-I_\ScPhi -I_{\text{YM}} -I_{\theta}
}
\nn\\ &=&
\Energy + \tfrac{1}{4\pi^2\xR}\tr\int\Bigl\{
\partial_\rE\bigl\lbrack\tfrac{r\rE}{\cfR}A_0\bigl(F_{0\rE}-\tfrac{r}{\rE\lvk\xR}F_{r\fT}\bigr)\bigr\rbrack
+\partial_r\bigl\lbrack\tfrac{r\rE}{\cfR}A_0\bigl(F_{0r}+\tfrac{r}{\rE\lvk\xR}F_{\rE\fT}\bigr)\bigr\rbrack
\nn\\ && 
+\partial_\fT\bigl\lbrack\tfrac{r}{\rE\cfR}A_0\bigl(F_{0\fT}+\tfrac{r\rE}{\lvk\xR}F_{r\rE}\bigr)\bigr\rbrack
\Bigr\} dr d\rE d\fT\,.
\label{eqn:Hamiltonian}
\eear
For a stationary configuration that satisfies the equations of motion and also satisfies $A_0=-\ScPhi$, the energy can be written as a sum of squares of the BPS equations plus total derivatives:
\bear
\Estatic &=&
\tfrac{1}{8\pi^2}\tr\int\tfrac{1}{\cfR}
\bigl\lbrack
(F_{r\rE}-\tfrac{r}{\lvk\xR\rE}D_\fT\ScPhi)^2
+\tfrac{1}{\rE^2}(F_{r\fT}+\tfrac{r\rE}{\lvk\xR}D_\rE\ScPhi)^2
+\tfrac{1}{\rE^2}(F_{\rE\fT}-\tfrac{r\rE}{\lvk\xR}D_r\ScPhi)^2
\bigr\rbrack
 r \rE dr d\rE d\fT
\nn\\ &+&
\tfrac{1}{4\pi^2\xR}\tr\int\bigl\{
\partial_\rE\bigl(\tfrac{r\rE}{\cfR}\ScPhi F_{0\rE}\bigr)
+\partial_r\bigl(\tfrac{r\rE}{\cfR}\ScPhi F_{0r}\bigr)
+\partial_\fT\bigl(\tfrac{r}{\rE\cfR}\ScPhi F_{0\fT}\bigr)
\bigr\}  dr d\rE d\fT
\nn\\ &+&
\tfrac{1}{4\pi^2\lvk\xR^2}\tr\int\bigl\{
\partial_\rE\bigl(\tfrac{r^2}{\cfR}\ScPhi F_{\fT r}\bigr)
+\partial_r\bigl(\tfrac{r^2}{\cfR}\ScPhi F_{\rE\fT}\bigr)
+\partial_\fT(\tfrac{r^2}{\cfR}\ScPhi F_{r\rE}\bigr)
\bigr\}  dr d\rE d\fT
\label{eqn:Estatic}
\eear
Equation \eqref{eqn:Estatic} assumes \eqref{eqn:EOMf}-\eqref{eqn:EOMphi}, but not \eqref{eqn:Drphi} (other than $A_0=-\ScPhi$).
The term on the RHS of the first line vanishes when the BPS equations \eqref{eqn:Drphi} are satisfied.
Substituting \eqref{eqn:Drphi} into \eqref{eqn:Estatic}, we find
\bear
\Ebps &=&
\tfrac{1}{4\pi^2\xR}\tr\int\bigl\{
\partial_\rE\bigl(\ScPhi F_{\fT r}\bigr)
+\partial_r\bigl(\ScPhi F_{\rE\fT}\bigr)
+\partial_\fT(\ScPhi F_{r\rE}\bigr)
\bigr\}  dr d\rE d\fT\,,
\label{eqn:Etopo}
\eear
which depends only on the behavior of the fields at infinity and reduces to the VEV $\Vph$ times the magnetic charge of the soliton [regarded as a monopole in the metric \eqref{eqn:effds}].

We note that \eqref{eqn:Drphi} also lead to another set of $2^{nd}$ order differential equations:
\bear
0 &=&
D_n F_{mn} + D_r F_{m r} -\tfrac{1}{r} F_{\alpha r} 
-i\tfrac{r^2}{\lvk^2\xR^2}\lbrack D_m\ScPhi,\ScPhi\rbrack
\,,\label{eqn:EOMfAlt}\\
0 &=& D_n F_{r n}-i\tfrac{r^2}{\lvk^2\xR^2}\lbrack D_r\ScPhi,\ScPhi\rbrack
\,,\label{eqn:EOMfrAlt}\\
0 &=&D_n D_n\ScPhi +D_r D_r\ScPhi +\tfrac{1}{r}D_r\ScPhi
\,,\label{eqn:EOMphiAlt}
\eear
where $m,n=1,2.$
Equations \eqref{eqn:EOMfAlt}-\eqref{eqn:EOMphiAlt} are the stationary equations for a Yang-Mills field $A$, minimally coupled to an adjoint scalar $\ScPhi$, on a space with metric \eqref{eqn:effds}. These equations presumably have additional solutions that do not solve \eqref{eqn:EOMf}-\eqref{eqn:EOMphi}.

\subsection{Derivation of the BPS equations}
\label{subsec:DeriveBPS}
In this subsection we explain how \eqref{eqn:phifBPSnon} was derived.
(The rest of the paper does not rely on this subsection, and it may be skipped safely.)
We wish to find the equations that describe the ``W-boson'' that appeared in \eqref{eqn:W2kqp} in terms of the low-energy fields of 4+1D SYM on $\Cig\times\R^{2,1}$, where $\R^{2,1}$ corresponds to directions $0,1,2$, and $\Cig$ is the ``cigar'' defined in \secref{subsec:cigar}.
That ``W-boson'' is {\it not} the W-boson of the 4+1D SYM on $\Cig$, but rather the W-boson of a dual 4+1D SYM on the $\R^{2,1}\times (\C/\Z_\lvk)$ background that appeared in \eqref{eqn:fib}. But, anyway, to derive the BPS equations it is convenient to start in six dimensions.

Let us first discuss the equations on the Coulomb branch of the $(2,0)$-theory.
The contents of the low-energy theory is a free tensor multiplet with $2$-form field $B$ (and anti-self-dual field strength $H=dB=-{}^*H$), five scalar fields $\Phi^6,\dots,\Phi^{10}$, and chiral fermions $\psi$ in the spinor representation $\rep{4}\otimes\rep{4}$ of $SO(5,1)\times SO(5)$. We assume 
$$
\Phi^6=\Phi^7=\Phi^8=\Phi^9=0
$$
and only allow $\Phi^{10}\equiv\phi$ to be nonzero.
The BPS equations are derived from the SUSY transformation of the fermions.
Let $\epsilon$ be a constant SUSY parameter, which we represent as a $32$-component spinor on which the 10+1D Dirac matrices $\Gamma^I$ ($I=0,\dots,10$) can act. The BPS conditions on $\epsilon$ are:
\begin{itemize}
\item
Invariance of $\epsilon$ under simultaneous rotations by $2\pi/\lvk$ in the planes $4-5$ and $6-7$;
\item
Invariance of an M$5$-brane along directions $0,\dots,5$ under a SUSY transformation of 10+1D SUGRA with parameter $\epsilon$; and
\item
Invariance of an M$2$-brane along directions $0,3,10$ under a SUSY transformation of 10+1D SUGRA with parameter $\epsilon$.
\end{itemize}
Therefore, the equations are (we set $10\equiv\ten$ in Dirac matrices):
\be\label{eqn:epsilonBPS}
\epsilon=\Gamma^{012345}\epsilon=\Gamma^{03\ten}\epsilon=\Gamma^{4567}\epsilon\,.
\ee
To get the BPS equations we require that the fermions $\psi$ of the tensor multiplet of the $(2,0)$-theory be invariant under any SUSY transformation with a parameter $\epsilon$ that satisfies \eqref{eqn:epsilonBPS}:
\be\label{eqn:delpsi}
0=\delta\psi \equiv (H_{\mu\nu\sigma}\Gamma^{\mu\nu\sigma} -\partial_\mu\phi\Gamma^{\mu\ten})\epsilon\,.
\ee
There are four linearly independent solutions to \eqref{eqn:epsilonBPS}, and substituting these into \eqref{eqn:delpsi} we find the equations:
\be\label{eqn:BPS-Hphi}
H_{03\mu}=\partial_\mu\phi\,,\qquad
H_{0ij}=0\,,\qquad (i,j=1,2,4,5\,,\qquad\mu=0,\dots,5)\,.
\ee
The other components of $H$ are determined by anti-self-duality $H=-{}^* H$.

We now convert the 5+1D BPS equations \eqref{eqn:BPS-Hphi} to 4+1D equations on $\Cig\times\R^{2,1}$ using \eqref{eqn:Hf} and the change of variables \eqref{eqn:tzC}.
To avoid ambiguity, we momentarily denote by $x_3'$ and $\theta'$ the coordinates before the change of variables, so that the change of variables is given by
$$
x_3 = x_3'\,,\qquad
\Cigth = \theta'-\frac{x_3'}{\lvk\xR}\,.
$$
We then find:
\be\label{eqn:Hpartialphi}
0 = H_{03'r}-\partial_r\phi 
= H_{03'\theta'}-\partial_{\theta'}\phi 
 = \partial_{3'}\phi 
 = \partial_0\phi
\,,\qquad
0 = H_{03' i}-\partial_i\phi 
\,,\qquad (i=1,2)\,,
\ee
and
$$
0 
= H_{012}
= H_{0ir}
= H_{0i\theta'}=H_{0i\Cigth}
= H_{0r\theta'}=H_{0r\Cigth}
\,,\qquad (i=1,2)
\,.
$$
The dual relations are
$$
0 
= H_{3'r\theta'}
= H_{3'i\theta'}
= H_{3'ir}
= H_{3'12}
\,,\qquad (i=1,2)
\,,
$$
which become in $(x_3,\Cigth)$ coordinates:
\be\label{eqn:H3pijr}
0 
=H_{3r\Cigth}
=H_{3i\Cigth}
=H_{3ir}-\tfrac{1}{\lvk\xR}H_{\Cigth i r}
=H_{312}-\tfrac{1}{\lvk\xR}H_{\Cigth 12}
\,,\qquad (i=1,2)
\,.
\ee
Next we use the anti-self-duality conditions
$$
H_{03'r} = \tfrac{1}{r}H_{\theta'12} = \tfrac{1}{r}H_{\Cigth 12}\,,\qquad
H_{03'1} = \tfrac{1}{r}H_{r\theta'2} = \tfrac{1}{r}H_{r\Cigth 2}\,,\qquad
H_{03'2} = -\tfrac{1}{r}H_{r\theta'1} = -\tfrac{1}{r}H_{r\Cigth 1}\,,
$$
and the relations \eqref{eqn:H3pijr}
to write
\be
H_{03'r} =  \tfrac{1}{r}H_{\Cigth 12} = \tfrac{\lvk\xR}{r}H_{312}\,,\quad
H_{03'1} = \tfrac{1}{r}H_{r\Cigth 2}= \tfrac{\lvk\xR}{r}H_{32r}\,,\quad
H_{03'2} =  -\tfrac{1}{r}H_{r\Cigth 1}= -\tfrac{\lvk\xR}{r}H_{31r}\,.
\ee
Combining with \eqref{eqn:Hpartialphi}, we end up with the BPS equations
\be
\partial_r\phi = H_{03'r} =   \tfrac{\lvk\xR}{r}H_{312}\,,\qquad
\partial_1\phi = H_{03'1} =  \tfrac{\lvk\xR}{r}H_{32r}\,,\qquad
\partial_2\phi = H_{03'2} = -\tfrac{\lvk\xR}{r}H_{31r}\,,
\ee
and further combining with \eqref{eqn:Hf} we have
\be
\partial_r\phi =  \tfrac{\lvk\xR}{r}f_{12}\,,\qquad
\partial_1\phi =  \tfrac{\lvk\xR}{r}f_{2r}\,,\qquad
\partial_2\phi =  -\tfrac{\lvk\xR}{r}f_{1r}\,.
\ee
Altogether, we have
\be\label{eqn:BPSab}
\partial_r\phi =  \tfrac{\lvk\xR}{r}f_{12} = f_{0r}\,,\qquad
\partial_1\phi =  \tfrac{\lvk\xR}{r}f_{2r} = f_{01}\,,\qquad
\partial_2\phi =  -\tfrac{\lvk\xR}{r}f_{1r} = f_{02}\,.
\ee
The equations \eqref{eqn:phifBPSnon} are the nonabelian extension of \eqref{eqn:BPSab}, and the fact that they imply the equations of motion \eqref{eqn:EOMf}-\eqref{eqn:EOMphi} shows that no additional terms are needed.

\subsection{The moduli space}
\label{subsec:ModSpace}

We are interested in solutions to \eqref{eqn:DiphiF} that correspond to a monopole on the space with metric \eqref{eqn:effds} with $m$ units of monopole charge. We focus on $m=1$, but the comments we make in this section apply to any number $m$ of monopole charge.
Recall that the moduli space of $m$ BPS $SU(2)$ monopoles on $\R^3$ is hyper-K\"ahler and can be described as the space of solutions to Nahm's equations \cite{Nahm:1979yw}, written in terms of three $m\times m$ anti-hermitian matrices $T^i$ which depend on a parmeters $s$:
\be\label{eqn:Nahm}
\frac{dT^i}{ds} = \tfrac{1}{2}\epsilon_{ijk} [T^i,T^j]\,,\qquad
-1\le s\le 1\,,\qquad i,j,k=1,2,3,\qquad T^i(s)\in\mathfrak{u}(m)\,,
\ee
with prescribed boundary conditions (Nahm poles) at $s=\pm 1$, and a reality condition $T(s)^* = T(-s)$.
It was given a nice string-theoretic interpretation in \cite{Diaconescu:1996rk} (using previous results on the moduli space of instantons \cite{Witten:1995gx,Douglas:1995bn}), was related to the moduli space of 2+1D gauge theories with $8$ supercharges in \cite{Hanany:1996ie}, and was further generalized to singular monopoles in \cite{Cherkis:1998hi}-\cite{Moore:2014gua}.

Our setting has only $4$ supercharges -- $16$ are preserved by the $(2,0)$-theory, half are broken by the geometry, and another half is broken by the Q-ball.
Our moduli space of solutions is therefore only K\"ahler and not hyper-K\"ahler.
We can show this explicitly using an adaptation of the Hamiltonian (Marsden-Weinsten) reduction technique of \cite{Hitchin:1986ea}.

We start with the space of all possible $SU(2)$ gauge field and scalar field configurations $(A_1, A_2, A_r,\ResPhi)$ on the $r\ge 0$ portion of space, subject to the boundary conditions
\be\label{eqn:ResPhiInfinitybc}
|\ResPhi|\rightarrow\Vph\qquad\text{at $x_1^2+x_2^2+r^2\rightarrow\infty$.}
\ee
At $r=0$ we note that \eqref{eqn:DiphiF} implies $F_{12}=0$ [see the left-most equation of \eqref{eqn:Drphi}], and so $A_1 dx_1 + A_2 dx_2$ reduces to a flat connection on the $r=0$ plane.
We can therefore pick a gauge so that $A_1 = A_2 = 0$ at $r=0$.
We still have the freedom to perform a gauge transformation with a gauge parameter $\lambda$ that approaches a constant (independent of $x_1, x_2$) at $r=0$ but with a possibly nonconstant $\partial_r\lambda$. We use this gauge freedom to set $A_r=0$ at $r=0$ as well. We therefore require:
\be\label{eqn:Abctip}
A_1=A_2=A_r=0\qquad\text{ at $r=0$.}
\ee
We denote the space of $(A_1, A_2, A_r,\ResPhi)$ configurations with the boundary conditions \eqref{eqn:ResPhiInfinitybc}-\eqref{eqn:Abctip} by $\APhiSpace$.
The infinite dimensional space $\APhiSpace$ is K\"ahler with a complex structure defined so that $A_1 + i A_2$ and $A_r + i r\ResPhi$ (evaluated at any point $x_1, x_2, r$) are holomorphic, and with a symplectic K\"ahler form given by
\be\label{eqn:SymplecticForm}
\omega = \tr\int (\frac{1}{r}\delta A_1\wedge \delta A_2 + \delta\ResPhi\wedge \delta A_r)dx_1 dx_2 dr.
\ee
The associated K\"ahler metric is
\be\label{eqn:BigMetric}
\tr\int\left\lbrack\frac{1}{r}(\delta A_1^2+\delta A_2^2+\delta A_r^2) + r\delta\ResPhi^2\right\rbrack dx_1 dx_2 dr.
\ee
The combination $A_r + i r\ResPhi$ was chosen so that the two middle equations of \eqref{eqn:Drphi} will be the real and imaginary parts of a holomorphic equation $(D_1+i D_2)\ResPhi = -\frac{i}{r}(F_{1r}+i F_{2r})$.

We are interested in the moduli space $\ModSpace_m$ of solutions to \eqref{eqn:DiphiF} with the boundary conditions \eqref{eqn:ResPhiInfinitybc}-\eqref{eqn:Abctip}, modulo gauge transformations with gauge parameter $\lambda$ that approaches a constant at $r=0$ and at $x_1^2+x_2^2+r^2\rightarrow\infty$, and such that
$$
\Vph m = \tr\int \sqrt{g} g^{ij} B_i D_j\ResPhi d^3 x = 
\tr\int\left\lbrack F_{12} D_r\ResPhi +F_{2r}D_1\ResPhi -F_{1r} D_2\ResPhi\right\rbrack dr dx_1 dx_2\,.
$$
We note that the metric \eqref{eqn:BigMetric} does not lead to the physical metric on the moduli space $\ModSpace_m$ (that is, the metric determined from the energy of a slowly time-varying configuration that corresponds to motion on $\ModSpace_m$), but rather to the metric that would result from the action of minimally coupled scalar and gauge fields, leading to the equations of motion \eqref{eqn:EOMfAlt}-\eqref{eqn:EOMphiAlt}. This metric is more directly related to the derived problem of 3D monopoles on the space with metric \eqref{eqn:effds}.

For any Lie-algebra valued gauge parameter $\lambda$ (that is a constant at $r=0$) we define the ``moment-map'':
\be\label{eqn:MomentMap}
\mu_\lambda = \tr\int\lambda\left(D_r\ResPhi-\frac{1}{r}F_{12}\right)dx_1 dx_2 dr\,.
\ee
When $\mu_\lambda$ is set to the Hamiltonian on the (infinite dimensional) symplectic manifold with symplectic form $\omega$, the generated flow (``time evolution'') corresponds to gauge transformations with gauge parameter $\lambda$.
The moduli space $\ModSpace_m$ is then equivalent to the Hamiltonian reduction of $\APhiSpace$ by these moment-maps (for all allowed $\lambda$'s). It is the subset of $\APhiSpace$ for which $\mu_\lambda=0$ for all admissible $\lambda$, modulo the equivalence relations corresponding to the gauge transformations generated by all the $\lambda$'s.
Since the gauge transformations preserve the complex structure (acting in an affine-linear way on the complex variables $A_1 + i A_2$ and $A_r + ir\ResPhi$) and the symplectic form, the arguments of \cite{Hitchin:1986ea} show that the resulting (finite dimensional) moduli space $\ModSpace_m$ is K\"ahler.

One can shed more light on the form of the metric \eqref{eqn:BigMetric} as follows.\footnote{The reasoning presented in this paragraph was pointed out to us by Sergey Cherkis.} One can derive \eqref{eqn:DiphiF} by reducing to $\Mxxr$ the instanton equations on $\R\times\Mxxr$ that are invariant under translations in $\R$ [where $\Mxxr$ was defined as the 3D space with metric \eqref{eqn:effds}]. The metric on $\R\times\Mxxr$ is taken to be $ds^2 = dx_4^2 + r^2 (dx_1^2 + dx_2^2 + dr^2)$, but since instanton equations are conformally invariant, we can replace this metric with the conformally equivalent metric $\tfrac{1}{r^2}|d(x_4+\tfrac{i}{2}r^2)|^2+|d(x_1 + i x_2)|^2$. The latter is clearly a K\"ahler manifold, as it describes the product of a 2D surface, parameterized by complex coordinate $x_4 + \tfrac{i}{2}r^2$ and a copy of $\C$, parameterized by $x_1 + i x_2$, and so the instanton moduli space is K\"ahler. Requiring invariance under translations in $\R$ is a holomorphic constraint, and so the space of $\R$-invariant solutions is also K\"ahler.

The metric on $\ModSpace_m$ is induced from the metric \eqref{eqn:BigMetric} on $\APhiSpace$ as follows.
Let $(A,\ResPhi)$ be a solution of \eqref{eqn:DiphiF}, and let $(\delta A,\delta\ResPhi)$ be a deformation to a nearby solution.
We need to fix the right gauge so that \eqref{eqn:BigMetric} will be minimal among gauge equivalent deformations.
This is equivalent to the gauge condition
\be\label{eqn:GaugeCondDef}
0 = r^2[\ResPhi,\delta\ResPhi] + D_1\delta A_1 + D_2\delta A_2 + D_r\delta A_r -\frac{1}{r}\delta A_r\,.
\ee
Now take a constant $r_0\gg 1/\sqrt{\Vph}$ and consider a portion of the moduli space comprising of solutions whose bulk of the energy is concentrated in the vicinity of $r_0$, allowing a spread of  $O(1/\Vph r_0)$ away from $r_0$. Then $(A,r_0\ResPhi)$ is an approximate solution of the flat space monopole equations, and if we approximate the explicit $r$ and $1/r$ factors in \eqref{eqn:SymplecticForm}-\eqref{eqn:MomentMap} by $r_0$ and $1/r_0$, we get the corresponding K\"ahler form, metric, and moment map of \cite{Hitchin:1986ea}, in one of the complex structures of the corresponding hyper-K\"ahler moduli space.
Set $\ResPhi_0\equiv r_0\ResPhi$. Then $(A_1, A_2, A_3,\ResPhi_0)$ approximately solve the BPS problem on $\R^3$, which we will refer to as the ``hyper-K\"ahler problem". In this context the $\R^3$ coordinates are taken to be $x_1, x_2$ and $\prx_3\equiv r-r_0$.

Now consider the case $m=1$. There are three moduli corresponding to the ``position'' of the monopole $(\Ca_1,\Ca_2,\Ca_3)$, with $\Ca_3\equiv\Ca-r_0$. (Note that this ``position'' is not necessarily the maximum of energy density for finite $\Ca$, but it is so in the limit $\Ca\rightarrow\infty$.) The combination $\Ca_1+i\Ca_2$ is holomorphic in the complex structure of $\ModSpace_1$, and the ``missing'' modulus $\theta$ that combines with $\Ca_3$ to form a holomorphic $\Ca_3+i\theta$ can be recovered as follows. First recall that for the hyper-K\"ahler problem, if we perform a large gauge transformation with gauge parameter $\Lambda = \exp(i\theta\ResPhi_0/r_0\Vph)$, where $0\le\theta\le \pi$, we obtain a different solution that still satisfies the correct boundary conditions at infinity of $\R^3$. The infinitesimal version $\lambda = (\delta\theta)\ResPhi_0/r_0\Vph$ solves the hyper-K\"ahler gauge condition, which we can recover from \eqref{eqn:GaugeCondDef} by dropping the last term on the RHS, as $r_0\rightarrow\infty$. Plugging the corresponding deformations $\delta A_i = D_i\lambda$, $\delta A_r=D_r\lambda$ and $\delta\ResPhi=0$ into \eqref{eqn:BigMetric}, we find that the metric on the $\theta$ direction behaves as $(\delta\theta)^2/r_0^2\Vph$. 
In our case, we also expect a modulus that corresponds to a large gauge transformation, but setting $\lambda$ to be proportional to $\ResPhi$, say $\lambda=\epsilon\ResPhi$, would not work, because: (i) $\ResPhi$ does not vanish at $r=0$, and (ii) the gauge condition \eqref{eqn:GaugeCondDef} requires
\be\label{eqn:lambdaGC}
0 = -r^2[\ResPhi,[\ResPhi,\lambda]] + D_1^2\lambda + D_2^2\lambda + D_r^2\lambda -\tfrac{1}{r}D_r\lambda\,,
\ee
but $\lambda=\epsilon\ResPhi$ does not satisfy \eqref{eqn:lambdaGC}. The sign of the rightmost term of \eqref{eqn:lambdaGC} is in conflict with what the equation of motion \eqref{eqn:EOMphi} requires it to be.
Instead, we need to look for a solution to \eqref{eqn:lambdaGC} with $\lambda=\epsilon\Psi$ such that $\Psi$ approaches a constant, say $\sigma^3$, at $r=0$ and approaches $\ResPhi/\Vph$ at infinity. In addition, $\Psi$ should map the boundary of the $r\ge 0$ space (the $x_1-x_2$ plane at $r=0$ together with a hemisphere at infinity) to $S^2$ in such a way as to have winding number $m=1$.
Gauge transformations by $\Lambda=\exp(i\theta\Psi)$ then correspond to a circular direction $0\le\theta<\pi$ in moduli space. The corresponding deformations are
$$
\delta A_1 = \delta\theta D_1\Psi\,,\qquad
\delta A_2 = \delta\theta D_2\Psi\,,\qquad
\delta A_r = \delta\theta D_r\Psi\,,\qquad
\delta\ResPhi = -i\delta\theta [\Phi,\Psi]\,.
$$
The metric on this direction is given by
$$
(\delta\theta)^2
\tr\int\left\{\frac{1}{r}\bigl\lbrack
(D_1\Psi)^2 +(D_2\Psi)^2 + (D_r\Psi)^2\bigr\rbrack
-r[\Phi,\Psi]^2\right\} dx_1 dx_2 dr\,,
$$
which can be integrated by parts, using \eqref{eqn:lambdaGC} (for $\lambda=\epsilon\Psi$), to give a surface integral on the boundary of the $r\ge 0$ space:
$$
(\delta\theta)^2
\int\frac{1}{r}\left\lbrack \partial_1\tr{(\Psi^2)} dx_2 dr + \partial_2\tr{(\Psi^2)} dx_1 dr +\partial_r\tr{(\Psi^2)} dx_1 dx_2\right\rbrack\,.
$$
This integral depends on the subleading terms in the behavior of $\Psi^2$ near the boundary, which, unfortunately, we do not know.

Now, consider the mode associated with translations. In the hyper-K\"ahler limit the associated deformation that satisfies the gauge condition \eqref{eqn:GaugeCondDef} is
$$
\delta A_1 = (\delta\Ca)F_{31}\,,\qquad
\delta A_2 = (\delta\Ca)F_{32}\,,\qquad
\delta A_3 = 0\,,\qquad
\delta\ResPhi_0 = (\delta\Ca)D_3\ResPhi_0\,,
$$
where we have augmented the translation by $\delta\Ca$ in the $x_3$ direction by a gauge transformation with gauge parameter $(\delta\Ca)A_3$. Plugging into \eqref{eqn:BigMetric} we get a metric $(\delta\Ca)^2\Vph$.
Rescaling by $\Vph$, so far we have the approximate metric
\be\label{eqn:dsCatheta}
ds^2\sim d\Ca^2 + \frac{d\theta^2}{\Vph^2\Ca^2}\,.
\ee
In general, the modulus $\Ca$ is defined from the boundary conditions of the solution $(A,\ResPhi)$.
Like the hyper-K\"ahler counterpart, for $r\rightarrow\infty$ the solution to \eqref{eqn:DiphiF} reduces, up to a gauge transformation, to the field of an abelian monopole centered at, say, $(0,0,\Ca)$. We will discuss the abelian solution and present its exact form in \secref{subsec:Abel}, but for now suffice it to say that the modulus $\Ca$ can be read off from the asymptotic form. The metric that we found above in \eqref{eqn:dsCatheta} would be consistent with a K\"ahler manifold if  $\tfrac{1}{2}\Vph\Ca^2+i\theta$ is a holomorphic coordinate.
From the discussion above, we find the asymptotic form of the metric on moduli space as
$$
ds^2 \sim d\Ca_1^2+d\Ca_2^2+d\Ca^2+\frac{d\theta^2}{\Vph^2\Ca^2}\,,\qquad
\Ca\rightarrow\infty.
$$
and the asymptotic K\"ahler form is
$$
k\sim d\Ca_1\wedge d\Ca_2 +  \frac{d\Ca}{\Vph\Ca}\wedge d\theta\,,\qquad
\Ca\rightarrow\infty.
$$
Beyond these observations, we do not have a simple description of the moduli space $\ModSpace_1$, and as we have seen, unlike the moduli space of $\R^3$ BPS monopoles, in our case the $x_3$ coordinate of the ``center'' (corresponding to $\tr T^3$ in Nahm's equations) does not decouple. Moreover, the Bogomolnyi equations that describe monopoles on $\R^3$ can be obtained as a limit of \eqref{eqn:DiphiF} (see \secref{subsec:LVexp} for more details) when $r\rightarrow \infty$. 
Thus, we expect to recover the moduli space of BPS monopoles with fixed center of mass at the boundary $r\rightarrow\infty$ of the moduli space of \eqref{eqn:DiphiF}.

\section{Analysis of the BPS equations}
\label{sec:analysisBPS}

In this section we will present several observations regarding the solution of the BPS equations \eqref{eqn:Drphi}.
It is convenient to regard the BPS equations as Bogomolnyi monopole equations \eqref{eqn:DiphiF} on a curved space with metric \eqref{eqn:effds}.
We are looking for a solution of unit monopole charge. We also require axial symmetry (i.e., independence of $\fT$), since we can assume that the string of the $(2,0)$-theory, which the solution describes, sits at the origin of the $x_1-x_2$ plane.
The fields are therefore functions of two variables, $r$ and $\rE$, only.
The Bogomolnyi monopole equations on $\R^3$ have the renowned Prasad-Sommerfield solution \cite{Prasad:1975kr} for one $SU(2)$ monopole, and the general solution was given by Nahm \cite{Nahm:1979yw}. It was given a string-theoretic interpretation in \cite{Diaconescu:1996rk}. The extension to hyperbolic space is also known \cite{Atiyah:1984fe}, but we are unaware of an extension of Nahm's technique to the space given by the metric \eqref{eqn:effds}, and standard techniques that exploit the integrability of the $\R^3$ problem do not work in our case. 
We were unable to find an exact solution, but we can make a few observations.
In \secref{subsec:Manton} we will reduce the number of independent fields from twelve to two by adapting a method developed in \cite{Manton:1977ht,Forgacs:1980yv} for finding axially symmetric (generally multi-monopole) solutions of the Bogomolnyi equations on $\R^3$.
We will then present the asymptotic form of the solution far away from the origin. In this region the solution reduces to a $U(1)$ monopole whose fields we write down explicitly. We then show that the solution can be encoded in a harmonic map from $AdS_3$ to $AdS_2$.
We conclude in \secref{subsec:LVexp} with an expansion up to second order in inverse VEV.

\subsection{Manton gauge}
\label{subsec:Manton}

We adopt an ansatz proposed in \cite{Manton:1977ht} for axially symmetric solutions.
Adapted from $\R^3$ to our metric \eqref{eqn:effds} we look for a solution in the form:
\be\label{eqn:Manton}
\Phi = \tfrac{1}{2}(\Phi_1\sigma_1+\Phi_2\sigma_2)\,,\qquad
A = -[(\eta_1\sigma_1+\eta_2\sigma_2)d\fT +W_2\sigma_3 d\rE  +W_1\sigma_3 dr],
\ee
where $\sigma_1$, $\sigma_2$, $\sigma_3$ are Pauli matrices, and $\Phi_1$, $\Phi_2$, $\eta_1$, $\eta_2$, $W_1$, $W_2$ are scalar fields.
The BPS equations then reduce to
\bear
\partial_\rE\Phi_1-W_2\Phi_2 &=& -\tfrac{1}{r\rE}(\partial_r\eta_1-W_1\eta_2)\,,\label{eqn:FHP1}\\
\partial_\rE\Phi_2+W_2\Phi_1 &=& -\tfrac{1}{r\rE}(\partial_r\eta_2+W_1\eta_1)\,,\label{eqn:FHP2}\\
\eta_2\Phi_1-\eta_1\Phi_2 &=& \tfrac{\rE}{r}(\partial_\rE W_1-\partial_r W_2)\,,\label{eqn:FHP3}\\
\partial_r\Phi_1-W_1\Phi_2 &=& \tfrac{1}{r\rE}(\partial_\rE\eta_1-W_2\eta_2)\,,\label{eqn:FHP4}\\
\partial_r\Phi_2+W_1\Phi_1 &=& \tfrac{1}{r\rE}(\partial_\rE\eta_2+W_2\eta_1)\,,\label{eqn:FHP5}
\eear
Next, we adapt to our metric the technique developed in \cite{Forgacs:1980yv}, solving \eqref{eqn:FHP1}-\eqref{eqn:FHP3} by setting
\be\label{eqn:MantonPhi}
\Phi_1 = -\frac{1}{r}\Ff^{-1}\partial_r\Fpsi\,,\quad
\Phi_2 = \frac{1}{r}\Ff^{-1}\partial_r\Ff\,,\quad
\eta_1 = \rE\Ff^{-1}\partial_\rE\Fpsi\,,\quad
\eta_2 =-\rE\Ff^{-1}\partial_\rE\Ff\,,
\ee
and
\be\label{eqn:MantonW}
W_1 = -\Ff^{-1}\partial_r\Fpsi\,,\quad
W_2 = -\Ff^{-1}\partial_\rE\Fpsi\,,
\ee
where $\Ff$ and $\Fpsi$ are as yet undetermined real functions of $r$ and $\rE$.
We plug the ansatz \eqref{eqn:MantonPhi}-\eqref{eqn:MantonW} into \eqref{eqn:FHP4}-\eqref{eqn:FHP5} and get:
\bear
0 &=&
\Ff\Fpsi_{rr}
+\Ff\Fpsi_{\rE\rE}
-2\Ff_r\Fpsi_r
-2\Ff_\rE\Fpsi_\rE
+\tfrac{1}{\rE}\Ff\Fpsi_\rE
-\tfrac{1}{r}\Ff\Fpsi_r
\,,\label{eqn:Epsi}\\
0 &=&
\Ff_r^2
+\Ff_\rE^2
-\Fpsi_r^2
-\Fpsi_\rE^2
-\Ff \Ff_{rr}
-\Ff \Ff_{\rE\rE}
+\tfrac{1}{r}\Ff \Ff_r
-\tfrac{1}{\rE}\Ff \Ff_\rE\,,
\label{eqn:Ef}
\eear
where subscripts $(\cdots)_r$ and $(\cdots)_\rE$ denote derivatives with respect to $r$ and $\rE$, respectively.

\subsection{The abelian solution}
\label{subsec:Abel}

We can trivially solve \eqref{eqn:Epsi} by setting $\Fpsi=0$.
The remaining equation \eqref{eqn:Ef} then states that $\log\Ff$ is a harmonic function on $AdS_3$.
Alternatively, the solution describes a $U(1)$ monopole on the $(x_1, x_2, r)$ space with metric \eqref{eqn:effds}. 
It is easiest to construct the solution starting from 5+1D.
Let us take the center of the monopole to be $(0,0,\Ca)$, which will then have to be a singular point for $\Ff$.
In the abelian limit, the fields of the $(2,0)$ theory that are relevant to our problem reduce to a free anti-self-dual $3$-form field $H=-{}^*H$ and a free scalar field $\phi$. We start by solving \eqref{eqn:BPS-Hphi} on $\R^{5,1}$, which in particular implies that $\phi$ is harmonic. Consider a solution that describes the $H$ and $\phi$ fields that emanate from a $(2,0)$-string centered at 
$(x_1,x_2,x_4,x_5) = (0,0,\Ca\cos\theta, \Ca\sin\theta)$.
The scalar field is given by
\be\label{eqn:stringsol}
\phi=\Vph+\frac{1}{x_1^2 + x_2^2 + (x_4-\Ca\cos\theta)^2 + (x_5-\Ca\sin\theta)^2}\,.
\ee
But the solution that we need must be indepedent of $\theta$, so we ``smear'' \eqref{eqn:stringsol} to obtain the requisite field:
\be\label{eqn:abphisol}
\phi(x_1, x_2, r) =\Vph+\frac{1}{2\pi}\int_0^{2\pi}
\frac{d\theta}{\rE^2 + (r\cos\theta-\Ca)^2 + r^2\sin^2\theta}
= \Vph+\frac{1}{\sqrt{(\rE^2 +r^2 + \Ca^2)^2 - 4 \Ca^2 r^2}}\,.
\ee
The $U(1)$ gauge field is now easy to calculate from \eqref{eqn:Drphi} and we find
\be\label{eqn:abAsol}
A = 
\bigl(
\frac{\rE^2+\Ca^2-r^2}{2\sqrt{(\rE^2 +r^2 + \Ca^2)^2 - 4\Ca^2 r^2}}-1
\bigr)\frac{x_2 dx_1 -x_1 dx_2}{\rE^2}\,,
\ee
where we picked a gauge for which $A_r=0$.
It is easy to find the associated $(\Ff,\Fpsi)$ fields. We have $\Fpsi=0$ and
\be\label{eqn:AbelFf}
\Ff = \exp\int\phi(r,\rE) r dr
= e^{-\frac{1}{2}\Vph r^2}\left(\rE^2 +r^2  - \Ca^2 + 
\sqrt{(\rE^2 +r^2 + \Ca^2)^2 - 4\Ca^2 r^2}\right).
\ee
Equation \eqref{eqn:abAsol} exhibits a Dirac string singularity that extends from $r=a$ to $r=\infty$ at $x_1=x_2=0$.
The abelian solution must describe the asymptotic behavior of the nonabelian solution when either $r\rightarrow\infty$ or $\rE\rightarrow\infty$ (or both).

\subsection{Relation to harmonic maps from $AdS_3$ to $AdS_2$}
\label{subsec:hmapsAdS}

The equations \eqref{eqn:Epsi}-\eqref{eqn:Ef} can be derived from the action
\be\label{eqn:fpsiI}
I = \int\frac{\rE}{r\Ff^2}(\Ff_r^2+\Ff_\rE^2+\Fpsi_r^2+\Fpsi_\rE^2)dr d\rE\,.
\ee
We can therefore give a simple geometrical meaning to the equations of motion \eqref{eqn:Epsi}-\eqref{eqn:Ef} by considering an auxiliary $AdS_3$ space parameterized by $(r,\rE,\fT)$ with metric
$$
ds^2 = \frac{1}{r^2}(dr^2 + d\rE^2 + \rE^2 d\fT^2)\,
$$
and interpreting the functions $\Ff(r,\rE)$ and $\Fpsi(r,\rE)$ as describing an axisymmetric map from $AdS_3$ to the two-dimensional $(\Ff,\Fpsi)$ ``target-space.'' If we further endow this target-space with the $AdS_2$ metric
\be\label{eqn:AdS2}
ds^2 = \frac{1}{\Ff^2}(d\Ff^2+d\Fpsi^2)\,,
\ee
it is easy to see that the equations of motion derived from \eqref{eqn:fpsiI} describe harmonic maps
\be\label{eqn:HmapAdS}
(\Ff,\Fpsi): AdS_3\mapsto AdS_2\,.
\ee
The connection between $AdS_2$ (the ``pseudosphere'') and axisymmetric solutions to monopole equations on $\R^3$ was first noted in \cite{Forgacs:1980yv}.
The harmonic map \eqref{eqn:HmapAdS} is required to have a singularity along a Dirac-like string, as we saw in \secref{subsec:Abel}.

To reproduce the abelian solution of \secref{subsec:Abel}, we set $\Fpsi=0$ and find that $\log\Ff$ is a harmonic function on $AdS_3$, as stated at the beginning of \secref{subsec:Abel}. To present its Dirac string more clearly, it is convenient to use instead of the Poincar\'e coordinates on $AdS_3$, a coordinate system with the point $r=\Ca$ at the origin. The change from $(r,\rE,\fT)$ to the new coordinate system $(\AdSchi,\AdSth,\fT)$ is given by:
$$
\frac{\rE}{r} = \sinh\AdSchi\,\sin\AdSth\,,\quad
\frac{\rE^2+r^2-\Ca^2}{2\Ca r} = \sinh\AdSchi\,\cos\AdSth\,,
$$
and the coordinates are defined in the range
$$
0\le\AdSchi<\infty\,,\qquad 0\le\AdSth\le\pi\,,\qquad 0\le\fT<2\pi.
$$
The metric in terms of $(\AdSchi,\AdSth,\fT)$ is
$$
ds^2 = d\AdSchi^2+\sinh^2\AdSchi (d\AdSth^2 +\sin^2\AdSth\,d\fT^2),
$$
and the inverse coordinate transformations are:
$$
r = \Ca\frac{\cosh\AdSchi+\sinh\AdSchi\,\cos\AdSth}{1+\sinh^2\AdSchi\,\sin^2\AdSth}
\,,\qquad
\rE=\Ca\frac{\cosh\AdSchi+\sinh\AdSchi\,\cos\AdSth}{1+\sinh^2\AdSchi\,\sin^2\AdSth}\sinh\AdSchi\,\sin\AdSth\,.
$$

In $(\AdSchi,\AdSth,\fT)$ coordinates we have, up to an unimportant constant,
\bear
\log\Ff  &=& -\tfrac{1}{2}\Vph\Ca^2\left(\frac{\cosh\AdSchi+\sinh\AdSchi\,\cos\AdSth}{1+\sinh^2\AdSchi\,\sin^2\AdSth}\right)^2
+\log\left(\frac{\cosh\AdSchi+\sinh\AdSchi\,\cos\AdSth}{1+\sinh^2\AdSchi\,\sin^2\AdSth}\right)
\nn\\ &&
+\log\sinh\AdSchi
+\log(1+\cos\AdSth)\,.
\label{eqn:LogFfNewCo}
\eear
The singularity in the last term at $\AdSth=\pi$ represents the Dirac string.

\subsection{Comments on (lack of) integrability}
\label{subsec:Integrability}

The classic Bogomolnyi equations for monopoles on $\R^3$ admit the well-known Nahm solutions \cite{Nahm:1979yw}, which also have a nice string-theoretic interpretation \cite{Diaconescu:1996rk}. The rich properties of these solutions essentially stem from an underlying integrable structure. One way to describe the structure is to map a solution of the Bogomolnyi equations to a holomorphic vector bundle over minitwistor space \cite{Hitchin:1982gh,Hitchin:1983ay}. (Minitwistor space is the space of oriented straight lines on $\R^3$, and it has a complex structure.) 
The Bogomolnyi equations arise as the integrability condition for an auxiliary set of differential equations for an auxiliary $2$-component field $\psi$, that require $\psi$'s gauge-covariant derivative along a line in $\R^3$ to be related to multiplication by the scalar field $\ResPhi$, and also require $\psi$ to be holomorphic in the directions transverse to the line. This technique can be extended to other metrics, such as $AdS_3$ (whose corresponding minitwistor space also possesses a complex structure and is equivalent to $\CP^1\times\CP^1$). But this technique fails for the metric \eqref{eqn:effds}, whose space of geodesics is not complex, and the monopole equations \eqref{eqn:DiphiF} cannot be expressed as the integrability condition for an auxiliary system of linear differential equations, at least not in an obvious way.

Another way to see where integrability fails is to focus on axially-symmetric solutions as in \cite{Forgacs:1980yv}.
Defining the symmetric $\SL(2,\R)$ matrix
$$
\gE\equiv\frac{1}{\Ff}\begin{pmatrix}
1 & -\Fpsi \\
-\Fpsi\,\,\,\, & (\Ff^2+\Fpsi^2) \\
\end{pmatrix}\,,
$$
the equations of motion \eqref{eqn:Epsi}-\eqref{eqn:Ef} can then be recast as
\be\label{eqn:sigmaeom}
0=\nabla^\alpha(\nabla_\alpha\gE\gE^{-1})\,,
\ee
where the covariant derivatives are with respect to another auxiliary metric,
\be\label{eqn:dsErnst}
ds^2 = dr^2+d\rE^2+(\frac{\rE^2}{r^2})d\fT^2\,,
\ee
 and $\gE(r,\rE)$ is, of course, assumed to be independent of $\fT.$
It is possible \cite{Forgacs:1980yv} to recast axially symmetric solutions of the Bogomolnyi equations on $\R^3$ in the form \eqref{eqn:sigmaeom} -- the metric in that case would be the Euclidean metric
$$
ds^2 = dr^2+d\rE^2+\rE^2 d\fT^2\,,
$$
and the connection with the $\sigma$-model \eqref{eqn:sigmaeom} leads to an integrable structure.
To describe the integrable structure we switch to complex coordinates,
$$
\xi\equiv r + i\rE\,,\qquad
\bxi\equiv r - i\rE\,,
$$
and write \eqref{eqn:sigmaeom} as the integrability condition for a system of first order linear differential equations for a two-component field $\Psi(\xi,\bxi)$:
$$
\Psi_\xi = \frac{1}{1+\gamma}\gE_\xi\gE^{-1}\Psi\,,\qquad
\Psi_\bxi = \frac{1}{1-\gamma}\gE_\bxi\gE^{-1}\Psi\,,
$$
where $(\cdots)_\xi$ and $(\cdots)_\bxi$ are derivatives with respect to $\xi$ and $\bxi$, and the function $\gamma(\xi,\bxi)$ has to be suitably chosen (so that the integrability condition $(\Psi_\xi)_\bxi = (\Psi_\xi)_\bxi$ will be automatically satisfied). There are, in fact, infinitely many choices for the function $\gamma$, but it has to be a solution of 
$$
\gamma_\xi = \frac{\gamma}{\xi-\bxi}(\frac{1+\gamma}{1-\gamma})\,,\qquad
\gamma_\bxi = -\frac{\gamma}{\xi-\bxi}(\frac{1-\gamma}{1+\gamma})\,,
$$
which are compatible (see \cite{Korotkin:1994au} for review).
This construction is easy to extend to any metric of the form
$$
ds^2 = dr^2 + d\rE^2 + \Lambda(r,\rE)^2 d\fT^2\,,
$$
as long as $\Lambda(r,\rE)$ is harmonic (in the metric $dr^2+d\rE^2$). 
In our case $\Lambda = \rE/r$ is not harmonic, so the standard integrability structure is not present.

One can also attempt to extend the technique of \cite{Diaconescu:1996rk}, to ``probe'' the solution with a string that extends in an extra dimension, say $x_8$. It is not hard to construct BPS string solutions that preserve some supersymmetry, compatible with that of the M$5$-branes and the twist. For example, in the M-theory variables we can take an M$2$-brane along a holomorphic curve given by
$x_4 + i x_5 = C_0 e^{\frac{i}{\lvk\xR}(x_3 + i x_8)}$,
where $C_0$ is a constant.
This would translate in type-IIA to a string whose $x_8$ coordinate varies logarithmically with $r$.
However, this string does not preserve any common SUSY with the soliton.
We were unable to find an exact solution to \eqref{eqn:DiphiF}, and in fact, the appearance of polylogarithms in the expansion at large VEV (see \secref{subsec:LVexp}) suggests that even if a closed form exists, it is very complicated.
We therefore resorted to an asymptotic expansion for large VEV, described below, and to numerical analysis, which we describe in \appref{app:Numerical}.

\subsection{Large VEV expansion}
\label{subsec:LVexp}

In this section we will discuss the asymptotic expansion of the solution to the BPS equations \eqref{eqn:Drphi} for large VEV $\Vph$.
Since the only dimensionless combination that governs the behavior of the solution is $\Vph\Ca^2$, we can just as well discuss fixed $\Vph$ and large $\Ca$, which means that the core of the monopole solution is far from the tip.
Let us set\footnote{We hope that no confusion will arise with the coordinate $x_3$ that was used in \secref{subsec:geom}.
That coordinate plays no role here, and the only coordinates relevant for this section are $x_1, x_2$ and $r=\Ca+x_3$.}
$x_3\equiv r-\Ca$ and rescale $\phi=\Ca\ScPhi/\lvk\xR$, so that equations \eqref{eqn:Drphi} can be rewritten as
\be\label{eqn:BPSrescaled}
(1+\tfrac{x_3}{\Ca})D_i\phi = \tfrac{1}{2}\epsilon_{ijk}F_{jk}\,,
\ee
where in this section $i,j,k=1,2,3$ refer to $x_1, x_2, x_3$ with Euclidean metric
$$
ds^2 = dx_1^2 + dx_2^2 + dx_3^2\,.
$$
In the limit $\Ca\rightarrow\infty$, the equations \eqref{eqn:BPSrescaled} reduce to Bogomolnyi's equations, and the one-monopole solution is \cite{Prasad:1975kr}:
\be\label{eqn:phiAOb0}
A_i^{a\,(0)} = \epsilon_{iaj}x_j\kPS(\psR)\,,\qquad
\phi^{a\,(0)}=x_a\hPS(\psR)\,,
\ee
where
\be\label{eqn:hkPS}
\hPS\equiv\tfrac{1}{\psR}\coth \psR-\tfrac{1}{\psR^2}\,,\qquad
\kPS\equiv\tfrac{1}{\psR\sinh\psR}-\tfrac{1}{\psR^2}\,,
\ee
and here
$\psR\equiv \sqrt{\sum_{i=1}^3 x_i^2}$.
We set
$$
\invCa\equiv\frac{1}{\Ca}\,,\qquad
\vec{\vb}\equiv(0,0,\invCa)\,,
$$
so that $\tfrac{x_3}{\Ca}=\vec{\vb}\cdot\vec{x}$, and \eqref{eqn:BPSrescaled} can be written as:
\be\label{eqn:BPSvb}
(1+\vec{\vb}\cdot\vec{x})D_i\phi = \tfrac{1}{2}\epsilon_{ijk}F_{jk}\,.
\ee
We can now expand around the Prasad-Sommerfield solution:
$$
A_i = A_i^{(0)} + \invCa A_i^{(1)} + \invCa^2 A_i^{(2)}+\cdots\,,\qquad
\phi = \phi^{(0)} + \invCa\phi^{(1)} +\invCa^2\phi^{(2)}+\cdots\,,
$$
where we set the $0^{th}$ order terms to the Prasad-Sommerfield solution \eqref{eqn:phiAOb0}.

At order $O(\invCa)$ we write all possible terms that are allowed by spherical symmetry
and we keep only the terms that are also invariant under the parity symmetry
\be\label{eqn:Parityvbx}
\phi^a(\vec{x},\vec{\vb})\rightarrow
-\phi^a(-\vec{x},-\vec{\vb})\,,\qquad
A_i^a(\vec{x},\vec{\vb})\rightarrow -A_i^a(-\vec{x},-\vec{\vb})\,.
\ee
The general expression is then
\bear
\invCa\phi^{a(1)}
 &=& 
\vb_a\Fa_{1,1}(\psR)+x_a (\vb_k x_k) \Fa_{1,2}(\psR)
\,,\label{eqn:phiOb1x}\\
\invCa A_i^{a(1)} &=& 
x_a\epsilon_{ijk}x_j\vb_k\Fa_{1,3}(\psR)
+x_i\epsilon_{ajk}x_j\vb_k\Fa_{1,4}(\psR)
+\epsilon_{aij}\vb_j\Fa_{1,5}(\psR)
\,,
\label{eqn:AOb1x}
\eear
and we note the identity
\be\label{eqn:Idxepsx}
x_{[i}\epsilon_{a]jk}x_j\vb_k
=\tfrac{1}{2}\epsilon_{aij}\vb_j\psR^2
-\tfrac{1}{2}(\vb_k x_k)\epsilon_{aij}x_j\,,
\ee
which is the reason why we did not include a term of the form $(\vb_k x_k)\epsilon_{aij}x_j\Fa_{1,6}$ in \eqref{eqn:AOb1x}.
The coefficients $\Fa_{1,1},\ldots,\Fa_{1,5}$ are unknown functions of $\psR$.

We also have the freedom to apply an infinitesimal $O(\invCa)$ gauge transformation which takes the form
$$
\delta\phi^a = \epsilon_{abc}\lambda^b\phi^c\,,\qquad
\delta A_i^a =\partial_i\lambda^a-\epsilon_{abc}A_i^b \lambda^c
$$
with
$$
\lambda^a = \epsilon_{abc}x_b\vb_c\Ga_{1,1}(\psR)\,.
$$
This gives
\bear
\delta\phi^a &=& 
-x_a\vb_k x_k\Ga_{1,1}\hPS +\vb_a \psR^2\Ga_{1,1}\hPS
\,,\label{eqn:phigg1}\\
\delta A_i^a &=&
-\epsilon_{iab}\vb_b\Ga_{1,1}
+\tfrac{1}{\psR}x_i\epsilon_{abc}x_b\vb_c\Ga_{1,1}'
+x_a\epsilon_{ibc}x_b\vb_c\Ga_{1,1}\kPS\,.
\label{eqn:Agg1}
\eear
Using this gauge transformation we can set one of the parameters in \eqref{eqn:phiOb1x}-\eqref{eqn:AOb1x} to zero.
We choose to set
\be\label{eqn:gaugefix1,5}
\Fa_{1,5}=0.
\ee
We end up with the general form of the $O(\invCa)$ correction:
\bear
\invCa\phi^{a(1)}
 &=& 
\vb_a\Fa_{1,1}(\psR)+x_a (\vb_k x_k) \Fa_{1,2}(\psR)
\,,\label{eqn:phiOb1}\\
\invCa A_i^{a(1)} &=& 
x_a\epsilon_{ijk}x_j\vb_k\Fa_{1,3}(\psR)
+x_i\epsilon_{ajk}x_j\vb_k\Fa_{1,4}(\psR)
\,.
\label{eqn:AOb1}
\eear
Plugging \eqref{eqn:phiAOb0} and \eqref{eqn:phiOb1}-\eqref{eqn:AOb1} into \eqref{eqn:BPSvb} and comparing terms of order $O(\invCa)$ we get:
\bear
\hPS\kPS-\tfrac{1}{\psR}\hPS' &=&
\tfrac{1}{\psR}(\Fa_{1,2}'+\Fa_{1,3}')
-\kPS\Fa_{1,2}
-\kPS\Fa_{1,3}
+(\kPS-\hPS)\Fa_{1,4}
\,,\label{eqn:ODEFaI}\\
0 &=&
\tfrac{1}{\psR}\Fa_{1,1}'+(1+\psR^2\kPS)\Fa_{1,3}+\psR^2\hPS\Fa_{1,4}
\,,\label{eqn:ODEFaII}\\
0 &=&
\psR\Fa_{1,3}'+\kPS\Fa_{1,1}-\Fa_{1,2}+3\Fa_{1,3}+(1+\psR^2\kPS)\Fa_{1,4}
\,\label{eqn:ODEFaIII}\\
-\hPS(1+\psR^2\kPS) &=&
\kPS\Fa_{1,1}
+(1+\psR^2\kPS)\Fa_{1,2}
+\Fa_{1,4}
\,,\label{eqn:ODEFaIV}
\eear
These are ordinary inhomogeneous linear differential equations in $\Fa_{1,1},\dots,\Fa_{1,4}$.
Note that $\Fa_{1,4}$ can be eliminated from \eqref{eqn:ODEFaIV}, so the general solution is given be an arbitrary solution of the full equations \eqref{eqn:ODEFaI}-\eqref{eqn:ODEFaIV} plus a linear combination of three linearly independent solutions of the homogeneous equations:
\bear
0&=&
\tfrac{1}{\psR}(\Fa_{1,2}'+\Fa_{1,3}')
-\kPS\Fa_{1,2}
-\kPS\Fa_{1,3}
+(\kPS-\hPS)\Fa_{1,4}
\,,\label{eqn:homogODEFaI}\\
0 &=&
\tfrac{1}{\psR}\Fa_{1,1}'+(1+\psR^2\kPS)\Fa_{1,3}+\psR^2\hPS\Fa_{1,4}
\,,\label{eqn:homogODEFaII}\\
0 &=&
\psR\Fa_{1,3}'+\kPS\Fa_{1,1}-\Fa_{1,2}+3\Fa_{1,3}+(1+\psR^2\kPS)\Fa_{1,4}
\,\label{eqn:homogODEFaIII}\\
0&=&
\kPS\Fa_{1,1}
+(1+\psR^2\kPS)\Fa_{1,2}
+\Fa_{1,4}
\,,\label{eqn:homogODEFaIV}
\eear
The general solution to \eqref{eqn:ODEFaI}-\eqref{eqn:ODEFaIV} that is nonsingular at $\psR=0$ is 
\bear
\Fa_{1,1} &=& 
-\tfrac{\psR}{2\sinh\psR}
+c_1(\tfrac{\psR\cosh^2\psR}{\sinh^3\psR} - \tfrac{\cosh\psR}{\sinh^2\psR})
+c_2(\tfrac{3\psR}{\sinh\psR} - \tfrac{3\psR^2\cosh\psR}{\sinh^2\psR} + \tfrac{\psR^3\cosh^2\psR}{\sinh^3\psR})
\,,\label{eqn:Fa1,1}\\
\Fa_{1,2} &=& 
\tfrac{1}{2\psR^2}+\tfrac{1-2\cosh\psR}{2\psR\sinh\psR}
+c_1(\tfrac{1}{\psR^4} - \tfrac{\cosh^2\psR}{\psR\sinh^3\psR} + \tfrac{\cosh\psR - 1}{\psR^2\sinh^2\psR})
+c_2(-\tfrac{\psR\cosh^2\psR}{\sinh^3\psR} + \tfrac{(2\cosh\psR - 3)}{\psR\sinh\psR} 
+ \tfrac{(3\cosh\psR - 1)}{\sinh^2\psR})
\,,\nn\\ &&\label{eqn:Fa1,2}\\
\Fa_{1,3} &=& 
\tfrac{1}{2\psR^2}-\tfrac{1}{2\psR}\coth\psR
+c_1(-\tfrac{1}{\psR^4} + \tfrac{\cosh\psR}{\psR\sinh^3\psR})
+c_2(\tfrac{\psR\cosh\psR}{\sinh^3\psR} + \tfrac{\cosh\psR}{\psR\sinh\psR} - \tfrac{2}{\sinh\psR})
\,,\label{eqn:Fa1,3}\\
\Fa_{1,4} &=& 
c_1(-\tfrac{1}{\psR^3\sinh\psR} - \tfrac{\cosh\psR}{\psR^2\sinh^2\psR} + \tfrac{1+\cosh^2\psR}{\psR\sinh^3\psR})
+c_2(\tfrac{\psR(1 + \cosh^2\psR)}{\sinh^3\psR} + \tfrac{3}{\psR\sinh\psR} - \tfrac{5\cosh\psR}{\sinh^2\psR})
\,,\label{eqn:Fa1,4}
\eear
where $c_1, c_2$ are undetermined constants.
Note that the functions \eqref{eqn:Fa1,1}-\eqref{eqn:Fa1,4} have a regular power series expansion at $\psR=0$ with nonnegative even powers of $\psR$ only.
We note that there is another homogeneous solution that we discarded because it is singular at $\psR=0$:

\be
\begin{array}{ll}
\Fa_{1,1} = 
c_3(\tfrac{\cosh^2\psR}{\sinh^3\psR})
\,,\qquad&
\Fa_{1,2} = 
-c_3(\tfrac{\coth\psR}{\psR^4}+\tfrac{1}{\psR^3\sinh^2\psR}+\tfrac{\cosh^2\psR}{\psR^2\sinh^3\psR})
\,,\\
\Fa_{1,3} = 
c_3(\tfrac{\coth\psR}{\psR^4}+\tfrac{1}{\psR^3\sinh^2\psR}+\tfrac{\cosh\psR}{\psR^2\sinh^3\psR})
\,,\qquad&
\Fa_{1,4} =
c_3(\tfrac{\cosh\psR}{\psR^3\sinh^2\psR}+\tfrac{1+\cosh^2\psR}{\psR^2\sinh^3\psR})
\,.\\
\end{array}
\ee
We are now left with two unknown parameters $c_1,c_2$, but one can be a adjusted to zero by a shift of the center of the zeroth order solution,
$\vec{x}\rightarrow\vec{x}+c_0\vec{\vb}$, followed by a suitable gauge transformation to fix back the $\Fa_{1,6}=0$ gauge. This allows us to set $c_1=0$.
The parameter $c_2$ is undetermined at this point, since it depends on the proper boundary conditions at $\psR=\infty$ and at $x_3 =-1/\invCa$.


Now we move on to order $O(\invCa^2)$. The general ansatz at this order is:
\bear
\invCa^2\phi^{a(2)}
 &=& 
\vb^2 x_a\Fa_{2,1}(\psR)
\nn\\ &&
+[\vb_a (\vb_k x_k)-\tfrac{1}{3}\vb^2 x_a] \Fa_{2,3}(\psR)
+x_a[(\vb_k x_k)(\vb_m x_m)-\tfrac{1}{3}\vb^2\psR^2] \Fa_{2,4}(\psR)
\,,
\label{eqn:phiOb2x}\\
\invCa A_i^{a(2)} &=& 
\vb^2\epsilon_{iak}x_k\Fa_{2,2}(\psR)
+x_a\epsilon_{ijk}x_j\vb_k(\vb_m x_m)\Fa_{2,5}(\psR)
+x_i\epsilon_{ajk}x_j\vb_k(\vb_m x_m)\Fa_{2,6}(\psR)
\nn\\ &&
+\epsilon_{aij}[\vb_j(\vb_m x_m)-\tfrac{1}{3}\vb^2 x_j]\Fa_{2,7}(\psR)
+(\vb_i\epsilon_{ajk}x_j\vb_k - \tfrac{1}{3}\vb^2\epsilon_{aji}x_j)\Fa_{2,8}(\psR)
\,,
\label{eqn:AOb2x}
\eear
where we have separated the different terms according to whether they can be expressed in terms of the spin-$0$ combination $\vb^2\equiv\vb_k\vb_k$ or the spin-$2$ combination $\vb_k\vb_m-\tfrac{1}{3}\vb^2\delta_{km}$.
We again used the identity \eqref{eqn:Idxepsx} to eliminate the term $\epsilon_{aij} x_j (\vec{\vb}\cdot\vec{x})^2$, and we also note the identity
$
\vb_{[i}\epsilon_{a]jk}x_j\vb_k
=\tfrac{1}{2}\epsilon_{aij}\vb_j(\vb_k x_k)
-\tfrac{1}{2}\vb^2\epsilon_{aij}x_j\,,
$
which we used to eliminate a term of the form $\vb_a\epsilon_{ijk}x_j\vb_k\Fa_{2,9}$.
At order $O(\invCa^2)$ the possible gauge parameters are of the form
$$
\lambda^a = \epsilon_{abc}x_b\vb_c(\vb_k x_k)\Ga_{2,1}(\psR)\,,
$$
and we use the corresponding gauge transformation to gauge fix $\Fa_{2,8}=0$.

Our parameters $\Fa_{2,1}, \Fa_{2,2}$ correspond to spin-$0$ terms, while $\Fa_{2,3},\dots,\Fa_{2,7}$ correspond to spin-$2$ terms.
The spin-$2$ equations are:
\bear
0 &=& 
\tfrac{1}{\psR}\Fa_{2,4}'-\tfrac{1}{\psR}\Fa_{2,5}'
-\kPS\Fa_{2,4}+\kPS\Fa_{2,5}+(\hPS-\kPS)\Fa_{2,6}
\nn\\ && \qquad\qquad
+\tfrac{1}{\psR}\Fa_{1,2}'-\kPS\Fa_{1,2}-\hPS\Fa_{1,4}-\Fa_{1,2}\Fa_{1,4}-\Fa_{1,3}\Fa_{1,4}
\,,
\label{eqn:ODEOb2s2I}\\
0 &=& 
\tfrac{1}{\psR}\Fa_{2,7}'+\hPS\Fa_{2,7}
+\kPS\Fa_{2,3}+(1+\psR^2\kPS)\Fa_{2,4}-2\Fa_{2,6}
\nn\\ &&\qquad\qquad
+\kPS\Fa_{1,1}+\Fa_{1,2}+\psR^2\kPS\Fa_{1,2}-\Fa_{1,1}\Fa_{1,3}
\,,
\label{eqn:ODEOb2s2III}\\
0 &=& 
\tfrac{1}{\psR}\Fa_{2,3}'
-(1+\psR^2\kPS)\Fa_{2,5}+(1-\psR^2\hPS)\Fa_{2,6}+(\kPS-\hPS)\Fa_{2,7}
\nn\\ &&\qquad
+\tfrac{1}{\psR}\Fa_{1,1}'
+\psR^2\hPS\Fa_{1,4}+\Fa_{1,1}\Fa_{1,3}+\Fa_{1,1}\Fa_{1,4}
+\psR^2\Fa_{1,2}\Fa_{1,4}+\psR^2\Fa_{1,3}\Fa_{1,4}
\,,
\label{eqn:ODEOb2s2V}\\
0 &=& \psR\Fa_{2,5}'-\tfrac{1}{\psR}\Fa_{2,7}'
-\kPS\Fa_{2,3}+2\Fa_{2,4}+4\Fa_{2,5}+(2+\psR^2\kPS)\Fa_{2,6}+\kPS\Fa_{2,7}
\nn\\ &&\qquad\qquad
-\kPS\Fa_{1,1}+\Fa_{1,2}
+\Fa_{1,1}\Fa_{1,3}+\psR^2\Fa_{1,3}\Fa_{1,4}\,,
\label{eqn:ODEOb2s2VI}\\
0 &=&\Fa_{2,3}-\psR^2\Fa_{2,6}-\Fa_{2,7}-\psR^2\Fa_{1,1}\Fa_{1,3}-\psR^4\Fa_{1,3}\Fa_{1,4}
\,.\label{eqn:ODEOb2s2VII}
\eear

The spin-$0$ equations are:
\bear
0 &=&
\Fa_{2,1}'+\tfrac{1}{\psR}\Fa_{2,1}+\tfrac{2}{\psR}(1+\psR^2\kPS)\Fa_{2,2}
+\tfrac{1}{3}\Fa_{1,1}'+\tfrac{1}{3}\psR^2\Fa_{1,2}'
+\tfrac{2}{3}\psR\Fa_{1,2}-\tfrac{2}{3}\psR\Fa_{1,1}\Fa_{1,4}
\,,\label{eqn:ODEOb2s0IplusII}\\
0 &=&
\Fa_{2,2}'+(\tfrac{2}{\psR}+\psR\hPS)\Fa_{2,2}
+\tfrac{1}{\psR}(1+\psR^2\kPS)\Fa_{2,1}
\nn\\ &&
+\tfrac{1}{3}\psR\kPS\Fa_{1,1}+\tfrac{1}{3}\psR(1+\psR^2\kPS)\Fa_{1,2}
+\tfrac{1}{2}\psR\Fa_{1,1}\Fa_{1,3}-\tfrac{1}{3}\psR^3\Fa_{1,3}\Fa_{1,4}
\,.\label{eqn:ODEOb2s0IV}
\eear

We first solve the spin-$0$ equations.
The general solution is given by:
\bear
\Fa_{2,1} &=& 
\tfrac{\psR^2}{36\sinh^2\psR}+\tfrac{1}{6}\psR\coth\psR
+c_4\bigr(
\tfrac{1}{\sinh^2\psR}-\tfrac{1}{\psR}\coth\psR
\bigl)
+c_5\bigr(
\tfrac{1}{\psR\sinh^2\psR}
\bigl)
\,,\label{eqn:Fa2,1}\\
\Fa_{2,2} &=& 
\tfrac{\psR^2\cosh\psR}{36\sinh^2\psR}-\tfrac{\psR}{8\sinh\psR}
+c_4\bigr(
\tfrac{\cosh\psR}{\sinh^2\psR}-\tfrac{1}{\psR\sinh\psR}
\bigl)
+c_5\bigr(
\tfrac{\cosh\psR}{\psR\sinh^2\psR}
\bigl)
\,.\label{eqn:Fa2,2}
\eear
Since $c_5$ multiplies an $\psR$-odd and singular solution,
we set $c_5=0$. 
The unknown $c_4$ needs to be determined by the boundary conditions at $\psR=\infty$ and $x_3=-\Ca$.

Now, we move on to the spin-$2$ equations.
First we look for a solution of the homogeneous spin-$2$ part:
\bear
0 &=& 
\tfrac{1}{\psR}\Fa_{2,4}'-\tfrac{1}{\psR}\Fa_{2,5}'
-\kPS\Fa_{2,4}+\kPS\Fa_{2,5}+(\hPS-\kPS)\Fa_{2,6}
\,,
\label{eqn:ODEOb2s2Ih}\\
0 &=& 
\tfrac{1}{\psR}\Fa_{2,7}'+\hPS\Fa_{2,7}
+\kPS\Fa_{2,3}+(1+\psR^2\kPS)\Fa_{2,4}-2\Fa_{2,6}
\,,
\label{eqn:ODEOb2s2IIIh}\\
0 &=& 
\tfrac{1}{\psR}\Fa_{2,3}'
-(1+\psR^2\kPS)\Fa_{2,5}+(1-\psR^2\hPS)\Fa_{2,6}+(\kPS-\hPS)\Fa_{2,7}
\,,
\label{eqn:ODEOb2s2Vh}\\
0 &=& \psR\Fa_{2,5}'-\tfrac{1}{\psR}\Fa_{2,7}'
-\kPS\Fa_{2,3}+2\Fa_{2,4}+4\Fa_{2,5}+(2+\psR^2\kPS)\Fa_{2,6}+\kPS\Fa_{2,7}
\,,
\label{eqn:ODEOb2s2VIh}\\
0 &=&\Fa_{2,3}-\psR^2\Fa_{2,6}-\Fa_{2,7}
\,.\label{eqn:ODEOb2s2VIIh}
\eear
The general solution that is well behaved as $\psR\rightarrow\infty$ is:
\bear
\Fa_{2,3}^{(homog)} &=& 
c_6\{
\tfrac{4\psR}{\sinh\psR}
\}
+c_7\{
\tfrac{4}{\psR^4\sinh\psR}
\}
\,,\label{eqn:Fa2,3h}\\
\Fa_{2,4}^{(homog)} &=& 
c_6\{
\tfrac{6\cosh\psR-4}{\psR\sinh\psR}-\tfrac{2}{\sinh^2\psR}
\}
+c_7\{
-\tfrac{4(\cosh\psR+1)}{\psR^6\sinh\psR}
-\tfrac{2}{\psR^5\sinh^2\psR}
\}\,,\label{eqn:Fa2,4h}\\
\Fa_{2,5}^{(homog)} &=& 
c_6\{
-\tfrac{2\cosh\psR}{\sinh^2\psR}-\tfrac{4\cosh\psR-6}{\psR\sinh\psR}
\}
+c_7\{
-\tfrac{4(\cosh\psR+1)}{\psR^6\sinh\psR}
-\tfrac{2\cosh\psR}{\psR^5\sinh^2\psR}
\}\,,\label{eqn:Fa2,5h}\\
\Fa_{2,6}^{(homog)} &=& 
c_6\{
-\tfrac{2}{\psR\sinh\psR}+\tfrac{2\cosh\psR}{\sinh^2\psR}
\}
+c_7\{
\tfrac{8}{\psR^6\sinh\psR}
+\tfrac{2\cosh\psR}{\psR^5\sinh^2\psR}
\}\,,\label{eqn:Fa2,6h}\\
\Fa_{2,7}^{(homog)} &=& 
c_6\{
\tfrac{6\psR}{\sinh\psR}-\tfrac{2\psR^2\cosh\psR}{\sinh^2\psR}
\}
+c_7\{
-\tfrac{4}{\psR^4\sinh\psR}
-\tfrac{2\cosh\psR}{\psR^3\sinh^2\psR}
\}\,.\label{eqn:Fa2,7h}
\eear
Additionally, there are two more linearly independent solutions that grow exponentially as $\psR\rightarrow\infty$.
They are given by:
\bear
\Fa_{2,3}^{(homog)} &=& 
c_8\{
-\tfrac{2\cosh^2\psR}{\psR^2\sinh\psR}
+\tfrac{6\cosh\psR}{\psR^3}
-\tfrac{6\sinh\psR}{\psR^4}
\}\nn\\ &&
+c_9\{
-\tfrac{6\cosh\psR}{\psR^4} - \tfrac{2\cosh\psR}{\psR^2} + \tfrac{6\cosh^2\psR}{\psR^3\sinh\psR}
\}
\,,\label{eqn:Fa2,3hX}\\
\Fa_{2,4}^{(homog)} &=& 
c_8\{
\tfrac{6\sinh\psR}{\psR^6}
-\tfrac{3(1+2\cosh\psR)}{\psR^5}
+\tfrac{2\cosh^2\psR}{\psR^4\sinh\psR}
+\tfrac{1}{\psR^3\sinh^2\psR}
\}\nn\\ &&
+c_9\{
\tfrac{6(1 + \cosh\psR)}{\psR^6} - \tfrac{3(\cosh\psR + 2\cosh^2\psR)}{\psR^3\sinh\psR} 
+  \tfrac{2\cosh\psR}{\psR^4} - \tfrac{3}{\psR^4\sinh^2\psR}
\}
\,,\label{eqn:Fa2,4hX}\\
\Fa_{2,5}^{(homog)} &=& 
c_8\{
\tfrac{6\sinh\psR}{\psR^6}
-\tfrac{3(2+\cosh\psR)}{\psR^5}
+\tfrac{2\coth\psR}{\psR^4}
+\tfrac{\cosh\psR}{\psR^3\sinh^2\psR}
\}\nn\\ &&
+c_9\{
\tfrac{6\cosh\psR}{\psR^6} + \tfrac{6}{\psR^6} - \tfrac{3\cosh^2\psR}{\psR^5\sinh\psR} 
- \tfrac{6\coth\psR}{\psR^5} + \tfrac{2}{\psR^4} - \tfrac{3\cosh\psR}{\psR^4\sinh^2\psR}
\}
\,,\label{eqn:Fa2,5hX}\\
\Fa_{2,6}^{(homog)} &=& 
c_8\{
-\tfrac{12\sinh\psR}{\psR^6}
+\tfrac{9\cosh\psR}{\psR^5}
-\tfrac{2\cosh^2\psR}{\psR^4\sinh\psR}
-\tfrac{\cosh\psR}{\psR^3\sinh^2\psR}
\}\nn\\ &&
+c_9\{
-\tfrac{12\cosh\psR}{\psR^6} + \tfrac{9\cosh^2\psR}{\psR^5\sinh\psR}
 +  \tfrac{3\cosh\psR}{\psR^4\sinh^2\psR} - \tfrac{2\cosh\psR}{\psR^4}
\}
\,,\label{eqn:Fa2,6hX}\\
\Fa_{2,7}^{(homog)} &=& 
c_8\{
\tfrac{6\sinh\psR}{\psR^4}
-\tfrac{3\cosh\psR}{\psR^3}
+\tfrac{\cosh\psR}{\psR\sinh^2\psR}
\}\nn\\ &&
+c_9\{
\tfrac{6\cosh\psR}{\psR^4} - \tfrac{3\cosh^2\psR}{\psR^3\sinh\psR} - \tfrac{3\cosh\psR}{\psR^2\sinh^2\psR}
\}
\,.\label{eqn:Fa2,7hX}
\eear
Once we have a complete linearly independent set of solutions to the homogeneous problem, we can find the solution to the inhomogeneous problem by integration. When we perform the integrals we obtain complicated expressions that contain polylogarithms
$
\Li_n(z)\equiv\sum_{k=1}^\infty\tfrac{z^k}{k^n}\,.
$
For example, if we set $c_2=0$ in \eqref{eqn:Fa1,1}-\eqref{eqn:Fa1,4}, we get:
\bear
\Fa_{2,3}^{(inhomog)} &=& 
-\tfrac{9}{2\psR^4\sinh\psR}\Li_4(e^{-2\psR})
+\left[
\tfrac{3}{\psR^3}(\sinh\psR-\tfrac{2}{\sinh\psR})-(\tfrac{3}{\psR^4}+\tfrac{1}{\psR^2})\cosh\psR
\right]\Li_3(e^{-2\psR})
\nn\\ &&
-\bigl[
\tfrac{3}{\psR^2\sinh\psR}+(\tfrac{6}{\psR^3}+\tfrac{2}{\psR})\cosh\psR-\tfrac{6\sinh\psR}{\psR^2}
\bigr]\Li_2(e^{-2\psR})
\nn\\ &&
+\bigl[(\tfrac{6}{\psR^2}+2)\cosh\psR-\tfrac{6}{\psR}\sinh\psR\bigr]\log(1-e^{-2\psR})
-\tfrac{1}{\sinh\psR}\bigl(\tfrac{1}{2}+\tfrac{45}{2\psR^3}+\tfrac{2}{\psR}+\tfrac{59}{120}\psR\bigr)
\nn\\ &&
+\bigl(\tfrac{\psR^2}{8\sinh^2\psR}+\tfrac{45}{2\psR^4}+\tfrac{15}{2\psR^2}+\tfrac{2}{\psR}+2+\tfrac{2}{3}\psR\bigr)\cosh\psR
-\bigl(\tfrac{45}{2\psR^3}+\tfrac{2}{\psR}+2+\tfrac{2}{3}\psR\bigr)\sinh\psR
\,,\nn\\ &&\label{eqn:Fa2,3ihZ}
\eear
\bear
\Fa_{2,4}^{(inhomog)} &=& 
(\tfrac{9}{2\psR^6\sinh\psR}+\tfrac{9}{2\psR^6}\coth\psR+\tfrac{9}{4\psR^5\sinh^2\psR})\Li_4(e^{-2\psR})
\nn\\
&+&
\bigl\lbrack
\tfrac{3}{\psR^6}+(\tfrac{3}{\psR^6}+\tfrac{1}{\psR^4})\cosh\psR+\tfrac{15}{2\psR^5}\coth\psR
+\tfrac{6}{\psR^5\sinh\psR}+\tfrac{3}{\psR^4\sinh^2\psR}-\tfrac{3}{\psR^5}\sinh\psR
\bigr\rbrack\Li_3(e^{-2\psR})
\nn\\
&+&
\bigl\lbrack
\tfrac{6}{\psR^5}
+\tfrac{3}{2\psR^3\sinh^2\psR}
+(\tfrac{6}{\psR^5}+\tfrac{2}{\psR^3})\cosh\psR
+\tfrac{6}{\psR^4}\coth\psR
+\tfrac{3}{\psR^4\sinh\psR}
-\tfrac{6}{\psR^4}\sinh\psR
\bigr\rbrack\Li_2(E^{-2\psR})
\nn\\
&+&
\bigl\lbrack
\tfrac{6}{\psR^3}\sinh\psR
-\tfrac{3}{\psR^3}\coth\psR
-(\tfrac{6}{\psR^4}+\tfrac{2}{\psR^2})\cosh\psR
-\tfrac{6}{\psR^4}
\bigr\rbrack\log(1-e^{-2\psR})
\nn\\
&+&
(\tfrac{45}{2\psR^5}+\tfrac{2}{\psR^3}+\tfrac{2}{\psR^2}+\tfrac{2}{3\psR})\sinh\psR
+(\tfrac{45}{4\psR^4}+\tfrac{1}{\psR^2}+\tfrac{1}{4\psR}+\tfrac{37}{120})\tfrac{1}{\sinh^2\psR}
-\tfrac{\cosh\psR}{8\sinh^2\psR}
\nn\\
&+&
(\tfrac{45}{2\psR^5}+\tfrac{2}{\psR^3}+\tfrac{1}{2\psR^2}+\tfrac{59}{120\psR})\tfrac{1}{\sinh\psR}
+(\tfrac{45}{4\psR^5}+\tfrac{2}{\psR^3}-\tfrac{1}{2\psR^2}+\tfrac{1}{5\psR})\coth\psR
\nn\\
&-&
(\tfrac{45}{2\psR^6}+\tfrac{15}{2\psR^4}+\tfrac{2}{\psR^3}+\tfrac{2}{\psR^2}+\tfrac{2}{3\psR})\cosh\psR
-\tfrac{45}{2\psR^6}-\tfrac{2}{\psR^3}+\tfrac{1}{8\psR^2}
\,,
\label{eqn:Fa2,4ihZ}
\eear
\bear
\Fa_{2,5}^{(inhomog)} &=& 
(\tfrac{9}{2\psR^6\sinh\psR}+\tfrac{9}{2\psR^6}\coth\psR+\tfrac{9\cosh\psR}{4\psR^5\sinh^2\psR})\Li_4(e^{-2\psR})
\nn\\ &&
+(\tfrac{3}{\psR^6}+\tfrac{1}{\psR^4}+\tfrac{3}{\psR^6}\cosh\psR-\tfrac{3}{2\psR^5}\sinh\psR+\tfrac{6}{\psR^5}\coth\psR+\tfrac{15}{2\psR^5\sinh\psR}+\tfrac{3\cosh\psR}{\psR^4\sinh^2\psR})\Li_3(e^{-2\psR})
\nn\\ &&
+(\tfrac{6}{\psR^5}+\tfrac{2}{\psR^3}+\tfrac{6}{\psR^5}\cosh\psR-\tfrac{3}{\psR^4}\sinh\psR+\tfrac{3}{\psR^4}\coth\psR+\tfrac{6}{\psR^4\sinh\psR}+\tfrac{3\cosh\psR}{2\psR^3\sinh^2\psR})\Li_2(e^{-2\psR})
\nn\\
&-&
\bigl\lbrack
\tfrac{6}{\psR^4}
+\tfrac{2}{\psR^2}
+\tfrac{6}{\psR^4}\cosh\psR
-\tfrac{3}{\psR^3}\sinh\psR
+\tfrac{3}{\psR^3\sinh\psR}
\bigr\rbrack\log(1-e^{-2\psR})
\nn\\ &&
-\tfrac{45}{2\psR^6}-\tfrac{15}{2\psR^4}-\tfrac{2}{\psR^3}-\tfrac{3}{8\psR^2}-\tfrac{2}{3\psR}
-\bigl(
\tfrac{45}{2\psR^6}+\tfrac{2}{\psR^3}+\tfrac{1}{\psR^2}
\bigr)\cosh\psR
\nn\\ &&
+\bigl(
\tfrac{45}{4\psR^5}+\tfrac{2}{\psR^3}+\tfrac{1}{\psR^2}
\bigr)\sinh\psR
+\bigl(
\tfrac{45}{2\psR^5}+\tfrac{2}{\psR^3}+\tfrac{1}{2\psR^2}+\tfrac{13}{15\psR}
\bigr)\coth\psR
\nn\\ &&
+\bigl(
\tfrac{45}{4\psR^5}+\tfrac{2}{\psR^3}-\tfrac{1}{2\psR^2}-\tfrac{27}{40\psR}
\bigr)\tfrac{1}{\sinh\psR}
+\bigl(
\tfrac{45}{4\psR^4}+\tfrac{1}{\psR^2}+\tfrac{1}{4\psR}+\tfrac{37}{120}
\bigr)\tfrac{\cosh\psR}{\sinh^2\psR}
-\tfrac{1}{8\sinh^2\psR}
\,,\label{eqn:Fa2,5ihZ}
\eear
\bear
\Fa_{2,6}^{(inhomog)} &=& 
-(\tfrac{9}{\psR^6\sinh\psR}+\tfrac{9\cosh\psR}{4\psR^5\sinh^2\psR})\Li_4(e^{-2\psR})
\nn\\ &&
-\bigl\lbrack(\tfrac{6}{\psR^6}+\tfrac{1}{\psR^4})\cosh\psR-\tfrac{9}{2\psR^5}\sinh\psR+\tfrac{27}{2\psR^5\sinh\psR}+\tfrac{3\cosh\psR}{\psR^4\sinh^2\psR}\bigr\rbrack\Li_3(e^{-2\psR})
\nn\\ &&
-\bigl\lbrack
(\tfrac{12}{\psR^5}+\tfrac{2}{\psR^3})\cosh\psR
-\tfrac{9}{\psR^4}\sinh\psR
+\tfrac{9}{\psR^4\sinh\psR}
+\tfrac{3\cosh\psR}{2\psR^3\sinh^2\psR}
\bigr\rbrack\Li_2(e^{-2\psR})
\nn\\ &&
+\bigl\lbrack
(\tfrac{12}{\psR^4}+\tfrac{2}{\psR^2})\cosh\psR
-\tfrac{9}{\psR^3}\sinh\psR
+\tfrac{3}{\psR^3\sinh\psR}
\bigr\rbrack\log(1-e^{-2\psR})
\nn\\ &&
-(\tfrac{45}{4\psR^4}+\tfrac{1}{\psR^2}+\tfrac{1}{4\psR}+\tfrac{13}{30})\tfrac{\cosh\psR}{\sinh^2\psR}
+(\tfrac{45}{\psR^6}+\tfrac{15}{2\psR^4}+\tfrac{4}{\psR^3}+\tfrac{3}{\psR^2}+\tfrac{2}{3\psR})\cosh\psR
\nn\\ &&
-(\tfrac{135}{4\psR^5}+\tfrac{4}{\psR^3}+\tfrac{3}{\psR^2}+\tfrac{2}{3\psR})\sinh\psR
-(\tfrac{135}{4\psR^5}+\tfrac{4}{\psR^3}-\tfrac{13}{30\psR})\tfrac{1}{\sinh\psR}
\,,\label{eqn:Fa2,6ihZ}
\eear
\bear
\Fa_{2,7}^{(inhomog)} &=& 
\Fa_{2,3}^{(inhomog)}-\psR^2\Fa_{2,6}^{(inhomog)}-\psR^2\Fa_{1,1}\Fa_{1,3}-\psR^4\Fa_{1,3}\Fa_{1,4}
\,.\label{eqn:Fa2,7ihZ}
\eear
We note that the combinations of polylogarithms that appear here are the results of the integrals
$$
\int\psR^3\coth\psR\,d\psR =
-\tfrac{3}{4}\Li_4(e^{-2\psR})
-\tfrac{3}{2}\psR\Li_3(e^{-2\psR})
-\tfrac{3}{2}\psR^2\Li_2(e^{-2\psR})
+\psR^3\log(1 -e^{-2\psR})
+\tfrac{1}{4}\psR^4
\,,
$$
and
$$
\int\psR^2\coth\psR\,d\psR =
-\tfrac{1}{2}\Li_3(e^{-2\psR})
-\psR\Li_2(e^{-2\psR})
+\psR^2\log(1 -e^{-2\psR})
+\tfrac{1}{3}\psR^3
\,.
$$
When we turn on $c_2\neq 0$ we get additional terms, but they can be expressed as rational functions of $e^\psR$ and $\psR$ and do not cancel the polylogarithms. In any case, this demonstrates that a simple solution to the BPS equations \eqref{eqn:Drphi}, involving only basic functions, does not exist.

\section{Discussion}
We have studied a 2+1D system constructed from the compactification of the $(2,0)$-theory on $(\R^2\times S^1)/\Z_\lvk$.
In the large $\lvk$ limit, we have reduced it to 4+1D SYM on the ``cigar'' geometry, and we have developed the BPS equations that describe Q-ball solitons.
In terms of the effective FQHE low-energy action, these solitons are bound states of $\lvk$ quasi-particles (each of $1/\lvk$ charge).
We mapped the BPS equations to the Bogomolnyi equations $D\Phi = {}*F$ on a manifold with metric
\be\label{eqn:Mmetric}
ds^2 = x_3^2 (dx_1^2 + dx_2^2 + dx_3^2),
\ee
and we described a relation between axisymmetric solutions (in particular, the 1-monopole solution) and harmonic maps $\varphi:AdS_3\rightarrow AdS_2$.
It would be interesting to explore this system further. 
We note that other interesting extensions of the classic Bogomolnyi equations were discovered in \cite{Hashimoto:2005hy}, in the context of D$3$-brane probes of a Melvin space (which is in fact T-dual to the orbifold background in our work), where the D$3$-branes are oriented in such a way that noncommutative geometry with a variable parameter is generated.

Our problem is reminiscent of the problem of monopoles on $AdS_3$ [if $x_3^2$ is replaced with $1/x_3^2$ in \eqref{eqn:Mmetric}]. The latter is integrable, with known solutions, and in particular the one-monopole solution is not difficult to construct \cite{Red'kov:2003uz}. Like the case of monopoles on $AdS_3$, the monopole solutions on the space \eqref{eqn:Mmetric} contain as a limit the classic Prasad-Sommerfield solutions (by going to the outskirts $x_3\rightarrow\infty$). Indeed, in \secref{subsec:LVexp} we outlined an expansion around the Prasad-Sommerfield solution, up to second order in $1/x_3$, albeit with a few undetermined coefficients.

Monopole equations on a three-dimensional space can be recast as the dimensional reduction of instanton equations on a four-dimensional space, which can provide new insights. For example, instanton equations on Taub-NUT spaces can be reduced to Bogomolnyi's equations on $\R^3$ (with singularities) \cite{Kronheimer:1985}, which recently led to the discovery of new explicit solutions \cite{Cherkis:2007qa,Blair:2010kz}, using the techniques developed in \cite{Cherkis:2008ip,Cherkis:2009jm} for solving instanton equations on Taub-NUT spaces. It might therefore be interesting to explore instanton equations on circle fibrations over \eqref{eqn:Mmetric} and look for their applications in string theory. More recently, a set of partial differential equations on $G_2$-manifolds was discovered \cite{Cherkis:2014xua}, which can be reduced in special cases to Bogomolnyi's equations on $\R^3$. It would be interesting to explore whether the system studied in this paper and the related Bogomolnyi equations on \eqref{eqn:Mmetric} have an interesting $7$-dimensional origin.

In this paper we focused on the case of a single monopole, corresponding to a $(2,0)$-string wound once.
It would be interesting to generalize the discussion to the case of multiple $(2,0)$-strings, which corresponds to monopole charge higher than $1$ in the effective metric \eqref{eqn:Mmetric}. Techniques for analyzing the low-energy description of multiple $(2,0)$-strings have recently been developed in \cite{Haghighat:2013gba}-\cite{Haghighat:2013tka}.


\section*{Acknowledgements}
We are grateful to Oren Bergman, Sergey Cherkis, Petr Ho\v{r}ava, and Hongyu Xiong for very helpful discussions, comments, and suggestions.
This research was supported by the Berkeley Center of Theoretical Physics. The research of N.~Moore was also supported in part by the U.S. National Science Foundation under grant No. PHY-10-02399. H.~S.~Tan would like to acknowledge support form the Fundamental Research on Matter (FOM) which is part of the Netherlands Organization for Scientific Research. The research of N.~R.~Torres-Chicon was supported by the National Science Foundation Graduate Research Fellowship Program under Grant No. DGE-11-06400.

\begin{appendix}

\section{Recasting the BPS equations in terms of a single potential}
\label{app:singlePhi}

The action \eqref{eqn:fpsiI} is invariant under dilatations that act as
$$
\Ff(r,\rE)\rightarrow\Ff(\lambda r,\lambda\rE)\,,\qquad
\Fpsi(r,\rE)\rightarrow\Fpsi(\lambda r,\lambda\rE)\,.
$$
The components of the associated Noether current are given by
\bear
J^r &=&
\frac{\rE\Ff_r^2}{2\Ff^2}
+\frac{\rE^2\Ff_r\Ff_\rE}{r\Ff^2}
-\frac{\rE\Ff_\rE^2}{2\Ff^2}
+\frac{\rE\Fpsi_r^2}{2\Ff^2}
+\frac{\rE^2\Fpsi_r\Fpsi_\rE}{r\Ff^2}
-\frac{\rE\Fpsi_\rE^2}{2\Ff^2}
\,,
\nn\\
J^\rE &=&
\frac{\rE^2\Ff_\rE^2}{2r\Ff^2}
+\frac{\rE\Ff_r\Ff_\rE}{\Ff^2}
-\frac{\rE^2\Ff_r^2}{2r\Ff^2}
+\frac{\rE^2\Fpsi_\rE^2}{2r\Ff^2}
+\frac{\rE\Fpsi_r\Fpsi_\rE}{\Ff^2}
-\frac{\rE^2\Fpsi_r^2}{2r\Ff^2}
\,.
\nn
\eear
The equations of motion \eqref{eqn:Epsi}-\eqref{eqn:Ef} imply the conservation equation\footnote{But note that $(J^r,J^\rE)$ are not directly related to the stress-energy tensor derived from the original (``physical'') action in the original fields $A_i$ and $\ScPhi$. The ``physical'' conserved currents associated with dilatations generally vanishes on BPS configurations \cite{Domokos:2013xqa}.}
$$
(J^r)_r +(J^\rE)_\rE = 0,
$$
which implies that there exists a potential function $\TPhi$ such that
\be\label{eqn:JTPhi}
J^\rE = \TPhi_r\,,\qquad
J^r = -\TPhi_\rE\,.
\ee
To proceed, we think of the functions $\Ff$ and $\Fpsi$ as defining a change of coordinates from $(\Ff,\Fpsi)$ to $(r,\rE)$ [similar to \eqref{eqn:HmapAdS}, except with the $\phi$ coordinate absent].
In $(r,\rE)$ coordinates, the $AdS_2$ metric \eqref{eqn:AdS2} becomes:
\be\label{eqn:dsG}
ds^2 =\psG_{rr}dr^2+2\psG_{r\rE}dr d\rE + \psG_{\rE\rE}d\rE^2\,,
\ee
where the metric $\psG$ can be expressed, using \eqref{eqn:JTPhi}, as:
\bear
\psG_{rr} &=&
-\tfrac{r^2}{r^2+\rE^2}(\TPhi_{\rE\rE}+\TPhi_{rr}+\tfrac{1}{r}\TPhi_r+\tfrac{1}{\rE}\TPhi_\rE)
\,,\nn\\
\psG_{\rE\rE} &=&
-\tfrac{r^2}{r^2+\rE^2}(\TPhi_{\rE\rE}+\TPhi_{rr}-\tfrac{1}{r}\TPhi_r-\tfrac{1}{\rE}\TPhi_\rE)
\,,\nn\\
\psG_{r\rE} &=& \tfrac{r}{\rE(r^2+\rE^2)}(r\TPhi_r-\rE\TPhi_\rE)
\,.\nn
\eear
$\TPhi$ then satisfies a nonlinear differential equation that states that the Ricci scalar of \eqref{eqn:dsG} is $R=-2$.
In order to incorporate the Dirac string for $r<\Ca$, the function $\TPhi$ must diverge like $\log\rE$ as $\rE\rightarrow 0$ and $r<\Ca$.
Define $\nZ$ and $\nR$ by:
\be\label{eqn:nRnZdef0}
\nZ\equiv\tfrac{1}{2\Ca}(\rE^2+r^2-\Ca^2)\,,\qquad
\nR\equiv\sqrt{\rE^2+\nZ^2}=\tfrac{1}{2\Ca}\sqrt{
(\rE^2+r^2-\Ca^2)^2+4\Ca^2\rE^2}
\,.
\ee
For large $\Ca$, the solution to $\Ff$ and $\Fpsi$ is given by adapting the Prasad-Sommerfield solution as given by \cite{Forgacs:1980yv}:
\be\label{eqn:FfPS}
\Ff = \frac{\rE\sinh\nR}{\nR+\nR\cosh\nR\cosh\nZ-\nZ\sinh\nZ\sinh\nR}
\,,\qquad
\Fpsi=\frac{\nZ\cosh\nZ\sinh\nR-\nR\sinh\nZ\cosh\nR}{\nR+\nR\cosh\nR\cosh\nZ-\nZ\sinh\nZ\sinh\nR}\,,
\ee
where we have set the VEV to $\Vph=1$, and we have used $\nR$ as a substitute for the distance from the core of the monopole.
From this we find, in the large $\Ca$ limit,
\be
\TPhi \rightarrow
-\tfrac{1}{4}\rE^2+\tfrac{1}{2}\log\rE-\log\nR+\log\sinh\nR\,.
\label{eqn:TPhiLargeCa}
\ee
We also note that the abelian solution
$$
\Ff =\left(\frac{\nR-\nZ}{2\Ca}\right)e^{-\frac{1}{2}\Vph r^2}
\,,\qquad
\Fpsi = 0\,,
$$
can be derived from the potential
$$
\TPhi = 
\tfrac{1}{4}\Vph^2r^2\rE^2+\tfrac{1}{2}\Vph(2\Ca\nR+r^2-\rE^2)
+\log\left\lbrack\frac{2\Ca\nR}{(\nR-\nZ)(\Ca+\nR+\nZ)}\right\rbrack\,.
$$
Finally, we note that a change of variables,
$$
r + i\rE =\Ca e^\tauIsig\,,\qquad
r-i\rE = \Ca e^\btauIsig\,,
$$
converts the metric \eqref{eqn:dsG} to the more compact form:
\be\label{eqn:dsSqNewVars}
ds^2 = 
-4\cosh^2(\tfrac{\tauIsig-\btauIsig}{2})\TPhi_{\tauIsig\btauIsig}d\tauIsig d\btauIsig
+\coth(\tfrac{\tauIsig-\btauIsig}{2}) (\TPhi_\tauIsig d\tauIsig^2-\TPhi_\btauIsig d\btauIsig^2)
\,,
\ee
where $\TPhi_\tauIsig\equiv\partial\TPhi/\partial\tauIsig$, $\TPhi_\btauIsig\equiv\partial\TPhi/\partial\btauIsig$, and $\TPhi_{\tauIsig\btauIsig}\equiv\partial^2\TPhi/\partial\tauIsig\partial\btauIsig$.
The equation to solve is again $R=-2$, where $R$ is the Ricci scalar calculated from the metric \eqref{eqn:dsSqNewVars}, and the result is a rather length nonlinear partial differential equation for the single field $\TPhi$, which we will not present here.

\section{Numerical results}
\label{app:Numerical}

\begin{figure}[t]
\includegraphics{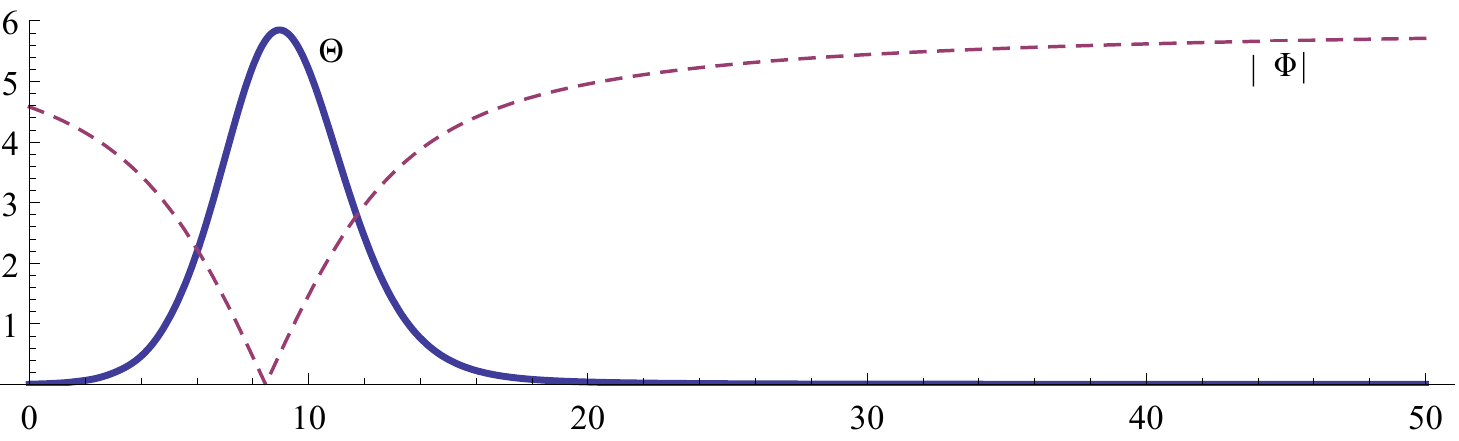}

\caption{
Results of a numerical analysis with parameters $\Cb=2.60$ and $\MaxUVn=22$. 
The graphs show the energy density $\Theta\equiv\Uex/\xV$ (solid line) and the gauge invariant absolute value of the scalar field $|\ResPhi|\equiv(\ResPhi^a\ResPhi^a)^{1/2}$ (dashed line) for VEV $\Vph=1$ and soliton center at $\Ca=1$.
The graphs are on the axis $\xU=0$ and the horizontal axis is $\xV$.
The vertical axis refers to $\Theta$, and the asymptotic value of $|\ResPhi|$ is $1$.
At $\xV=0$ the value of $\Theta$ is $1.5\times 10^{-3}$ and the value of $|\ResPhi|$ is $0.76$.
The value of the excess energy $\Eex$ for this configuration is less than $2\times 10^{-5}$ of $\Ebps$.
}
\label{fig:EnergyDensityAndAbsPhi}
\end{figure}

As a first step towards a numerical analysis of the solution to the BPS equations \eqref{eqn:DiphiF} we find it convenient to recast the equations in a different gauge from the one we used in the main text.
We begin by parameterizing the scalar field components as:
\be\label{eqn:Mansatzphi}
\phi^\alpha = x_\alpha(\fMZ+\hMZ)\,,
\quad
\phi^3 = \gMZ\,,
\ee
and the gauge field components as:
\be\label{eqn:MansatzA}
A_\beta^\alpha =
x_{(\beta}\epsilon_{\alpha)\gamma}x^\gamma\pMZ
+\tfrac{1}{2}\epsilon_{\alpha\beta}\qMZ
\,,\quad
A_r^\alpha = -r\epsilon_{\alpha\gamma}x_\gamma(\fMZ-\hMZ)\,,
\quad
A_\beta^3 = \epsilon_{\beta\gamma}x^\gamma\vMZ\,,
\quad
A_r^3 = 0\,.
\ee
with $\alpha,\beta,\gamma=1,2$, $\epsilon_{\alpha\beta}$ the anti-symmetric Levi-Civita symbol, and $\fMZ$, $\gMZ$, $\hMZ$, $\pMZ$, $\qMZ$, and $\vMZ$ functions of $(r,\rE)$ only.
Next, we fix the gauge by setting $\pMZ=0$. Defining
$$
\xU\equiv\rE^2\,,\qquad\xV\equiv r^2,
$$
the BPS equations \eqref{eqn:DiphiF} reduce (after rescaling $\phi$ by $\lvk\xR$) to:
\bear
0 &=&
\hMZ\vMZ-2\tfrac{\partial\hMZ}{\partial\xU}
\,,\label{eqn:EqW1p=0}\\
0 &=&
2\xU\tfrac{\partial\fMZ}{\partial\xU}+\xU\vMZ\fMZ+2\fMZ
+\tfrac{\partial\qMZ}{\partial\xV}+\tfrac{1}{2}\gMZ\qMZ
\,,\label{eqn:EqW2p=0}\\
0&=&
\xV(\hMZ-\fMZ)\gMZ
+\tfrac{1}{2}\qMZ\vMZ
+2\xV\tfrac{\partial\fMZ}{\partial\xV}
+2\xV\tfrac{\partial\hMZ}{\partial\xV}
-\tfrac{\partial\qMZ}{\partial\xU}
\,,\label{eqn:EqW3p=0}\\
0 &=&
\tfrac{\partial\vMZ}{\partial\xV}
-\tfrac{\partial\gMZ}{\partial\xU}
+\tfrac{1}{2}\hMZ\qMZ
\,,\label{eqn:EqW4p=0}\\
0 &=&
\xU\xV(\fMZ^2-\hMZ^2)
+\tfrac{1}{4}\qMZ^2
+2\vMZ
+2\xV\tfrac{\partial\gMZ}{\partial\xV}
+2\xU\tfrac{\partial\vMZ}{\partial\xU}
\,.\label{eqn:EqW5p=0}
\eear
Let us also set
\be\label{eqn:nRnZdef1}
\nZ\equiv\tfrac{1}{2\Ca}(\rE^2+r^2-\Ca^2)\,,\qquad
\nR\equiv\sqrt{\rE^2+\nZ^2}=\tfrac{1}{2\Ca}\sqrt{
(\rE^2+r^2-\Ca^2)^2+4\Ca^2\rE^2}
\,,
\ee
as in \eqref{eqn:nRnZdef0}.
The advantage of the ansatz \eqref{eqn:Mansatzphi}-\eqref{eqn:MansatzA} is that the abelian solution \eqref{eqn:abphisol}-\eqref{eqn:abAsol} can be written in the form:
\be\label{eqn:fghqvAb}
\fMZ =
\frac{\Vph}{2\nR}-\frac{1}{\Ca\nR^2}
\,,\quad
\gMZ =
\frac{\Vph\nZ}{\nR}-\frac{\nZ}{\Ca\nR^2}
\,,\quad
\hMZ =
\frac{\Vph}{2\nR}
\,,\quad
\qMZ =
\frac{\Ca^2+\xU-\xV}{\Ca\nR^2}
\,,\quad
\vMZ =
-\frac{1}{\nR^2}-\frac{\nZ}{\Ca\nR^2}
\,.
\ee
which has no singularities except at $r=\Ca$ (and in particular no Dirac string).

We now require that at either limit $r\rightarrow\infty$ or $\rE\rightarrow\infty$ the full solution should reduce to the abelian solution.
At the tip $r=0$ the solution is required to be regular. This allows us to determine $\qMZ$, $\hMZ$, and $\vMZ$ at the tip as follows.
Setting $\xV=0$ in \eqref{eqn:EqW1p=0}, \eqref{eqn:EqW3p=0}, and \eqref{eqn:EqW5p=0}, we get the ordinary differential equations
\be\label{eqn:ODEtip}
\hMZ\vMZ-2\tfrac{\partial\hMZ}{\partial\xU}
=
\tfrac{1}{2}\qMZ\vMZ
-\tfrac{\partial\qMZ}{\partial\xU}
=
\tfrac{1}{4}\qMZ^2
+2\vMZ
+2\xU\tfrac{\partial\vMZ}{\partial\xU}
=0\,,
\qquad(\xV=0)
\ee
which we can solve uniquely, given the known boundary conditions at $\xU\rightarrow\infty$. This is easily done by expressing $\qMZ$ and $\hMZ$ in terms of the function $(1+\xU\vMZ)$ and its derivatives, and changing variables to $\log\xU$. The result is that unique solution to \eqref{eqn:ODEtip} that satisfies the boundary conditions at $\xU\rightarrow\infty$ is
\be\label{eqn:qvhTip}
\qMZ = \frac{4\Ca}{\xU+\Ca^2}\,,\qquad
\vMZ=-\frac{2}{\xU+\Ca^2}\,,\qquad
\hMZ = \frac{\Vph\Ca}{\xU+\Ca^2}\,,
\qquad(\xV=0),
\ee
which is none other than the abelian solution \eqref{eqn:fghqvAb} at $\xV=0$.

We cannot determine $\fMZ$ and $\gMZ$ at $\xV=0$ so easily, and our strategy will be to find an approximate solution to \eqref{eqn:EqW1p=0}-\eqref{eqn:EqW5p=0} by the variational method, minimizing the energy of the field configuration within a certain class of trial functions of $(\xU,\xV)$.
For the energy we take the expression for the excess energy above the BPS bound for a stationary configuration of gauge field and minimally coupled adjoint scalar on a manifold given by the three dimensional metric \eqref{eqn:effds}, that is,
\bear
\Eex &\equiv&
\frac{1}{2}\tr\int \sqrt{g} g^{ij}(D_i\ResPhi-B_i)(D_j\ResPhi-B_j) d^3 x
\nn\\
&=&
\frac{1}{2}\tr\int
\left\lbrack
(r D_r\ResPhi- F_{12})^2 
+(r D_1\ResPhi- F_{2r})^2 
+(r D_2\ResPhi- F_{r1})^2 
\right\rbrack\rE d\rE (\tfrac{dr}{r})\,,
\label{eqn:Edef}
\eear
where $B_i$ and $\ResPhi$ were defined in \eqref{eqn:defResPhiB}, and the ``$\tr$'' is in the fundamental representation.
Note that $\Eex$ is different from the physical energy \eqref{eqn:Estatic}.
The integrand in \eqref{eqn:Edef} is $\cfR/r^2$ bigger than the integrand in the first term on the RHS of \eqref{eqn:Estatic}, but they are both minimized on the BPS configurations, and \eqref{eqn:Edef} gives more weight to the vicinity of $r=0$.
We can rewrite $\Eex$ in terms of the right-hand-sides of \eqref{eqn:EqW1p=0}-\eqref{eqn:EqW5p=0} as follows. Setting
\bear
\EqX_1 &=&
\hMZ\vMZ-2\tfrac{\partial\hMZ}{\partial\xU}
\,,\label{eqn:EqX1}\\
\EqX_2 &=&
2\xU\tfrac{\partial\fMZ}{\partial\xU}+\xU\vMZ\fMZ+2\fMZ
+\tfrac{\partial\qMZ}{\partial\xV}+\tfrac{1}{2}\gMZ\qMZ
\,,\label{eqn:EqX2}\\
\EqX_3 &=&
\xV(\hMZ-\fMZ)\gMZ
+\tfrac{1}{2}\qMZ\vMZ
+2\xV\tfrac{\partial\fMZ}{\partial\xV}
+2\xV\tfrac{\partial\hMZ}{\partial\xV}
-\tfrac{\partial\qMZ}{\partial\xU}
\,,\label{eqn:EqX3}\\
\EqX_4 &=&
\tfrac{\partial\vMZ}{\partial\xV}
-\tfrac{\partial\gMZ}{\partial\xU}
+\tfrac{1}{2}\hMZ\qMZ
\,,\label{eqn:EqX4}\\
\EqX_5 &=&
\xU\xV(\fMZ^2-\hMZ^2)
+\tfrac{1}{4}\qMZ^2
+2\vMZ
+2\xV\tfrac{\partial\gMZ}{\partial\xV}
+2\xU\tfrac{\partial\vMZ}{\partial\xU}
\,,\label{eqn:EqX5}
\eear
we get \eqref{eqn:Edef} in the form
\be
\Eex = 
\int\bigl(
\frac{1}{8}\xU^2\EqX_1^2
+\frac{1}{8}\EqX_2^2
+\frac{\EqX_3^2}{16\xV}
+\frac{1}{4}\xU\EqX_4^2
+\frac{\EqX_5^2}{16\xV}
\bigr)d\xU d\xV
\,.
\ee
We also note that the BPS bound on energy is given by
\bear
\Ebps &=& \tr\int \sqrt{g} g^{ij}(B_j D_i\ResPhi) d^3 x
\nn\\ &=&
\tr\int \left\lbrack
F_{12}D_r\ResPhi
+F_{2r}D_1\ResPhi
+F_{r1}D_2\ResPhi
\right\rbrack
 \rE d\rE dr
=\int d\lambdaBPS\,,
\eear
where the $1$-form $\lambda$ is defined by
\bear
\lambdaBPS &\equiv& 
\left\lbrack
\tfrac{1}{8}\xU\qMZ\vMZ(\fMZ+\hMZ)
+\tfrac{1}{16}\qMZ^2\gMZ
+\tfrac{1}{2}\vMZ\gMZ
+\tfrac{1}{2}\xU\gMZ\frac{\partial\vMZ}{\partial\xU}
-\tfrac{1}{4}\xU(\fMZ+\hMZ)\frac{\partial\qMZ}{\partial\xU}
\right\rbrack d\xU
\nn\\
&+&\left\lbrack
\tfrac{1}{8}\xU\gMZ\qMZ(\hMZ-\fMZ)
+\tfrac{1}{4}\xU(\hMZ^2-\fMZ^2)(1+\xU\vMZ)
-\tfrac{1}{4}\xU(\fMZ+\hMZ)\frac{\partial\qMZ}{\partial\xV}
+\tfrac{1}{2}\xU\gMZ\frac{\partial\vMZ}{\partial\xV}
\right\rbrack d\xV\,.
\nn
\eear
Requiring the asymptotic behavior for large $\xU$ and $\xV$ to be as in \eqref{eqn:fghqvAb}, we find
$$
\Ebps = 2\Vph\,.
$$

We construct our trial functions by modifying the abelian solution \eqref{eqn:fghqvAb}. 
But first we need to smooth out the singularity of that solution at $\xV=\Ca^2$ while preserving the asymptotic behavior at large $\xU$ and $\xV$, as well as the behavior \eqref{eqn:qvhTip} at $\xV=0$. 
For this purpose we define:
\be
\SmoothedR\equiv\sqrt{\xU+\xV+\Ca^2} = \sqrt{r^2+\rE^2+\Ca^2}
\ee
and then define smoothed versions of $\fMZ$, $\gMZ$, $\hMZ$, $\qMZ$, $\vMZ$:
\bear
\tfMZ &\equiv&
\frac{\Ca\Vph}{\SmoothedR^2}
+\frac{2\Ca(\Vph\Ca^2-2)}{\SmoothedR^4}
-\frac{2\Ca^3\Vph\xU}{\SmoothedR^6}
\,,\nn\\
\tgMZ &\equiv&
\Vph
-\frac{2}{\SmoothedR^2}
-\frac{2\Vph\Ca^2\xU}{\SmoothedR^4}
\,,\nn\\
\thMZ &\equiv&
\Vph\left(\frac{\Ca}{\SmoothedR^2}
+\frac{2\Ca^3}{\SmoothedR^4}
-\frac{2\Ca^3\xU}{\SmoothedR^6}
-\frac{2\Ca^5(\Ca^2+\xU)}{\SmoothedR^8}
\right)
\,,\nn\\
\tqMZ &\equiv&
\frac{4\Ca}{\SmoothedR^2}
-\frac{8\Ca\xV}{\SmoothedR^4}
\,,\nn\\
\tvMZ &\equiv&
-\frac{2}{\SmoothedR^2}
-\frac{8\Ca^2}{\SmoothedR^4}
+\frac{8\Ca^2(\Ca^2+\xU)}{\SmoothedR^6}
\,,\nn
\eear
so that for $\xV\rightarrow\infty$ at fixed $\xU$ we have
\bear
\tfMZ &=&
\frac{\Vph}{2\nR}-\frac{1}{\Ca\nR^2}
+O\left(\frac{1}{\xV^4}\right)
\,,\nn\\
\tgMZ &=&
\frac{\Vph\nZ}{\nR}-\frac{\nZ}{\Ca\nR^2}
+O\left(\frac{1}{\xV^3}\right)
\,,\nn\\
\thMZ &=&
\frac{\Vph}{2\nR}
+O\left(\frac{1}{\xV^4}\right)
\,,\nn\\
\tqMZ &=&
\frac{\Ca^2+\xU-\xV}{\Ca\nR^2}
+O\left(\frac{1}{\xV^3}\right)
\,,\nn\\
\tvMZ &=&
-\frac{1}{\nR^2}-\frac{\nZ}{\Ca\nR^2}
+O\left(\frac{1}{\xV^4}\right)
\,,\nn
\eear
and $\tfMZ$, $\tgMZ$, $\thMZ$, $\tvMZ$, $\tqMZ$ are smooth everywhere.
We also define
$$
\SmoothedRCb\equiv\sqrt{\xU+\xV+\Cb^2} = \sqrt{r^2+\rE^2+\Cb^2}\,,
$$
where $\Cb$ is a parameter to be determined dynamically by the variational principle.
We now pick a sufficiently large integer $\MaxUVn$ (we chose $\MaxUVn=20$ below), and construct trial functions in the form:
\bear
\fMZ &=& \tfMZ + \frac{1}{\SmoothedRCb^{5+2\MaxUVn}}\sum_{n,m\ge 0}^{n+m\le\MaxUVn}\fMZ_{m,n}\xU^m\xV^n
\,,\nn\\
\gMZ &=& \tgMZ + \frac{1}{\SmoothedRCb^{4+2\MaxUVn}}\sum_{n,m\ge 0}^{n+m\le\MaxUVn}\gMZ_{m,n}\xU^m\xV^n
\,,\nn\\
\hMZ &=& \thMZ + \frac{\xV}{\SmoothedRCb^{5+2\MaxUVn}}\sum_{n,m\ge 0}^{n+m\le\MaxUVn-1}\hMZ_{m,n}\xU^m\xV^n
\,,\nn\\
\qMZ &=& \tqMZ + \frac{\xV}{\SmoothedRCb^{4+2\MaxUVn}}\sum_{n,m\ge 0}^{n+m\le\MaxUVn-1}\qMZ_{m,n}\xU^m\xV^n
\,,\nn\\
\vMZ &=& \tvMZ + \frac{\xV}{\SmoothedRCb^{4+2\MaxUVn}}\sum_{n,m\ge 0}^{n+m\le\MaxUVn-1}\vMZ_{m,n}\xU^m\xV^n
\,,
\nn
\eear
where $\fMZ_{m,n}$, $\gMZ_{m,n}$, $\hMZ_{m,n}$, $\qMZ_{m,n}$, $\vMZ_{m,n}$ are constant coefficients to be determined.
These expressions are designed to preserve the boundary condition \eqref{eqn:qvhTip}, as well as the asymptotic behavior for large $\xU$ and $\xV$.
We then find the coefficients $\fMZ_{m,n}$, $\gMZ_{m,n}$, $\hMZ_{m,n}$, $\qMZ_{m,n}$, $\vMZ_{m,n}$ that minimize $\Eex$, using the Newton-Raphson method for given $\Cb$, and finally we optimize $\Cb$.
For example, we find for the dimensionless coefficient $\Vph\Ca^2=1$ and $\MaxUVn=22$ that the optimal $\Cb$ is $2.6\Ca$.
We define the energy density
\be
\Uex\equiv
\frac{1}{2}\tr\left\lbrack
(D_r\ResPhi)^2
+(D_1\ResPhi)^2
+(D_2\ResPhi)^2
\right\rbrack
+\frac{1}{2}r^2\tr\left\lbrack
F_{12}^2
+F_{1r}^2
+F_{2r}^2
\right\rbrack
\,,
\ee
for the exact solution we have
\be
\Uex=
\Ubps \equiv
r\tr\left\lbrack
F_{12}D_r\ResPhi
+F_{2r}D_1\ResPhi
+F_{r1}D_2\ResPhi
\right\rbrack\,.
\ee
The total energy is then 
$$
\Ebps = \frac{1}{4}\int\frac{1}{\xV}\Ubps d\xV d\xU\,.
$$
We present in \figref{fig:EnergyDensityAndAbsPhi} our\footnote{The graph was drawn by {\tt Mathematica}, Version 9.0, (Wolfram Research, Inc.).} numerical results for $\Theta\equiv\Uex/\xV$ as well as for the gauge invariant absolute value of the scalar field
$$
|\ResPhi|\equiv(\ResPhi^a\ResPhi^a)^{1/2} = \sqrt{\xU(\fMZ+\hMZ)^2+\gMZ^2}\,.
$$
The results are for $\Vph\Ca^2=1$, and it is interesting to note that for such a relatively small value of $\Vph\Ca^2$, the core of the soliton (where $|\ResPhi|=0$) is at $r\approx 2.9$ ($\xV=8.46$ in the graph of \figref{fig:EnergyDensityAndAbsPhi}), which is far from $\Ca=1$.

\end{appendix}


\bibliographystyle{my-h-elsevier}

\end{document}